\documentclass{jfm}
\usepackage{epstopdf, epsfig}

\usepackage{amssymb}

\usepackage[hyphens]{url}
\usepackage{adjustbox}
\usepackage{lipsum}
\usepackage{graphicx}
\usepackage[utf8]{inputenc}
\usepackage{bm}
\usepackage{latexsym} 
\usepackage{color}
\usepackage{amsmath}
\usepackage{amsfonts}
\usepackage{amssymb}
\usepackage{supertabular}
\usepackage{multirow}
\usepackage{setspace}
\usepackage{picture}
\usepackage[percent]{overpic}
\usepackage{textcomp}
\usepackage{multicol}
\usepackage{booktabs}
\usepackage{tabularx}
\usepackage{units}
\usepackage{tablefootnote}
\usepackage[USenglish]{babel}
\usepackage{etoolbox}
\usepackage{blindtext}
\usepackage{pifont}
\usepackage{diagbox}
\usepackage{appendix}
\usepackage{threeparttable}
\usepackage{lscape}
\usepackage{subcaption}
\usepackage{caption}
\usepackage{array}
\usepackage{hyperref}
\usepackage{floatrow}

\begin{document}

\graphicspath{{./plots/}}

\newtheorem{lemma}{Lemma}
\newtheorem{corollary}{Corollary}

\shorttitle{Flutter Instability in an Internal Flow Energy Harvester}
\shortauthor{L. P. Tosi, B. Dorschner and T. Colonius}

\title{Flutter Instability in an Internal Flow Energy Harvester}

\author{L. P. Tosi \aff{1}
  \corresp{\email{ltos@caltech.edu}},
  B. Dorschner \aff{1}
 \and T. Colonius \aff{1}}

\affiliation{\aff{1}Division of Engineering and Applied Science \\California Institute of Technology, Pasadena, CA 91125}

\maketitle

\begin{abstract}
Vibration-based flow energy harvesting enables robust, in-situ energy extraction for low-power applications, such as distributed sensor networks. 
Fluid-structure instabilities dictate a harvester's viability since the structural response to the flow determines its power output.  
Previous work on a flextensional-based flow energy harvester demonstrated that an elastic member within a converging-diverging channel is susceptible to the aeroelastic flutter.
This work explores the mechanism driving flutter through experiments and simulations.  
A model is then developed based on channel flow-rate modulation and considering the effects of both normal and spanwise flow confinement on the instability. Linear stability analysis of the model replicates flutter onset, critical frequency, and mode shapes observed in experiments.  The model suggests that flow modulation through the channel throat is the principal mechanism for the fluid-induced vibration.
The generalized  model presented can serve as the foundation of design parameter exploration for energy harvesters, perhaps leading to more powerful devices in the future, but also to other similar flow geometries where the flutter instability arises in an elastic member within a narrow flow passage. 
\end{abstract}

\begin{keywords}
flutter, fluid-structure interaction, fluid-induced vibration, leakage-flow instability, cantilever in channel flow, flow energy harvester
\end{keywords}

\section{Introduction}

{\it In-situ} energy harvesting in pipes could power sensors and actuators that improve efficiency and/or production in oil wells \citep{Sharma2002, rester1999,Wood2013} and irrigation systems \citep{Zhou2012, Bastiaanssen2000}.  They require power levels $O(10^{-3}-10^1)$ W dependent on data rates and system architecture \citep{Moschitta2014}. 
While turbines can achieve such power levels \citep{Tong2010}, they are susceptible to wear and friction within their bearing assemblies, and not favorable alternatives for decades of use without maintenance \citep{Guo2009,Doll2010,Tong2010}.
Hydrokinetic energy harvesters based on flow-induced vibration (FIV) avoid the use of bearings or gears altogether, shifting the primary failure to structural fatigue. 
FIV devices with power outputs of $O(10^{-7} - 10^{4})$ W \citep{Bernitsas2008, Zhu2011} and decades of maintenance-free operation may be feasible if designed to maintain internal stresses within material fatigue limits. 

Flow-induced vibration in these devices is driven by fluid-structure interaction (FSI) instabilities that provide high amplitude, oscillatory forces on a responsive structure. 
One such instability is aeroelastic flutter, which relies on a positive feedback between the natural modes of a vibrating structure and aerodynamic forces.  
Flow-energy harvester devices developed by the authors and collaborators have focused on flutter instabilities in an internal flow geometry \citep{Sherrit2014, Lee2015, Lee2016}.  After a number of design iterations \citep{Sherrit2014,Lee2016}, a device employing a flextensional transducer (figure \ref{fig:Flowpath}), where a cantilever exposed to the flow is mounted on a flexure containing piezoelectric elements, was found to provide a number of advantages in terms of power output and longevity.  
This paper aims to analyze the associated FSI mechanisms taking place in this and similar devices; the resulting model could be used as a basis for design, geometry, and scaling of devices in the future.

The stability of an elastic member within a constant channel, or as part of the channel, has been studied analytically, via reduced models, and numerically for many decades \citep{Johansson1959,Miller1960,Inada1988,Inada1990,Nagakura1991,Gurugubelli2014,Cisonni2017}.  A number of other applications fall under this canonical  problem, including wind instruments \citep{Sommerfeldt1988,Backus1963}, human snoring \citep{Balint2005,Tetlow2009} or vocalization \citep{Tian2014}, enhanced heat transfer systems \citep{Shoele2015,Hidalgo2015}. 
Modeling the structure displacement, velocities, and fluid forces approximated via simplified equations of motion appear as early as the 1960's \citep{Miller1960,Johansson1959}, where the divergence instability in channels within nuclear reactor cooling systems is addressed.  
More recent work has taken an inviscid approach to understanding the onset of flutter in a symmetric channel \citep{Guo2000}. 
Other similar formulations include a vortex sheet model to calculate flutter boundary \citep{Alben2015}, and a plane wake vortex sheet model in unconfined flows \citep{Alben2008}. The latter was extended to asymmetric channel flow to account for the effects of channel confinement \citep{Shoele2015}. 
Viscous formulations that account for the flow rate modulation due to change in the channel geometry were also devised and progressed at around the same time \citep{Nagakura1991,Paidoussis2003,Wu2005}.  
These employed fluid force terms applied to an elastic beam in channel flow that originally had been devised for %
for rigid plates in converging or diverging channels \citep{Inada1988}. 
This framework was also extended to cylindrical constant channels \citep{Fujita1999,Fujita2001,Fujita2007}.  
More recently, methods that include viscosity in compressible and incompressible potential flow have been devised to interrogate the confined beam flutter stability problem, also considering the addition of beam tension \citep{Jaiman2014}.  

The two-dimensional viscous flow problem has also been explored numerically to define the flutter boundary dependence on the fluid-to-structure mass ratio and Reynolds number for a relatively flexible cantilever (and a two-cantilever system) within a confined channel \citep{Gurugubelli2014,Gurugubelli2019}, as well as its dependence on throat-to-beam length ratio and Reynolds number \citep{Cisonni2017}.  
Two-dimensional channel geometries with small throat-to-beam length ratios have also been modelled and results tested against numerical simulations for a constant channel over a range of Reynolds numbers, geometry and material parameters \citep{Tosi2018}. 
This model, devised by the authors, has failed to correctly predict experimentally observed flutter onset of the flextensional device. 
In the present paper, we extend the model formulation to account for the three-dimensional effect from lateral beam confinement.  
The importance of considering the full geometry becomes apparent, for example, when comparing two-dimensional flag flutter to that of flutter in spanwise confined flags \citep{Doare2011theo,Doare2011expr}. 

The remainder of the paper is structured as follows.  
We define the details of the flextensional flow energy harvester in section \ref{sec:FEHDesign}. 
In section \ref{sec:experiments}, we present experiments that first characterize the flextensional properties as a function of set-screw torque, then the flutter boundary as a function of flow rate.  
A numerical simulation of the system, presented in section \ref{sec:NumericalSimulations}, is used to obtain insights into three-dimensional aspects of the flow field and the relevant fluid-structure mechanisms driving flutter.  
Those insights guide the model derived in section \ref{sec:Model}, which is based on the modulation of the channel flow rate by the beam displacement and velocity, and predicts the flutter instability on-set flow rate, frequency, and mode shapes near the plane-asymmetric diffuser separation angle of $\approx 7^{\circ}$.

\section{Flextensional flow-energy harvester}
\label{sec:FEHDesign} 
We begin by defining the design of the energy harvester that is tested experimentally, then simulated and modeled in subsequent sections.  

\subsection{Device description} 
\begin{figure}[H]
    \centering
    \begin{subfigure}[b]{0.7\textwidth}
        \centering
   	    \includegraphics[width=1\textwidth]{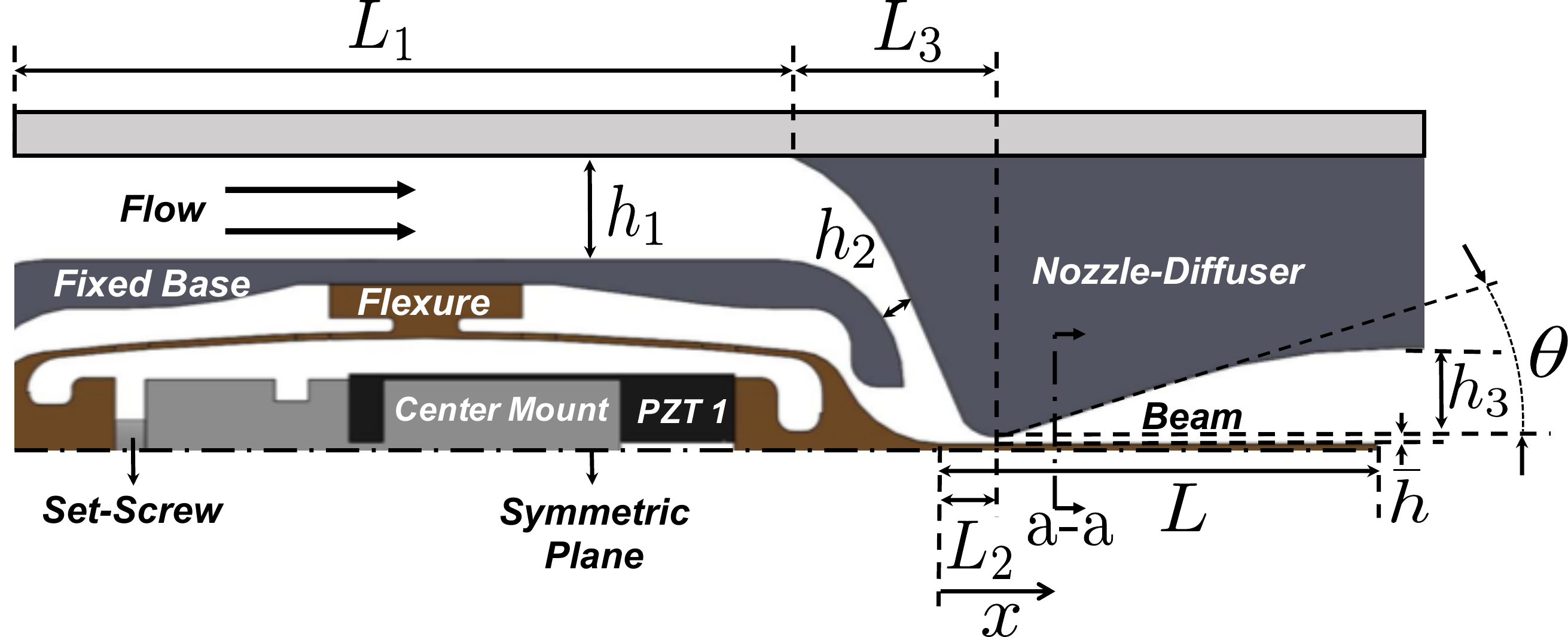}
        \caption{Axial cross-section of flow geometry illustrating flow path.} \label{fig:flowpatha}
    \end{subfigure}%
    ~ 
    \begin{subfigure}[b]{0.2\textwidth}
        \centering
   	    \includegraphics[width=1\textwidth]{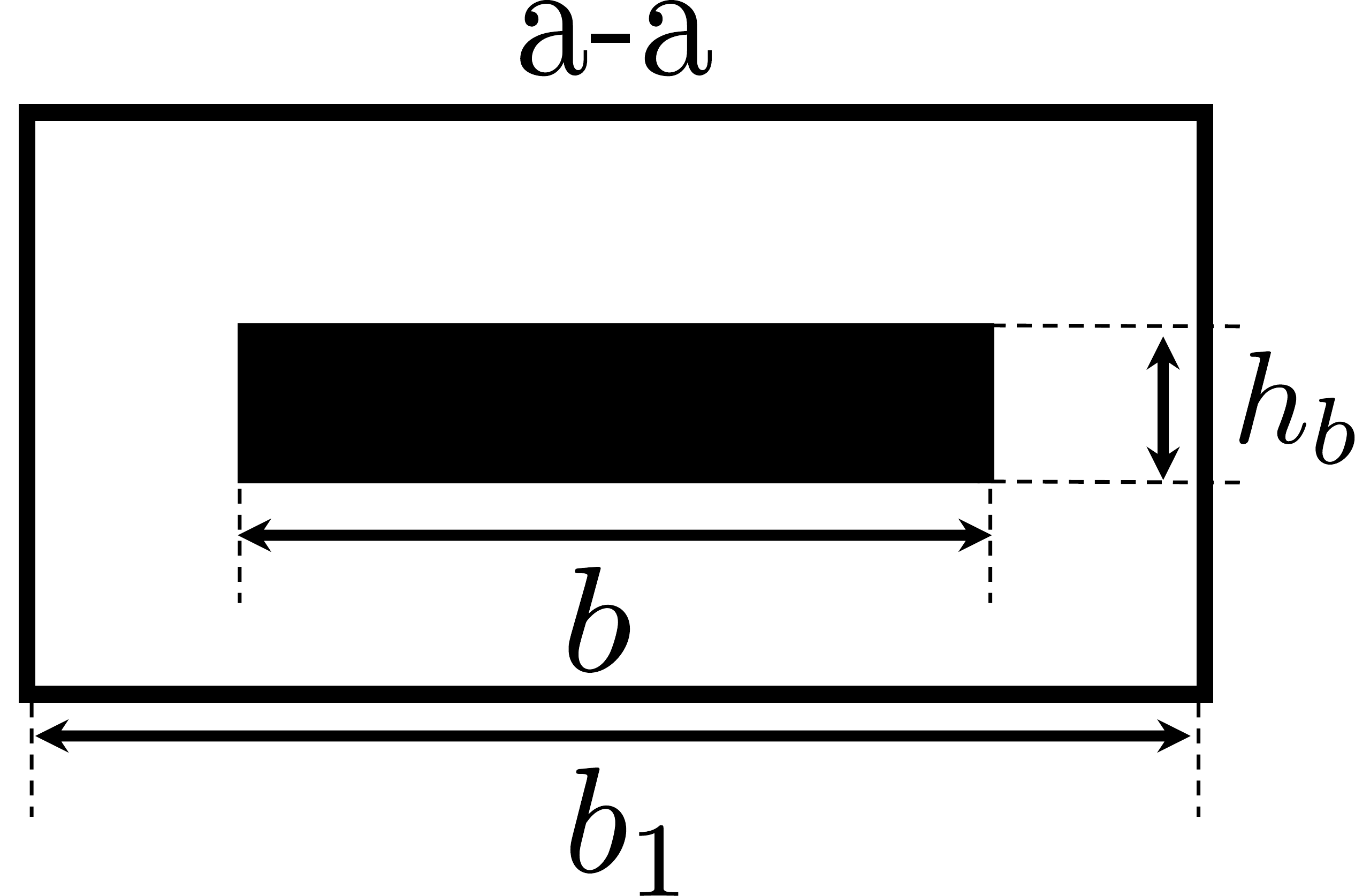}
        \caption{Cut section \textbf{a-a}} 
    \end{subfigure}
    \caption{Illustration of flow path and relevant geometry. Values and units are listed in table \ref{tab:FlowPathGeom}.}  
    \label{fig:Flowpath}
\end{figure}

A flow-energy harvester based on flextensional actuators (figure \ref{fig:Flowpath}) converts the motion of a cantilever excited by the flow into electricity via piezoelectric crystals \citep{Lee2015}.  
Flextensional structures are designed as actuators that convert compressive piezoelectric stresses to flexural displacements; here the device is used in reverse as a transducer to generate compressive piezoelectric stresses from flexure displacements. This produces more energy for the same displacement as compared to a piezoelectric bimorph transducer \citep{Sherrit2014,Sherrit2015}.

As seen in figure \ref{fig:Flowpath}, the flexure supports two piezoelectric stacks (PZT 1 shown, with a symmetric PZT 2) through a center mount that is attached  to the fixed base with a set-screw.  Torque applied to the set-screw pre-stresses the stacks, and changes the dynamical (and static) properties of the flexure.  
By adding or removing torque to the set-screw $\tau_S$, the effective stiffness $k_0$, damping $c_0$, and mass $m_0$ of the flexure can be altered. The measurement of flexure properties is discussed in section \ref{sec:flextureMeasurements}. 

The device works on the premise that flow can interact and excite the beam structure.  
In our experiments, the flow path begins from a round pipe inlet into the test section. The flow impinges on the fixed base and is directed onto the top and bottom paths as illustrated in figure \ref{fig:Flowpath}. 
The beam is centered along the channel, such that the flow path is symmetric.      
The figure illustrates the top channel, with dimensions listed in table \ref{tab:FlowPathGeom} in the appendix.  The flow is converging for $L_2 \approx 0.1 L$ along $x$ until it bypasses the constriction at the throat $\bar{h}$, and expands in a planar, $\theta = 19^{\circ}$ diffuser for $\approx 0.7 L$. In the remaining $0.2 L$, the diffuser tapers off into $<1^{\circ}$ exit at the end of the beam.  The total expansion is $\approx$15:1 from $\bar{h}$.  Our flowing experiments are carried out in air.

The flexure and the beam are made of a single aluminum stock, and comprise the moving structure.  
The fixed base is fastened with screws to both the test section and the flexure.  
The flexure behaves like a translational spring that transfers motion from the beam surface normal direction into compression and expansion of the piezoelectric stacks.  
The pre-stress from the set-screw and center mount ensure that the piezoelectric elements are always in compression:
as the flexure moves above the channel centerline, the bottom stack is compressed, and the top stack pre-stress is released, although maintained positive, and vice-versa when the beam moves below the channel centerline. The up and down motion gives rise to two voltage signals that are $180^{\circ}$ out of phase.
Vacuum grease and rubber inserts are used to seal and restrain the flow path to that shown in figure \ref{fig:Flowpath}.  
An electrical fitting is used to connect the piezoelectric stacks to the data acquisition card on the outside of the test section.

The piezoelectric stacks are composed of multiple thin, alternately poled, piezoelectric layers ``stacked'', or mechanically connected in series and electrically in parallel. 
They operate in what is known as the 33 mode, where the applied force is parallel to the poling direction.  
When a resistor is placed in parallel with the stack, its response to a step input force is that of an RC circuit with the capacitor having an initial voltage equivalent to the open circuit step-force voltage.  
The voltage $V(t)$ is measured across the resistor $R_e$ is given by
$V(t) = V_{in}\exp\left(-t/R_e C^*_p\right). $
If the time constant $\tau = R_e C^*_p$ is large enough, the system will act as a low-pass filter and any oscillating voltage upstream of the resistor (opposite to ground) at a frequency $f_{res}$ that satisfies,
\begin{equation} \label{eq:CondResitor}
\omega_{\mathrm{res}} \gg \frac{1}{\tau } = \omega_c,
\end{equation}
will not pass through the resistor. 
Hence, the voltage output will be measured as if the system was an open circuit. 
We implement this circuit and choose an $R_e$ large enough such that the resonances of the structure satisfy condition \ref{eq:CondResitor}.  
Specifically, we expect the stacks to act as strain gauges for high enough frequencies, where the output oscillating voltage is proportional to the flextensional displacement.

The combination of the flow path, the structure, the piezoelectric elements, and the electronics comprise the flow-energy harvester design.  In our current work, we are particularly interested in the coupling between the flextensional and beam structure to the flow path in the channel.

\subsection{Flextensional and beam parameters} \label{sec:flextureMeasurements}

\begin{figure}[h!]
    \centering
   \includegraphics[width=.65\textwidth]{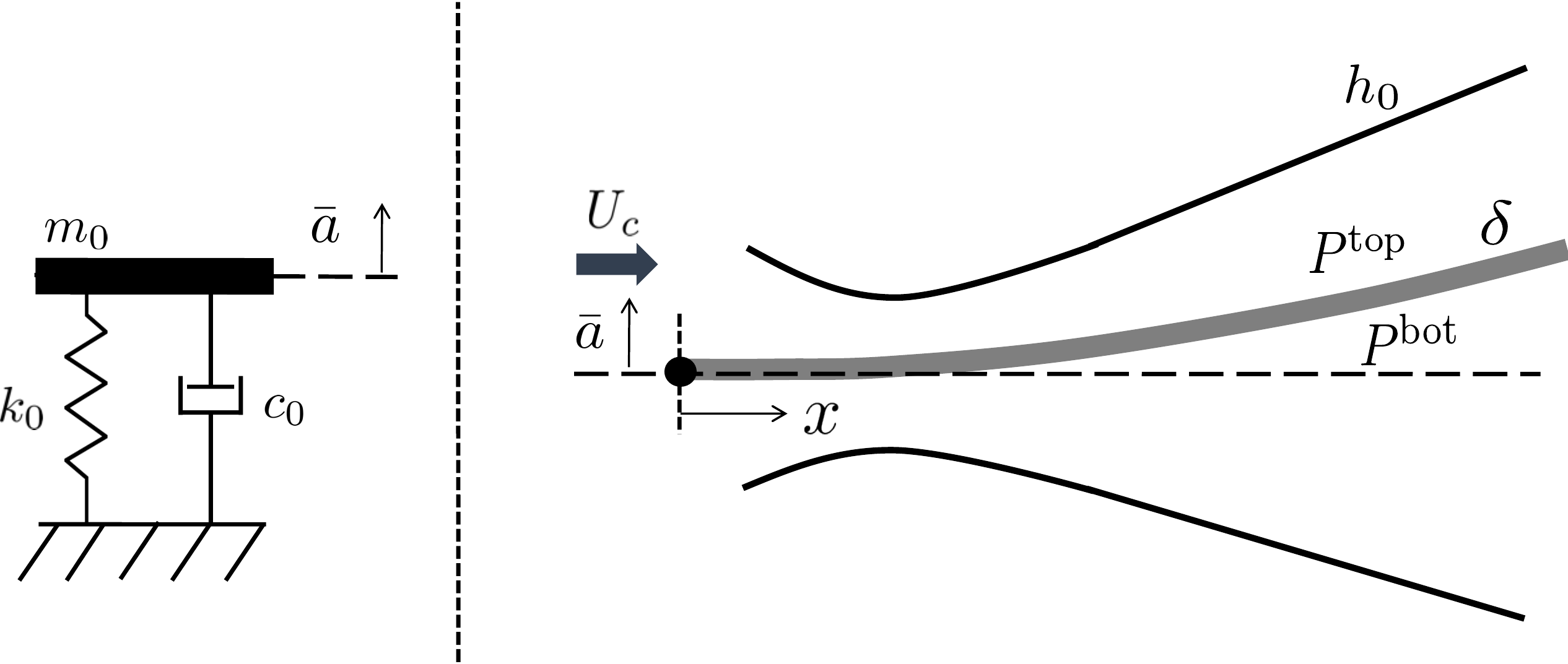}
    \caption{Illustration cantilever beam in a converging-diverging channel geometry (right) with simple harmonic boundary condition (left).}  
     \label{fig:NozzleDiffuserGeometry}
\end{figure}

Figure \ref{fig:NozzleDiffuserGeometry} divides the flow-energy harvester into two distinct parts: the flextensional dynamics on the left, which provides the boundary condition for the flow-driven beam dynamics on the right.      

\begin{table}[hbt!] 
\centering
\caption{Table of fluid-structure dimensional parameters.} \label{tab:dimensionpars}
\begin{tabular}{ c  c  c  }  \hline 
\textbf{Variable} & \textbf{Description} & \textbf{Dimension} \\ 
\hline
 $\delta$ & beam displacement & $l$ \\ 
 $x$ & beam length coordinate & $l$\\   
 $t$ & time & $t$\\  
 $p$ & pressure & $m*l^{-1}*t^{-2}$\\  
 $U_c$ & characteristic velocity & $l*t^{-1}$ \\  	
 $k_0$ & flexure stiffness & $m*t^{-2}$\\  	
 $c_0$ & flexure damping & $m*t^{-1}$\\	
 $m_0$ & flexure mass & $m$\\  				
 $L$ & beam length & $l$\\  			
 $h_b$ & beam thickness & $l$\\  		
 $b$ & beam width & $l$\\
 $\bar{h}$ & throat height & $l$\\
 $\rho_f$ & fluid density & $m*l^{3}$ \\
 $\mu_f$ & fluid viscosity & $m*l^{-1}*t^{-1}$ \\
 $\rho_s$ & beam density & $m*l^{-3}$\\  		
 $E$ & Young's modulus & $m*l^{-1}*t^{-2}$\\  	
 \hline					
 \end{tabular} 
\end{table}

Informed by finite element results of the flexture \citep{Tosi2019thesis}, a damped harmonic oscillator can be used to capture the flextensional fundamental mode dynamics.  In particular, 
\begin{equation} \label{eq:StructureBC_MovingEquation}
\frac{m_0}{b} \ddot{\bar{a}} + \frac{c_0}{b}\dot{\bar{a}} + \frac{k_0}{b} \bar{a} = f_r,
\end{equation}
where $\frac{m_0}{b}$, $\frac{c_0}{b}$, and $\frac{k_0}{b}$ are the flexure mass, damping, and stiffness constants \textit{per unit span}, and $\bar{a}$ is the displacement of the flextensional boundary.  
The force $f_r$ is equivalent to the total force per unit span acting on the flexure interface to the cantilever. It can be defined as the integrated pressure difference between top and bottom channels over the beam length, 
\begin{equation} \label{eq:IntegrateBoundaryForce}
    f_r = \int_0^L \Delta P dx.
\end{equation}
Here $\Delta P = P^{\mathrm{bot}} - P^{\mathrm{top}}$, and $P^{\mathrm{bot}}(x,t)$ and $P^{\mathrm{top}}(x,t)$ are the pressures acting on the bottom and top of the beam, per superscript.  
Values for $m_0$, $c_0$ and $k_0$ are inferred from measurements of the actual device in section~\ref{sec:FlexureCharacterization}.

The goal of our work is to understand the flutter instability when the system is near zero displacement.  
Hence, to describe the beam motion in transverse vibration, we consider the \emph{undamped}, Euler-Bernoulli beam equation per unit span \citep{Inman2008},
\begin{equation} \label{eq:EulerBernoulliBeam}
\rho_s h_b \frac{\partial^2}{\partial t^2} \delta(x,t) + 
\frac{\partial^2}{ \partial x^2} \left( \frac{E I}{b} \frac{\partial^2 }{\partial x^2} \delta(x,t)  \right) = 
\Delta P,
\end{equation}
where the $I$ is the area moment of inertia in equation \ref{eq:MomofIntertiaI}.
The beam is moving at its leading edge and free at its trailing edge, so the boundary conditions are
\begin{equation} \label{eq:StructureBC_ClampedFree}
\delta(0,t) = \bar{a}, \ \frac{\partial }{ \partial x} \delta(0,t) = 0, \ 
\frac{\partial^2 }{ \partial x^2} \delta(L,t)= 0, \ 
\frac{\partial^3 }{ \partial x^3} \delta(L,t)= 0.
\end{equation}
Equation \ref{eq:EulerBernoulliBeam} and data \ref{eq:StructureBC_ClampedFree} assume that the beam is thin relative to its length ($L>>h_b$); 
that the rotational inertia is negligible; 
and that beam extension and shear displacement are negligible when compared with the transverse displacement (the beam is inextensible). It follows that flow shear stresses do not impact the motion of the elastic member. 
Furthermore, the amplitude of oscillation is small relative to the beam length ($\left|\left| \delta \right|\right|_{\infty} \ll L$) such that $x$ is equivalent to the Lagrangian coordinate of the beam.
These simplifications are consistent with our system near the zero displacement equilibrium. 

Though equation \ref{eq:EulerBernoulliBeam} is undamped, external damping due to the fluid (i.e. Rayleigh damping) is accounted for within the pressure terms on its right-hand-side.  
Internal damping (i.e. internal to the solid), however, is not accounted within our equations of motion.  
A range of internal material damping formulations exist for a beam \citep{Banks1991}.  
Typically a strain-rate proportional form (Kelvin-Voigt) is used but time- or spatial- hysteretic formulations have also been deemed appropriate for certain materials and configurations.
Yet material coefficients corresponding to any of the aforementioned formulations are difficult to obtain and require specialized experimental equipment. %
The effect of neglecting internal material damping results in underestimating the energy dissipated by the \emph{beam} structure, leaving modes associated with the flexible beam specially susceptible to instabilities.  However, the experimental results in section \ref{sec:experimentalResults} show that the flextensional rigid-body motion at the cantilever base is responsible for the flutter bifurcation observed, rendering the flutter stability boundary observed largely independent of the beam modes.  However, this limitation of the present model should be noted if it is generalized to designs that rely on significant beam bending.
Results are verified and discussed in more detail within the numerical simulation and modeling in sections \ref{sec:3DDNSFEH} and \ref{sec:ModelResults}, respectively.   
  
In considering the flow separately in the top and bottom channels, we write the geometrical constraint,
 \begin{equation} \label{eq:BeamShapeConstraint}
\delta(x,t) = \delta^{\mathrm{top}}(x,t) = -\delta^{\mathrm{bot}}(x,t).
\end{equation}

Furthermore, we define $\delta = L \delta^*$, $t = \left( L/U_c \right) t^*$, $f_r = \rho_f U_c^2 L f_r^*$, and $\Delta P = \rho_f U_c^2 \Delta P^*$ where $*$ superscript represents non-dimensional quantities.  The non-dimensionalization of equations \ref{eq:StructureBC_MovingEquation} and \ref{eq:EulerBernoulliBeam} as such yields the fluid-structure non-dimensional groups in table \ref{tab:NDpars_Moving}. 
A gap-to-length ratio, $\hat{h} = \bar{h}/L$ arises in dimensional analysis when we define a second length-scale $\bar{h}$ associated with the channel geometry $h_0 = \bar{h}h_0^*$, as does the beam width-to-length ratio, $\hat{b} = b/L$.  
The relevant fluid-only dimensional groups depend on the form of the pressure term, as related to the velocity field.

This model is intended to describe the initial, small displacement behavior of the fluid-structure system as a function of fluid/structure parameters, and is appropriate for the stability analyses that follow in section \ref{sec:Model}.    

\begin{table}[hbt!] 
\centering
\caption{Table fluid-structure non-dimensional parameters for flextensional boundary and cantilevered beam.} \label{tab:NDpars_Moving}
\begin{tabular}{ c  c  c  }  \hline 
\textbf{Variable} & \textbf{Expression} &\textbf{Description}  \\ 
\hline
 $\hat{m}_{\mathrm{bc}}$ & $\frac{1 }{ \rho_f L^2 } \frac{m_0}{b} $ & boundary mass ratio \\ 
 $\hat{k}_{\mathrm{bc}}$ & $ \frac{ 1 }{\rho_f U_c} \frac{k_0}{b}$ & boundary stiffness ratio \\ 
 $\hat{c}_{\mathrm{bc}} $ & $\frac{1}{\rho_f U_c L} \frac{c_0}{b}$ & boundary damping ratio   \\ 
 $\hat{m}$ & $\frac{\rho_s h_b }{ \rho_f L }$ & beam mass ratio \\  
 $\hat{k}$ & $\frac{ E }{\rho_f U_c^2} \frac{I}{b L^3}$ & beam stiffness ratio \\\hline

\end{tabular} 
\end{table}

\section{Experiments} \label{sec:experiments}
From the design and parameter definitions in the previous section, we first experimentally measure the flextensional boundary parameters values in section \ref{sec:FlexureCharacterization}, then quantify the dynamics of the device as a function of flow rate in section \ref{sec:FLowExpr}.  
Rather than a complete experimental characterization of the dynamics however, three settings are selected (and measured) instead, i.e. a \emph{baseline, high, and low} set-screw torque values (shown in table \ref{tab:FlexureBoundaryProperties}), to span a range of flextensional mass, stiffness, and damping properties for comparison with the numerical simulation and modeling results in sections \ref{sec:NumericalSimulations} and \ref{sec:Model}, respectively.

\subsection{Flexure characterization} \label{sec:FlexureCharacterization}
Two experiments are carried out to quantify $m_0$, $c_0$, and $k_0$, all in still air at standard pressure and temperature (STP). First, the flexure stiffness $k_{0}$ is characterized through a static measurement of force $F_a$ for displacement $\bar{a}$, 
\begin{equation} \label{eq:StiffBCTerm}
k_0 = \frac{b f_r}{\bar{a}} = \frac{F_a}{\bar{a}},  
\end{equation}
derived from the steady equation \ref{eq:StructureBC_MovingEquation}.  
The second experiment measures the voltage output of the PZT stacks when the flextensional fundamental mode is excited.  This is done with an impulse force at the $x=0$ location (figure \ref{fig:Flowpath}).  
With the damped resonant frequency $\omega$ and exponential decay rate $\zeta$ measured from PZT voltage outputs, the solution to the homogeneous equation \ref{eq:StructureBC_MovingEquation} ($f_r = 0$) is used to map the dynamic voltage response to parameter properties as, 
\begin{equation} \label{eq:MassBCTerm}
m_0 = \frac{k_0}{\omega^2 + \zeta^2},
\end{equation} 
\begin{equation} \label{eq:DampBCTerm} 
c_0 = 2 \zeta m_0.
\end{equation} 
Equations \ref{eq:StiffBCTerm},  \ref{eq:MassBCTerm}, and \ref{eq:DampBCTerm} allow us to map experimentally measured quantities $F_a$, $\zeta$, and $\omega$ onto $k_0$, $c_0$, and $m_0$.  
\begin{table}[H] 
\centering
\caption{Table of calculated mean flextensional properties.} \label{tab:FlexureBoundaryProperties}
\begin{tabular}{ c c c c }  \hline 
\textbf{Variable}	&	\textbf{Flex. 1}	&	\textbf{Flex. 2}	&	\textbf{Flex. 3}	 \\ \hline
$\tau_S$ [Nm]   &   1.2     &   2       &   0.8     \\
$k_0$ [N/m]	    &	3.73E4	&	4.12E4	&   2.16E4	\\
$m_0$ [kg]	    &	0.0274	&	0.0204	&	0.0366 	\\
$c_0$ [kg/s]	&	0.135	&	0.314	&	0.281 	\\ \hline
\end{tabular}
\end{table} 

\subsection{Flow experiments}  \label{sec:FLowExpr}
Next, the flextensional energy harvester is tested in flowing conditions to quantify the critical properties at the flutter instability, which encompass those properties at or near the quasi-stable point where the systems transitions from the stable equilibrium into flutter and vice-versa.  
Parameters are systematically varied in two ways: first, the set-screw torque $\tau_S$ sets the structural parameters of the flextensional corresponding to the boundary conditions in figure \ref{fig:NozzleDiffuserGeometry} and to values in table \ref{tab:FlexureBoundaryProperties}.  
Second, for any of the three set-screw settings (i.e flex. 1, 2, and 3), an experiment is run where the flow rate is first increased \emph{past} the critical point, where the stable equilibrium to flutter transition is observed; then decreased \emph{past} the fold point where the flutter transition to a stable equilibrium is observed. This topology holds true for all three settings tested.  
The dynamics of the flowing system are assessed by measuring the voltage output from each piezoelectric stack, and by processing video images of the beam displacement.
In the \textit{increasing} flow rate branch, the critical point is described by the critical flow rate $Q_{\mathrm{cr}}$ where the system is not longer stable, and the critical frequency $f_{\mathrm{cr}}$, corresponding to the dominant oscillatory frequency at the nearest point $Q \geq Q_{\mathrm{cr}}$ where self-sustained oscillations can be observed in the measured outputs. 
The fold point in the \textit{decreasing} flow rate branch is characterized by the fold flow rate  $Q_{\mathrm{r}}$.
This data provides quantitative values by which we can compare numerical and analytical results in the subsequent sections.   

Two output data products are extracted from experiments: PZT voltages and beam displacement videos.  
The voltage data set is processed through peak extraction to obtain average amplitudes over the relevant time series, and fast Fourier transformed using Welch's method to obtain the signal frequency response corresponding to the highest peak.  No other processing technique or filtering was applied to the voltage signals, as the system is responding to oscillatory forcing that satisfies condition \ref{eq:CondResitor}. 
The video data set is decomposed and processed to characterize predominant vibration modes and their amplitudes.
From the video, the transverse displacement of a section of the elastic beam is measured using edge-detection through a Canny filter \citep{Canny1986} for the top and the bottom edges of the beam. 
The precision per-pixel is $\approx0.15$ mm or $\approx0.4$ \% of the beam length.
The extracted edges are averaged to estimate the beam center-line displacement.  The resulting space-time series is processed using the spectral proper orthogonal decomposition (SPOD), which allows the most energetic mode shapes at each frequency to be robustly extracted \citep{Schmidt2017b,Towne2018}.  Further details of the SPOD applied here are given in appendix \ref{sec:SPOD}.  In subsequent results, frequencies are labelled as $f$ with subscript 1 representing that of the highest power spectral density (PSD) value, and subsequent peaks following numerically.   

\subsubsection{Experimental Results} \label{sec:experimentalResults}

The video and voltage data sets are processed for the three flextensional settings over air flow rates ranging from 5 to 500 L/min.  
The dynamics observed as the flow rate increases are consistent for all three flextensional settings: small decaying beam displacement and voltage amplitude behavior prevails until a critical flow rate is reached. 
At the critical flow rate, both the beam and PZT voltage amplitudes significantly increase and display self-sustaining oscillations (i.e. limit-cycle).  

Figure \ref{fig:SPODProcessing_Flex1} shows a representative example for flextensional setting 1 when the system has reached the self-sustained oscillation regime. The data set is at flow rate $Q = 246$  L/min,  38 L/min above the flextensional setting 1 critical flow rate of 208 L/min.  The spectrum shows a clear peak at $f_1 = 197$ Hz, and the corresponding mode contains more than 99\% of the variance in the PSD of the beam displacement.  The phase diagram shows a limit-cycle behavior and the mode shape resembles the rigid body motion of the cantilever base, denoting excitation of the flexure itself.  
Given the predicted cantilever fundamental mode frequency of 346 Hz from classical Euler-Bernoulli beam theory for clamped-free cantilever boundary conditions (shown in appendix \ref{sec:EBbeamFreq}), we can reasonably associate the second peak at $f_2 = 341$ Hz to the beam fundamental mode. This is further validated from the mode shape shown: though the extracted transverse displacement data does not reach the cantilever base, the mode shape monotonically decreases as $x/L$ decreases without the appearance of a fixed node.  
The illustrated mode shape also contains over 99\% of the variance of the signal at $f_2$.  The phase diagram shows behavior typical of a lightly-damped resonance, where the mode amplitude and velocity are perturbed around their equilibrium points through sporadic forcing \citep{Schmidt2018}.  
The remaining peaks shown in the power spectrum plot are harmonics of $f_1$.  
Similar behavior was observed for flextensional settings 2 and 3 results, where $f_2 \approx 341$ Hz appears in all three settings.
\newcommand{\flexsettingnum}{1}
\begin{figure}[hbt!]
    \centering
    \begin{subfigure}{.7\textwidth}     %
        \centering
        \includegraphics[width=.8\textwidth]{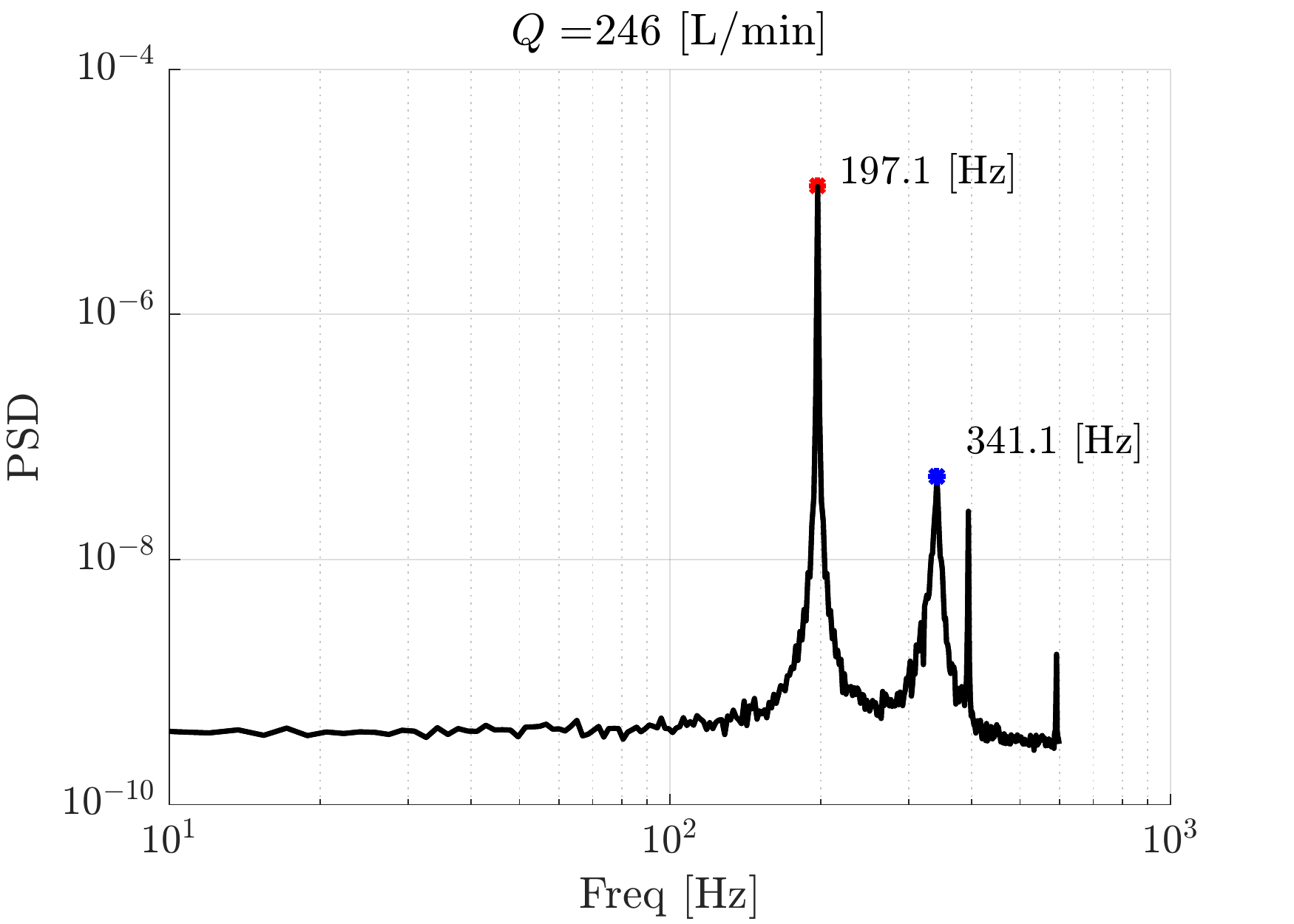}%
        \caption{Trace of cross-spectral density matrices at discrete frequencies. }
    \end{subfigure}%

    \begin{subfigure}{1\textwidth}     %
        \centering
        \includegraphics[width=.8\textwidth]{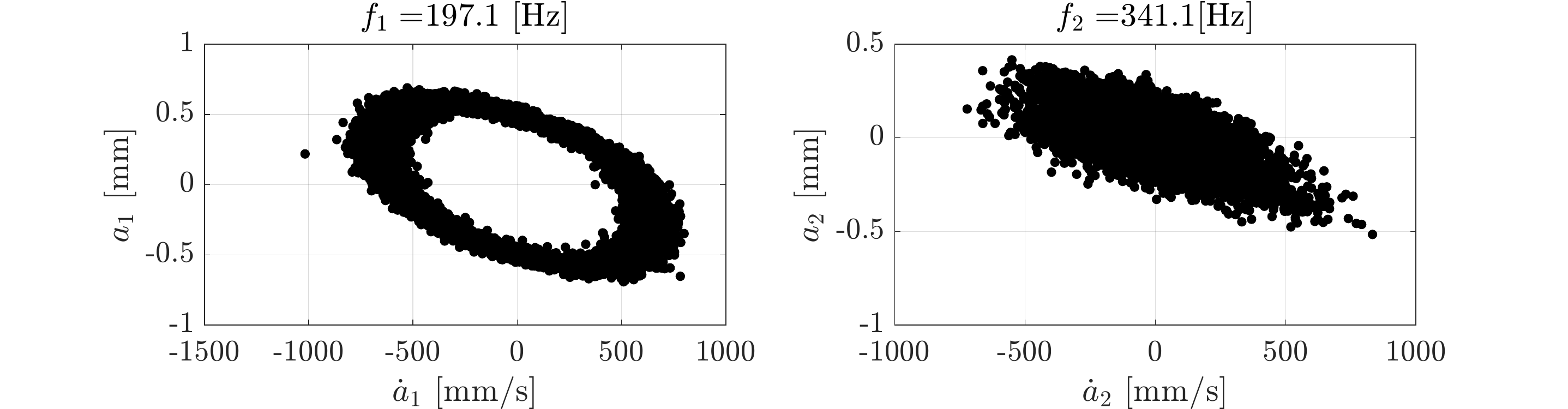}%
        \caption{Phase diagrams for mode 1 (left) and mode 2 (right).}
    \end{subfigure}%

    \begin{subfigure}{1\textwidth}     %
        \centering
        \includegraphics[width=.8\textwidth]{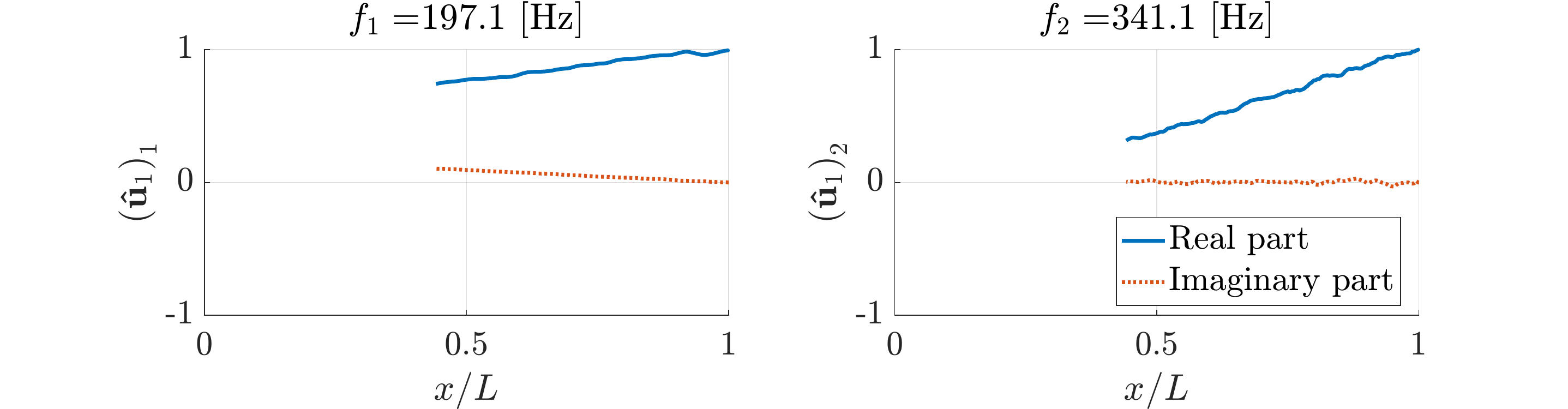}%
        \caption{SPOD mode shapes for mode 1 (left) and mode 2 (right).}
    \end{subfigure}%
    \caption{Representative SPOD data for flextensional setting  \flexsettingnum \ showing self-sustained oscillating regime of mode 1 and under-damped regime of mode 2. }  
    \label{fig:SPODProcessing_Flex1}
\end{figure}

\begin{figure}[H]
    \centering
    \begin{subfigure}[b]{0.31\textwidth}
        \centering
        \captionsetup{width=.8\linewidth}
   	    \includegraphics[width=1\textwidth]{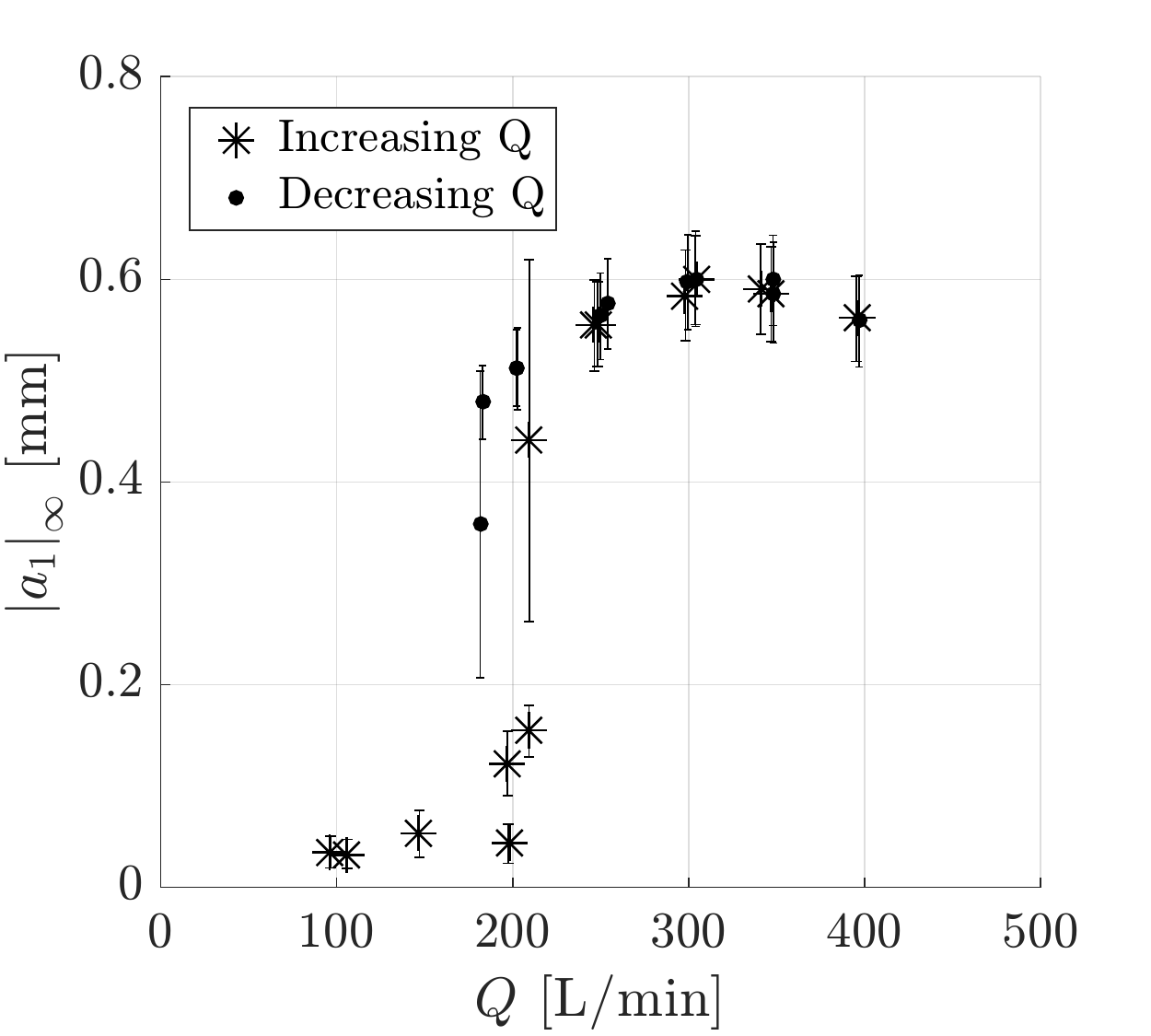}
        \caption{Flextensional 1.}  
    \end{subfigure}%
    ~ 
    \begin{subfigure}[b]{0.31\textwidth}
        \centering
        \captionsetup{width=.8\linewidth}
   	    \includegraphics[width=1\textwidth]{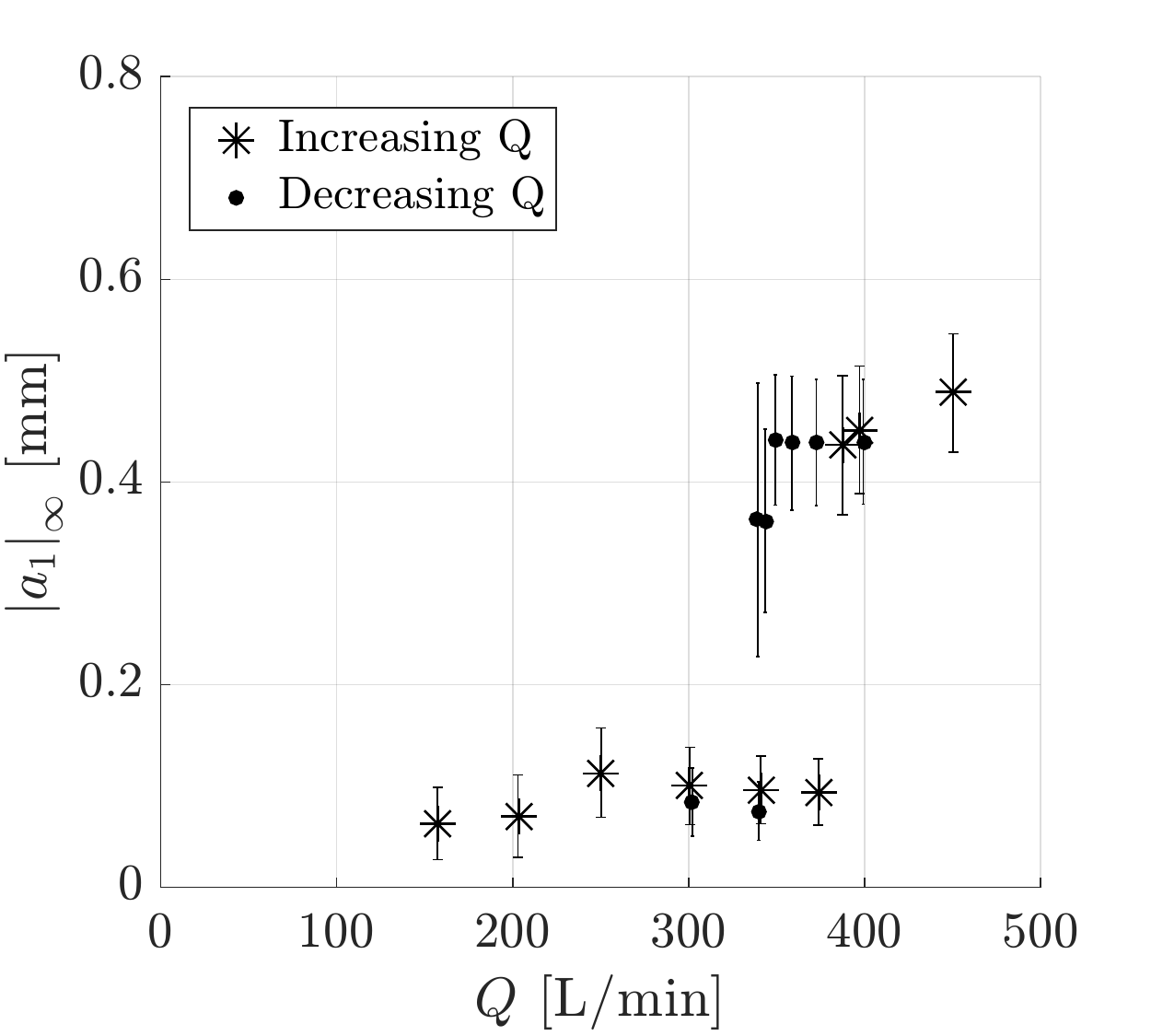} 
        \caption{Flextensional 2.}  
    \end{subfigure}
     ~ 
    \begin{subfigure}[b]{0.31\textwidth}
        \centering
        \captionsetup{width=.8\linewidth}
   	    \includegraphics[width=1\textwidth]{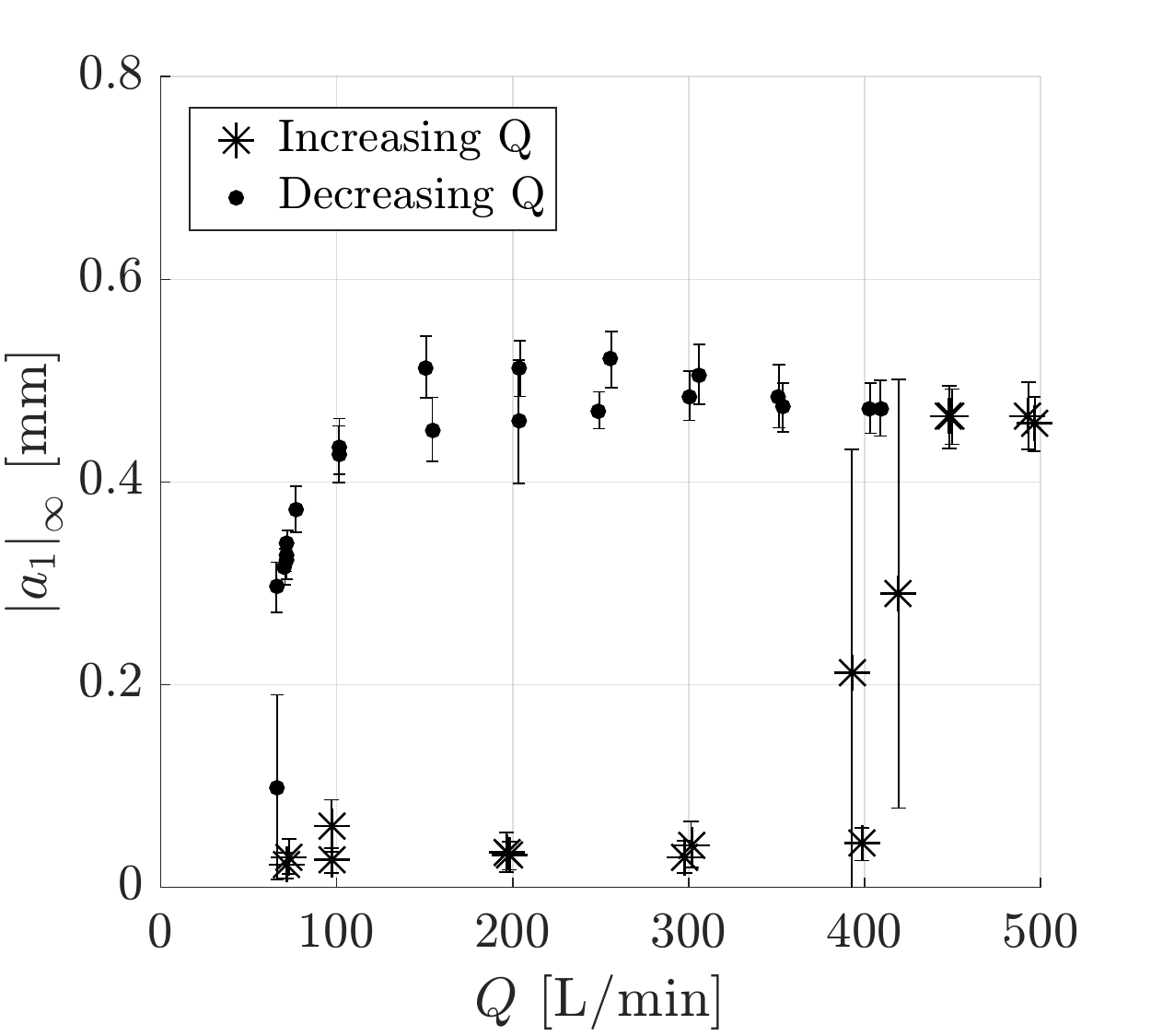} 
        \caption{Flextensional 3.}  
    \end{subfigure}
    
    \caption{Video data set showing mode 1 amplitude vs. flow rate for all flextensional settings. }  \label{fig:QvsDisp}
\end{figure}

\begin{figure}[H]
    \centering
    \begin{subfigure}[b]{0.31\textwidth}
        \centering
        \captionsetup{width=.8\linewidth}
   	    \includegraphics[width=1\textwidth]{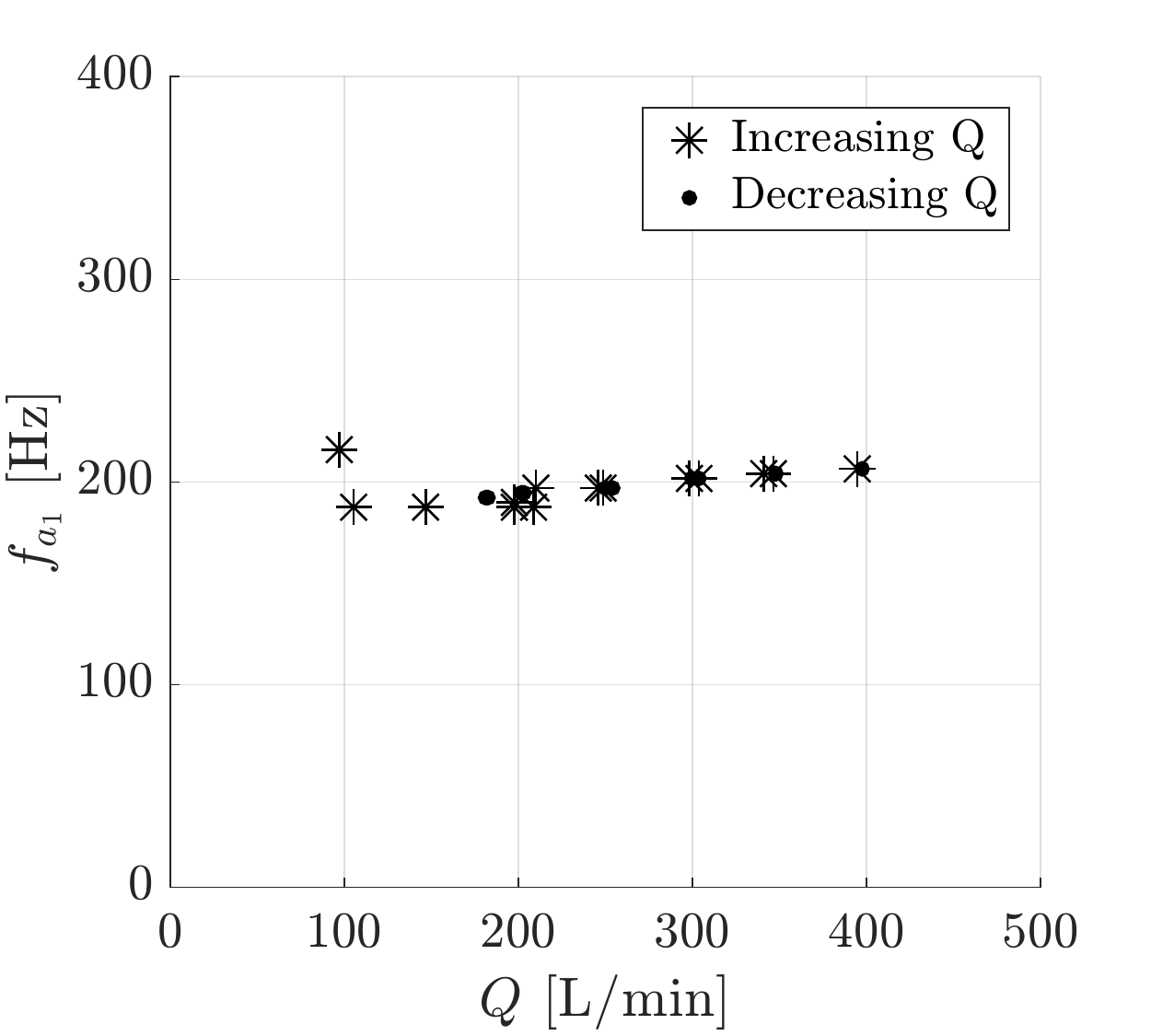}
        \caption{Flextensional 1.}  
    \end{subfigure}%
    ~ 
    \begin{subfigure}[b]{0.31\textwidth}
        \centering
        \captionsetup{width=.8\linewidth}
   	    \includegraphics[width=1\textwidth]{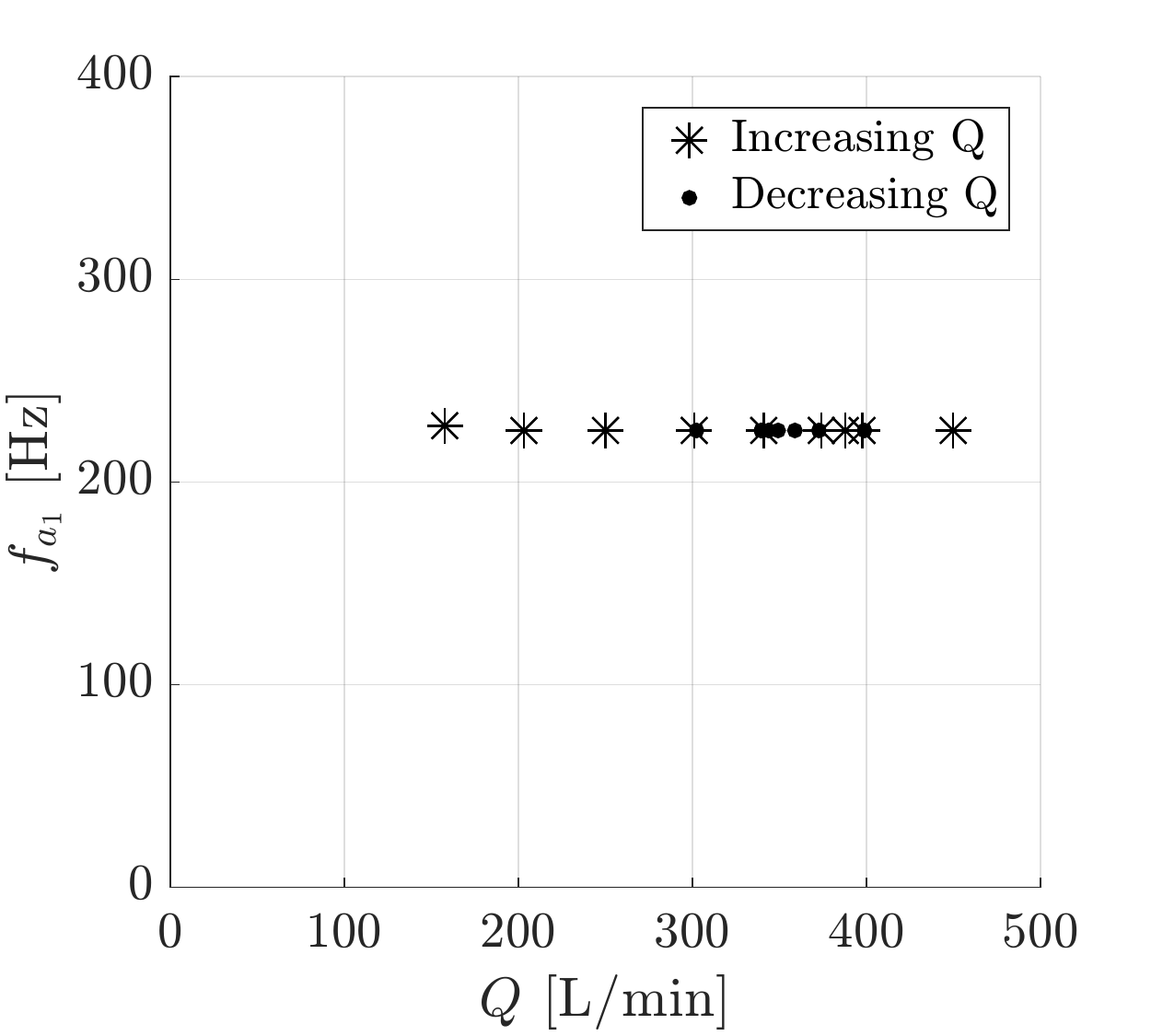} 
        \caption{Flextensional 2.}  
    \end{subfigure}
     ~ 
    \begin{subfigure}[b]{0.31\textwidth}
        \centering
        \captionsetup{width=.8\linewidth}
   	    \includegraphics[width=1\textwidth]{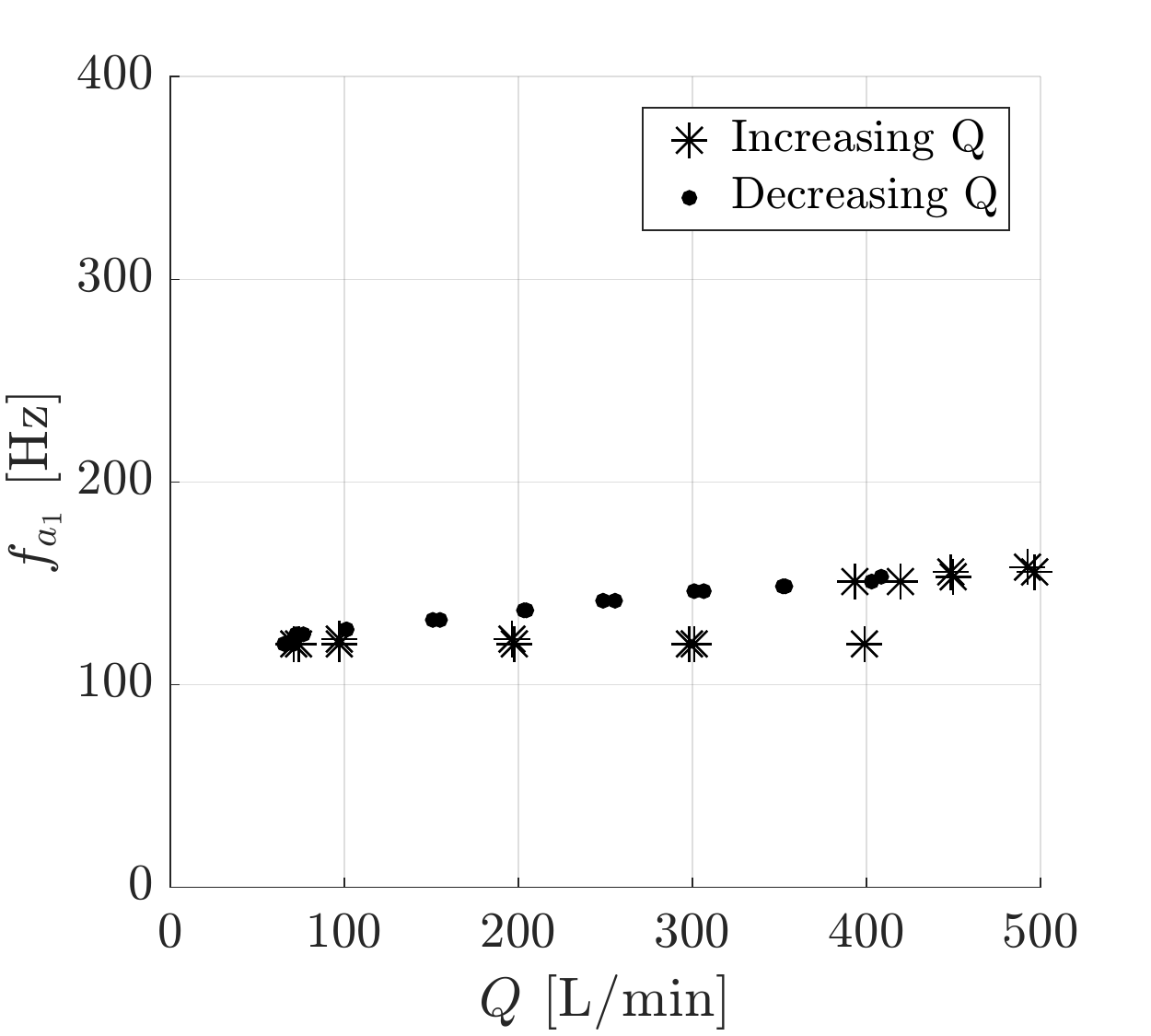} 
        \caption{Flextensional 3.}  
    \end{subfigure}
    
    \caption{Video data set showing mode 1 frequency response vs. flow rate for all flextensional settings. }  \label{fig:QvsfreqDisp}
\end{figure}

Figures \ref{fig:QvsDisp} and \ref{fig:QvsfreqDisp} display the dominant mode (mode 1) amplitude and frequency, respectively, as a function of flow rate for all three flextensional settings.  
The amplitudes are obtained by projecting the beam transverse position space-time series onto the two most energetic spatial SPOD modes of $f_1$ and $f_2$, per the method described in the appendix.  
The mean and standard deviation (markers and error bars, respectively) of the resulting modal amplitude time-series corresponding to the highest PSD frequency are plotted in figure  \ref{fig:QvsDisp}.
To look for hysteresis, tests are carried out by first increasing then decreasing the flow rate.  
Plots in figure \ref{fig:QvsDisp} show that the primary mode amplitude remains small (lightly-damped resonance) until a critical flow rate $Q_{\mathrm{cr}}$ is reached, which demarcates a transition to a high-amplitude, limit-cycle regime. 
Increasing the flow rate beyond $Q_{\mathrm{cr}}$, however, does not significantly increase the resulting amplitude. A slight amplitude decrease is sometimes seen at the highest flow rates, corresponding to the point when the beam appears to collide with the throat.    
As the flow rate is decreased through $Q_{\mathrm{cr}}$, a hysteresis loop becomes evident in all three flextensional settings, with its size ($\Delta Q = Q_{\mathrm{cr}} - Q_{\mathrm{r}}$) varying between each setting.  
The system recovers the small amplitude regime at  $Q < Q_{\mathrm{r}} < Q_{\mathrm{cr}}$, where $Q_{\mathrm{r}}$ is the fold flow rate.  
This hysteresis suggests that the system is undergoing a subcritical Hopf bifurcation at $Q_{\mathrm{cr}}$, giving rise to the bi-stable region captured in the data.
All three frequency responses in figure \ref{fig:QvsfreqDisp} appear constant until $Q_{\mathrm{cr}}$ is reached, at which point the frequencies tend to increase slightly with increasing flow rate.  The hysteretic behavior is most pronounced in the frequency data from  setting 2. 
Critical and fold properties for the observed bifurcation are summarized in table \ref{tab:FlexureCriticalProperties}.


\begin{figure}[H]
    \centering
    \begin{subfigure}[b]{0.31\textwidth}
        \centering
        \captionsetup{width=.8\linewidth}
   	    \includegraphics[width=1\textwidth]{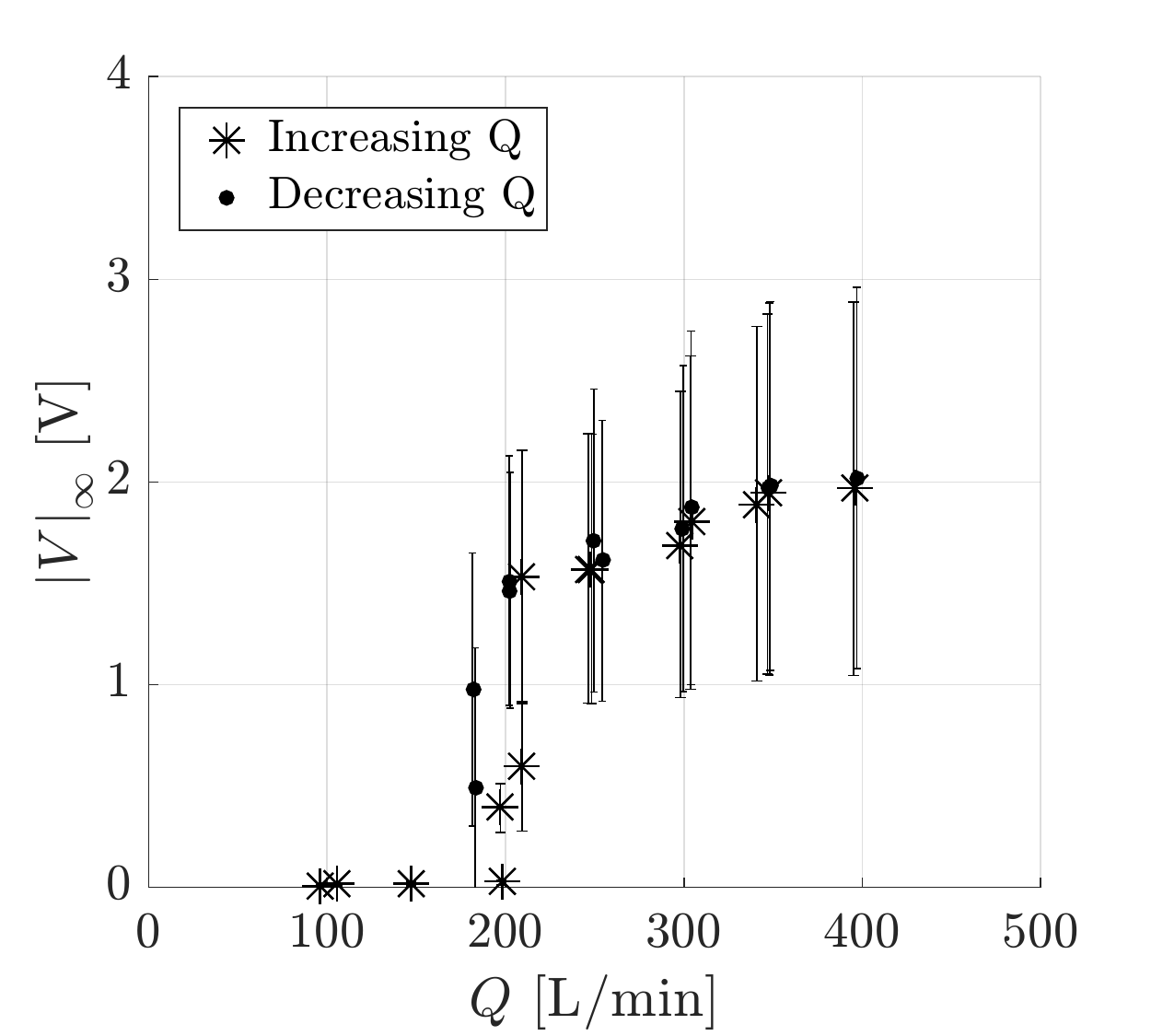}
        \caption{Flextensional 1.}  
    \end{subfigure}%
    ~ 
    \begin{subfigure}[b]{0.31\textwidth}
        \centering
        \captionsetup{width=.8\linewidth}
   	    \includegraphics[width=1\textwidth]{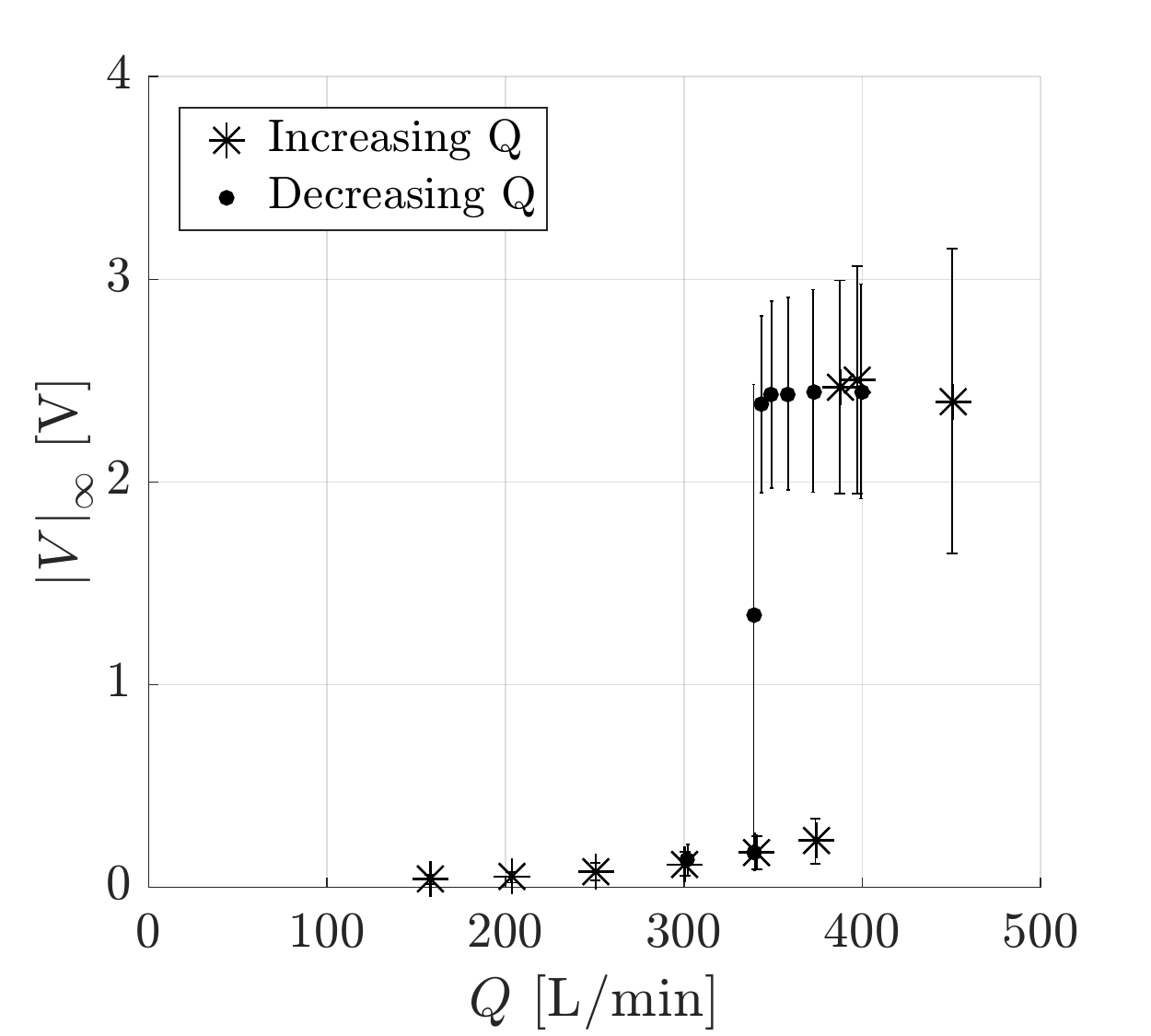} 
        \caption{Flextensional 2.}  
    \end{subfigure}
     ~ 
    \begin{subfigure}[b]{0.31\textwidth}
        \centering
        \captionsetup{width=.8\linewidth}
   	    \includegraphics[width=1\textwidth]{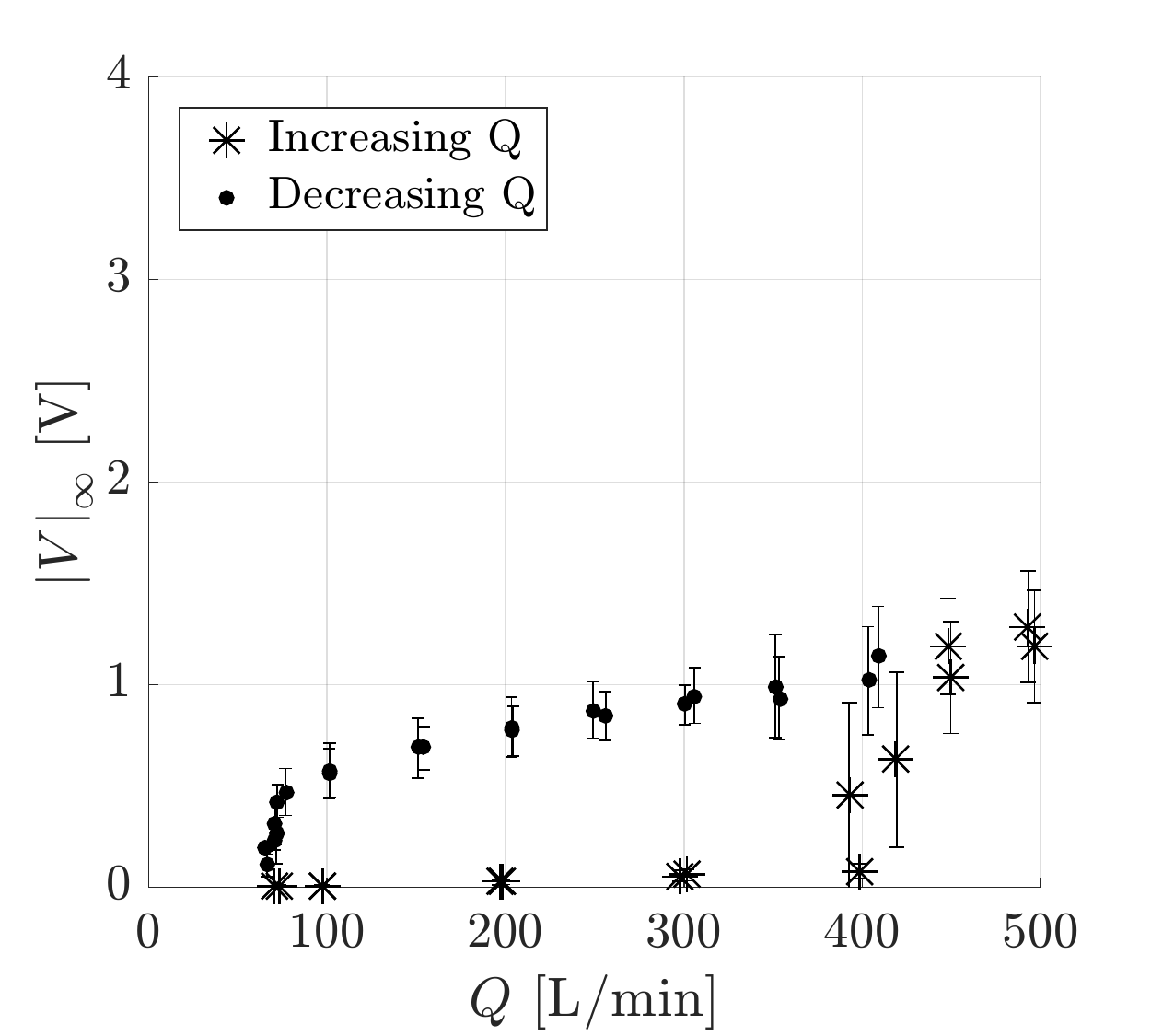} 
        \caption{Flextensional 3.}  
    \end{subfigure}
    
    \caption{PZT 1 voltage amplitude vs. flow rate for all flextensional settings.}  \label{fig:QvsVolt}
\end{figure}

\begin{figure}[H]
    \centering
    \begin{subfigure}[b]{0.31\textwidth}
        \centering
        \captionsetup{width=.8\linewidth}
   	    \includegraphics[width=1\textwidth]{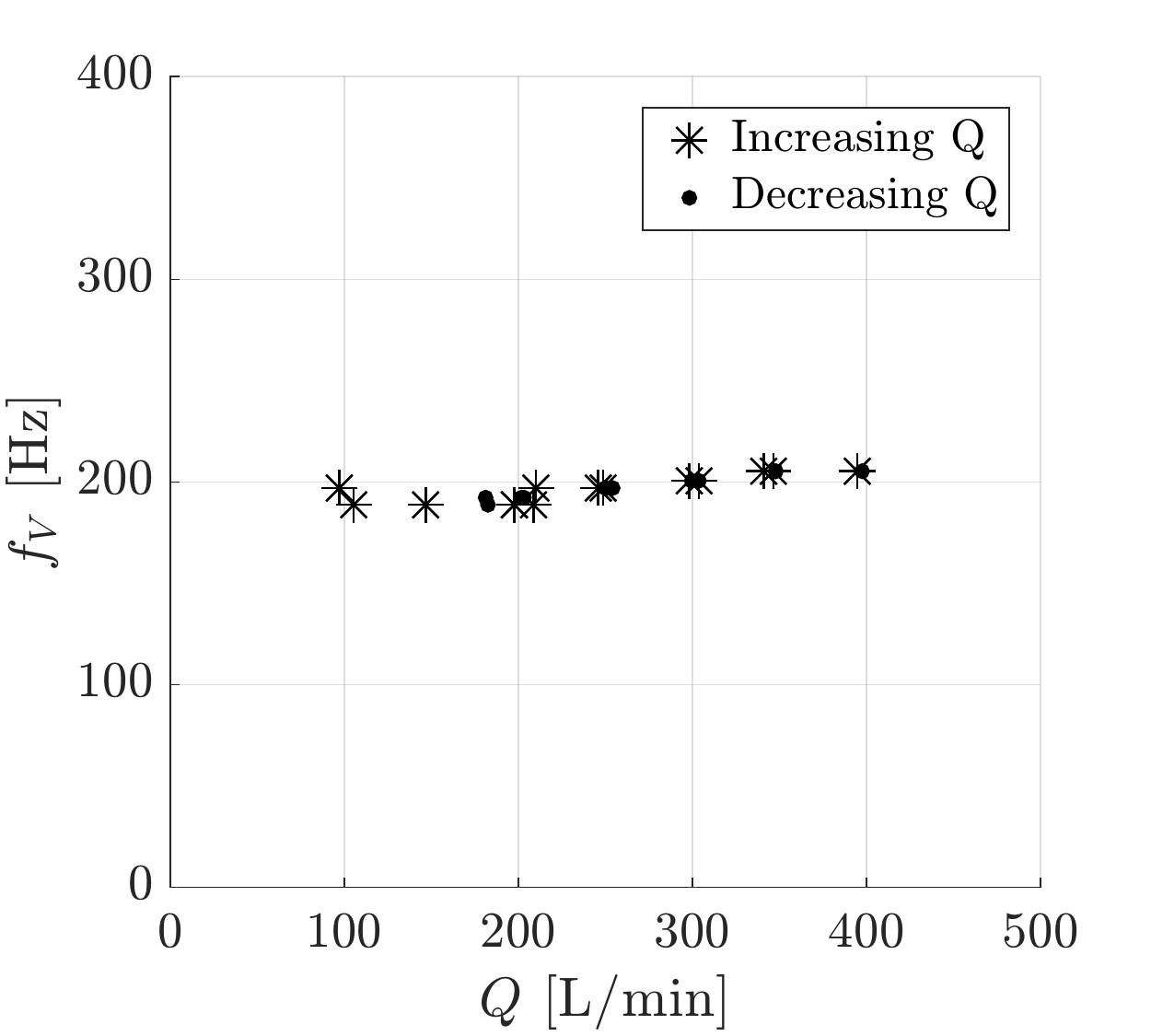}
        \caption{Flextensional 1.}  
    \end{subfigure}%
    ~ 
    \begin{subfigure}[b]{0.31\textwidth}
        \centering
        \captionsetup{width=.8\linewidth}
   	    \includegraphics[width=1\textwidth]{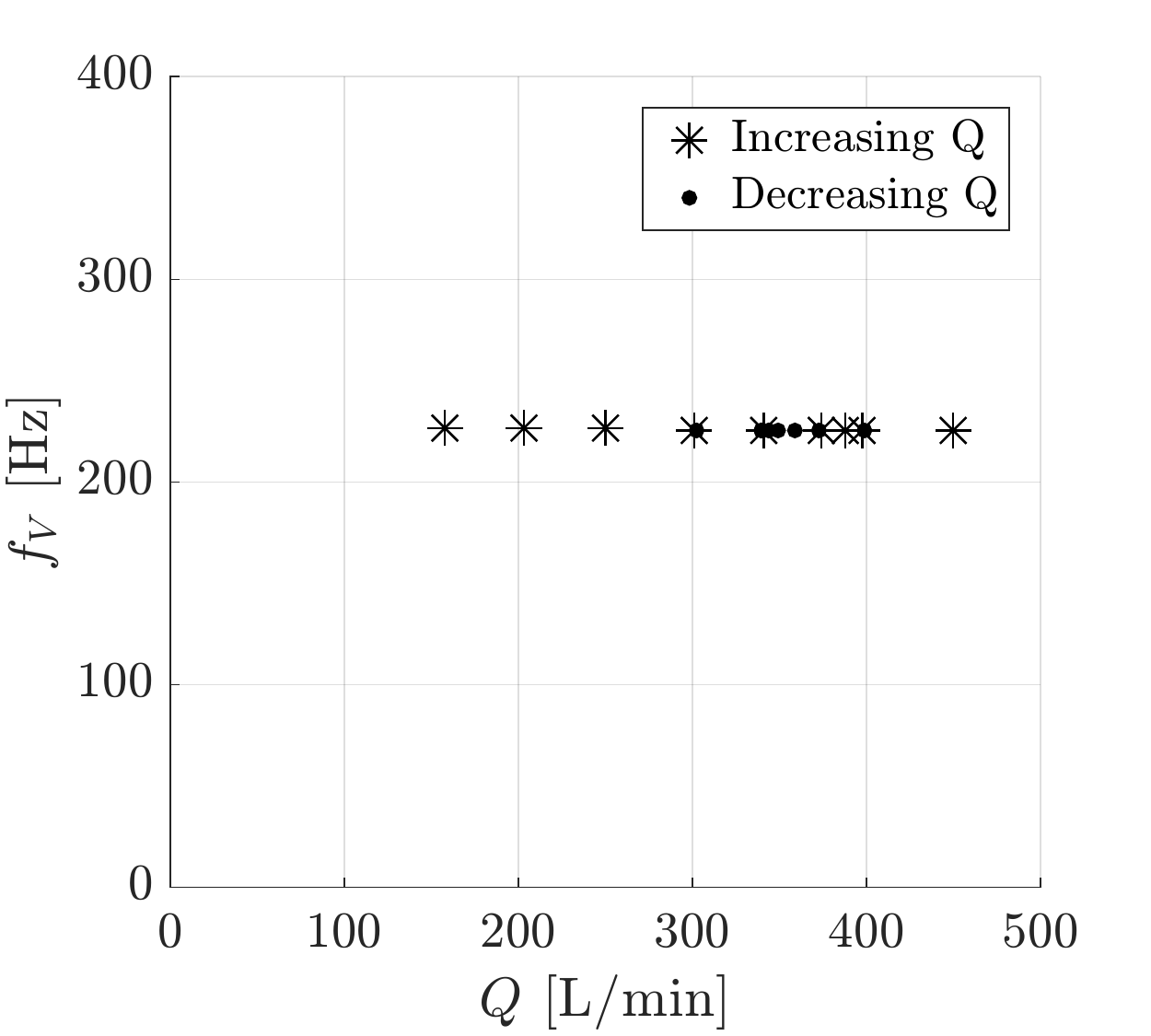} 
        \caption{Flextensional 2.}  
    \end{subfigure}
     ~ 
    \begin{subfigure}[b]{0.31\textwidth}
        \centering
        \captionsetup{width=.8\linewidth}
   	    \includegraphics[width=1\textwidth]{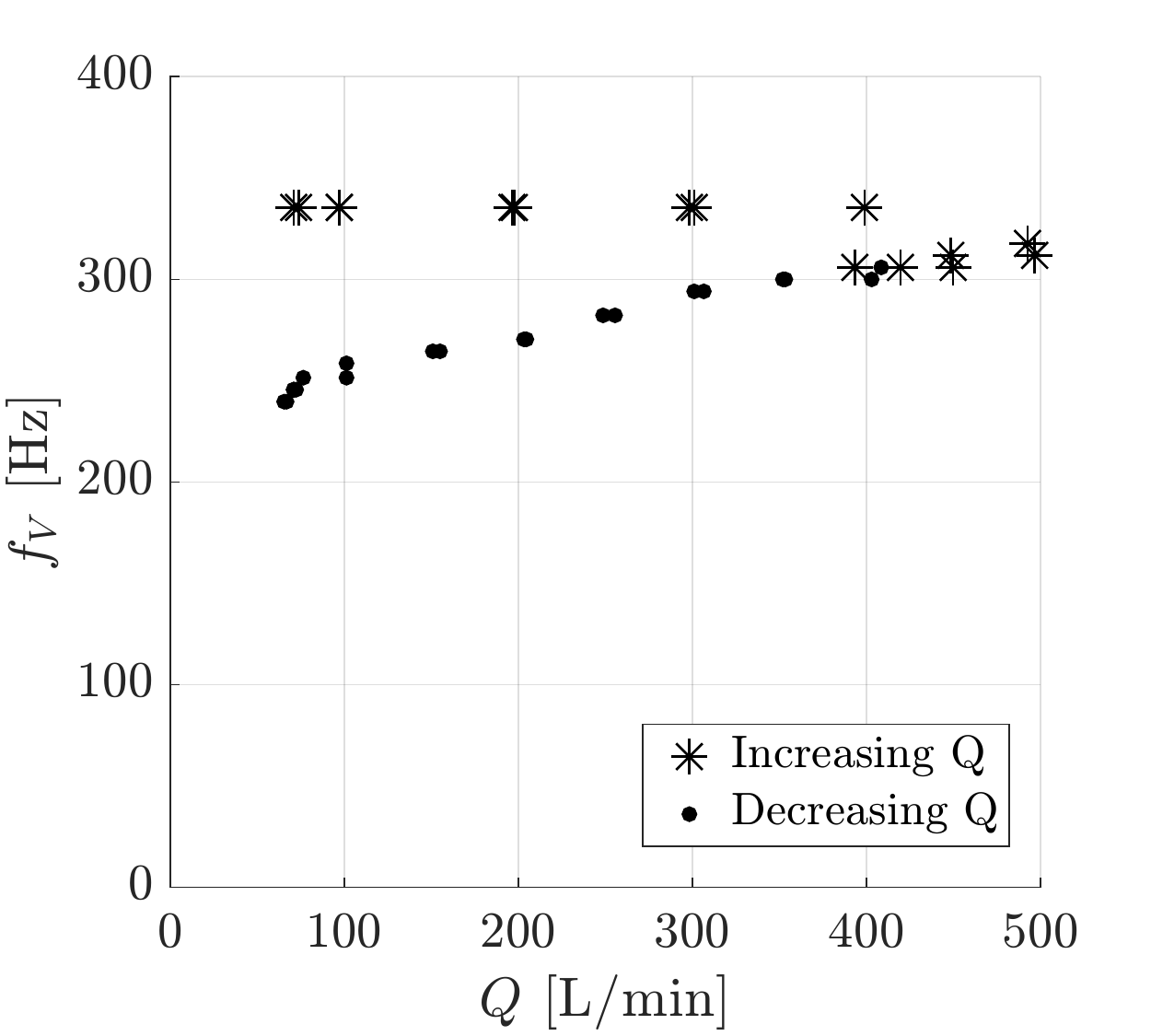} 
        \caption{Flextensional 3.}  
    \end{subfigure}
    
    \caption{PZT 1 frequency primary peak is voltage power spectral density vs. flow rate for all flextensional settings.}  \label{fig:QvsfreqVolt}
\end{figure}

Analogous amplitude and frequency plots for the PZT voltage are shown in figures \ref{fig:QvsVolt} and \ref{fig:QvsfreqVolt}, respectively, with critical and hysteresis results in agreement with video displacement data.  One discrepancy however, is observed in the flex. 3 frequency data.
Specifically, the plot shows the beam fundamental frequency as dominant until $Q_{\mathrm{cr}}$, at which point the voltage response frequency is double that of the video displacement frequency in figure \ref{fig:QvsfreqDisp}.    
This effect is caused by lightly pre-stressed piezoelectric elements, as this flexure configuration represents conditions with the least amount of torque applied on the set-screw.  The phenomenology is as follows: once the oscillation reaches the full extension at either the top or bottom of the flextensional stroke, the decompressed stack looses contact with the flexure structure.  This in turn causes a strong response that flips the sign of the voltage output, and appears as a frequency doubling through the discrete Fourier transform.  The nonlinear loss-of-contact behavior has been observed by 
\citep{Sherrit2009} as flextensional actuators loose their bond between stacks and the flexure.  Voltage amplitudes are also notably lower in flexure setting 3 than the other two flexure configurations.


\begin{table}[H] 
\centering
\caption{Table of critical and restoring (fold) values for flextensional settings.} \label{tab:FlexureCriticalProperties}
\begin{tabular}{c c c c c}  \hline 
	\textbf{Variables}	        &	\textbf{Flex. 1}&	\textbf{Flex. 2}	&	\textbf{Flex. 3}	&	\textbf{Description}	\\ \hline
$	Q_{\mathrm{cr}}$ [L/min]	&	208	            &	376	&	410	&	critical flow rate	\\
$	Q_{\mathrm{r}}$ [L/min]	    &	179	            &	334	&	73	&	fold flow rate	\\
$	f_{\mathrm{cr}}$ [Hz]	    &	186	            &	226	&	120	&	critical frequency	\\
 \hline
\end{tabular}
\end{table} 

Given the observed $Q_{\mathrm{cr}}$ values in table \ref{tab:FlexureCriticalProperties}, it is plausible that throat velocities may reach a considerable fraction of the sound speed when operating in air.  In appendix~\ref{sec:CompExp}, we estimate the Mach number at the channel throat $\mathcal{M}_{\mathrm{t}}$.  We discuss the potential limitations of the incompressible flow assumption made in the subsequent simulation section next.

\section{Numerical Simulations} \label{sec:NumericalSimulations}
Experimental results from the previous section show a rich set of dynamics and different regimes consistent with a subcritical Hopf bifurcation.  In this section, we use two-way coupled numerical simulations of a beam in the converging-diverging channel in order to investigate the three-dimensional flow field and provide insights into the flow patterns and instability mechanisms that drive the bifurcation. 
Experimental results also point us primarily to explore the flextensional mode dynamics, the only one that reaches the limit-cycle, for which the beam is essentially in rigid-body motion.  
We thus consider a rigid beam that is allowed to oscillate, via the lumped parameter model, using the experimentally-measured values for mass, stiffness, and damping ratios for flextensional setting 1.  We further discuss the validity of the rigid-body approximation below and well as in section \ref{sec:Model}.

\subsection{Numerical method}
Our
simulations are based on the lattice Boltzmann method (LBM), which originates
from kinetic theory and thus evolves discretized particle distribution
functions (populations) $f_i(\bm x,t)$, which are associated with discrete
velocities $\bm{c}_i, i=1,\ldots,Q$ and designed to recover the macroscopic
Navier-Stokes equations (NSE) in the hydrodynamic limit.  By organizing the set
of discrete velocities into a regular lattice, LBM eventually reduces to a
simple, efficient, and scalable stream-and-collide algorithm with the additional
advantage of exact propagation and local non-linearity, which is incorporated
through the collision operator.  In recent years, LBM has made significant
progress and early stability issues of the classical lattice
Bhatnagar–Gross–Krook (LBGK) model have been overcome.  While on one hand
explicit turbulence models have shown success for turbulent flows
\citep{chen2003extended,malaspinas2012consistent}, the class of parameter-free
entropic lattice Boltzmann schemes (ELBM) have shown accurate and robust
solutions for both resolved and under-resolved simulations for laminar,
transitional as well fully turbulent flows
\citep{Bosch2015,Bosch2015b,dorschner2016entropic,dorschner2017transitional}.
In particular, we use the multi-relaxation time (MRT) variant of ELBM (KBC) \citep{karlin2014gibbs},
which exploits the high dimensionality of the kinetic system and chooses the
relaxation of higher-order, non-hydrodynamic moments such that the entropy of
the post-collision state is maximized. The KBC model has been discussed in
various contributions and we will restrict ourselves to the main steps in case
of isothermal flow using the standard $D3Q27$ lattice.

We start from the general lattice Boltzmann equation for the population $f_i(\bm x,t )$:
\begin{equation}
\label{eq:f_equations}
	f_i(\bm{x+c_i}, t+1)=f_i^{\prime} = (1-\beta)f_i(\bm{x},t) + \beta f_i^{\text{mirr}}(\bm{x},t),
\end{equation}
where the streaming step is indicated by the left-hand side and the post-collision state 
$f^\prime_i$  on the right-hand side is given by a convex-linear combination of $f_i(\bm{x},t)$ and a mirror state 
$f_i^{\text{mirr}}(\bm{x},t)$.
We use natural moments to represent the population as a sum of the kinetic part $k_i$, the shear part $s_i$
and the remaining higher-order moments $h_i$:
\begin{equation}
	f_i = k_i + s_i + h_i.
\label{eq:popSplit}
\end{equation}
The mirror state can thus be represented as 
\begin{equation}
\label{eq:f_mirr}
	f_i^{\text{mirr}} = k_i + \left( 2 s_i^{\rm eq} -s_i \right) + \left( \left(1 - \gamma\right)  h_i + \gamma h_i^{eq}\right),
\end{equation}
where $s_i^{eq}$ and $h_i^{eq}$ denote $s_i$ and $h_i$ evaluated at equilibrium.

The equilibrium distribution function $f^{\rm eq}$ is defined as the minimum of the entropy function
\begin{equation}
	H(f) = \sum_{i=1}^Q f_i \ln \left( \frac{f_i}{W_i} \right) , \\
	\label{eq:feq_min}
\end{equation}
subject to the local conservation laws for mass and momentum
\begin{equation}
	\sum_{i=1}^Q \lbrace 1, \bm{c_i} \rbrace f_i = \lbrace \rho, \rho \bm{u} \rbrace,
	\label{eq:conservationLaws}
\end{equation}
and the weights $W_i$ are lattice-specific constants. 
By minimizing the $H$-function in the post-collision state 
one obtains the relaxation parameter 
\begin{equation}
	\gamma = \frac{1}{\beta} - \left( 2 - \frac{1}{\beta}\right) \frac{\left< \Delta s | \Delta h\right>}{\left< \Delta h | \Delta h\right>  },
	\label{eq:gamma_min_approx}
\end{equation}
where $\Delta s_i = s_i - s_i^{\text{eq}}$ and $\Delta h_i=h_i-h_i^{\text{eq}}$ are the deviation from equilibrium and 
the entropic scalar product is defined as $\left< X | Y \right> = \sum_i (X_i Y_i / f_i^{\text{eq}})$.
The KBC model recovers the Navier-Stokes equations in the hydrodynamic limit
for which the viscosity is related to the parameter $\beta$ as
\begin{equation}
	\nu=c_s^2 \left( \frac{1}{2\beta}- \frac{1}{2}\right),
	\label{eq:visc}
\end{equation}
where $c_s=1/\sqrt{3}$ is the lattice speed of sound. 

Finally, to include two-way coupling of the fluid with the cantilever beam, we
follow the procedure as outlined in
\citep{dorschner2015grad,dorschner2017entropic,dorschner2018fluid}, using
second-order Grad boundary conditions to account for the momentum transfer from
the fluid onto the beam and vice versa.  The beam velocity, needed to 
prescribe the boundary conditions, is obtained by solving Newton's equations
of motions using an Euler integration and the fluid force is evaluated by the
Galilean invariant momentum exchange method (see \cite{Wen2015}).  This
procedure has been validated extensively for various test-cases for one- and
two-way coupled simulations as well as fully coupled fluid-structure
interaction problems involving deforming geometries.

\subsection{Simulation of the flow energy harvester} \label{sec:3DDNSFEH}
\begin{figure}[!t]
	\centering					
	\includegraphics[width=0.7\textwidth]{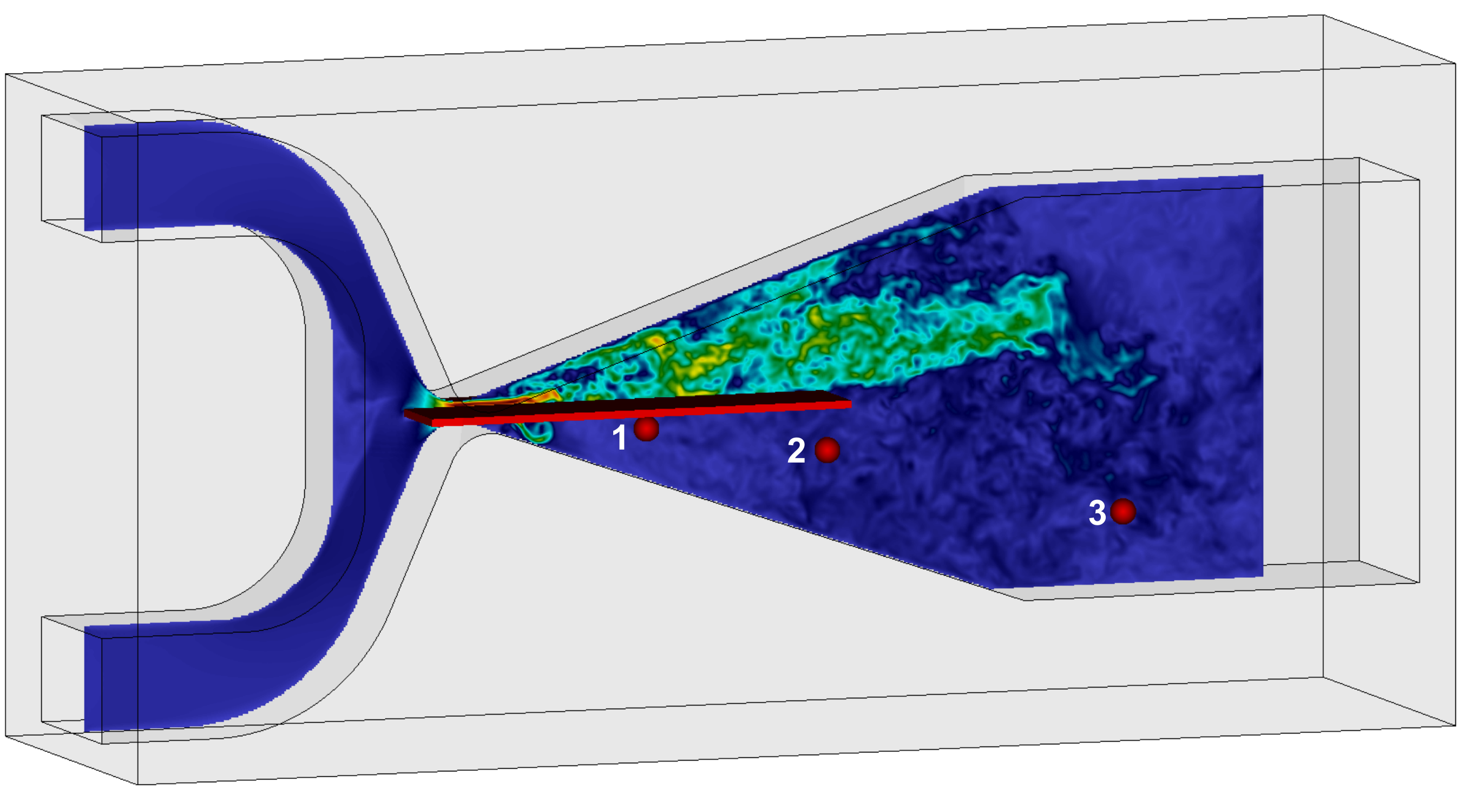}
    \caption{Instantaneous snapshot of the computational setup for $Q=208$
    L/min, showing a slice of velocity magnitude. Exemplary observer points
             are indicated by the red spheres.}
	\label{fig:lbm_setupSnapshot}
\end{figure}
\begin{figure}[!t]
	\centering					
	\includegraphics[width=0.9\textwidth]{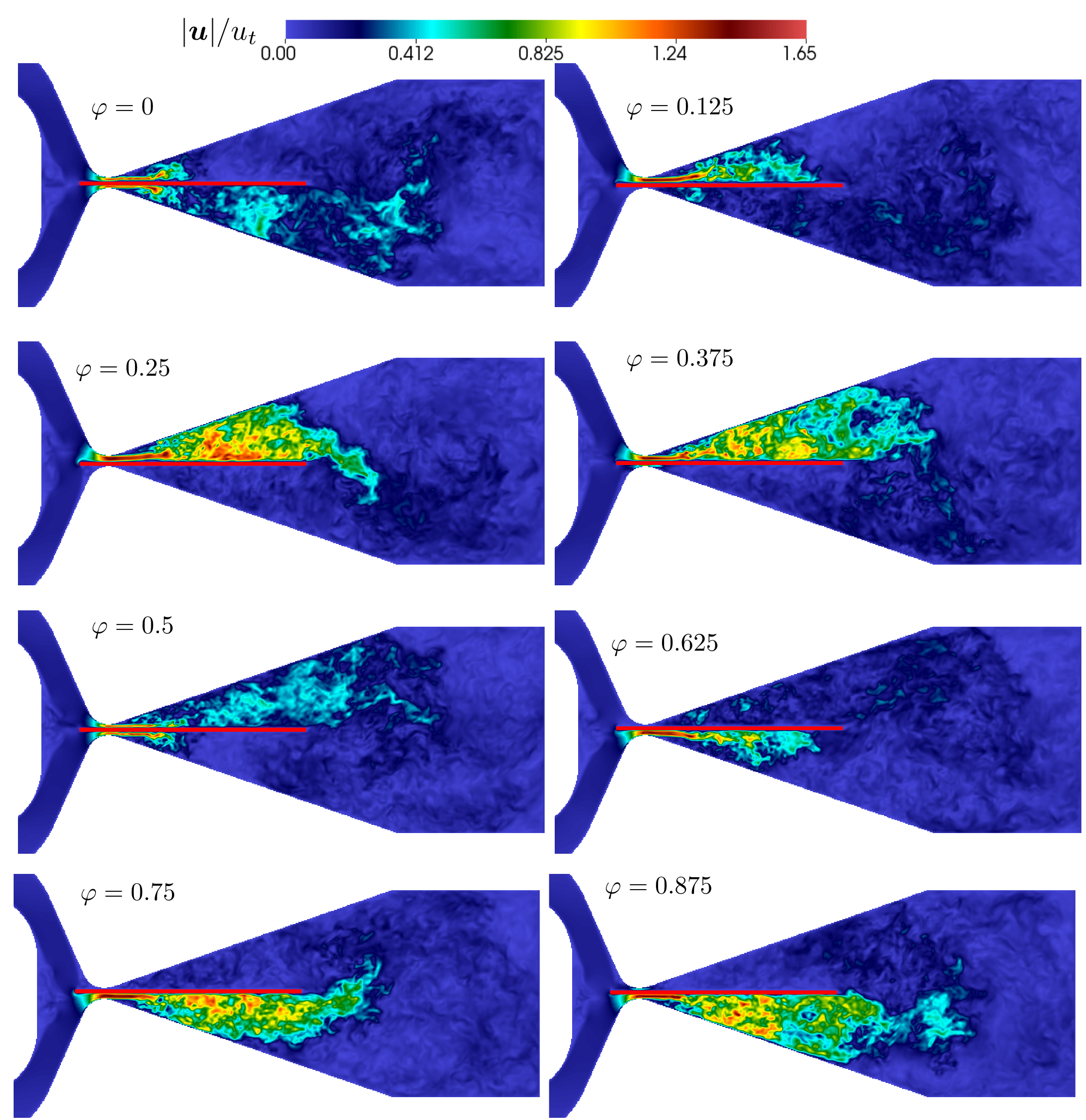}
    \caption{Flow evolution for one period of the energy harvester, showing a slice 
             of velocity magnitude.}
	\label{fig:feh_evolution_all}
\end{figure}
\begin{figure}[!t]
	\centering					
	\includegraphics[width=0.9\textwidth]{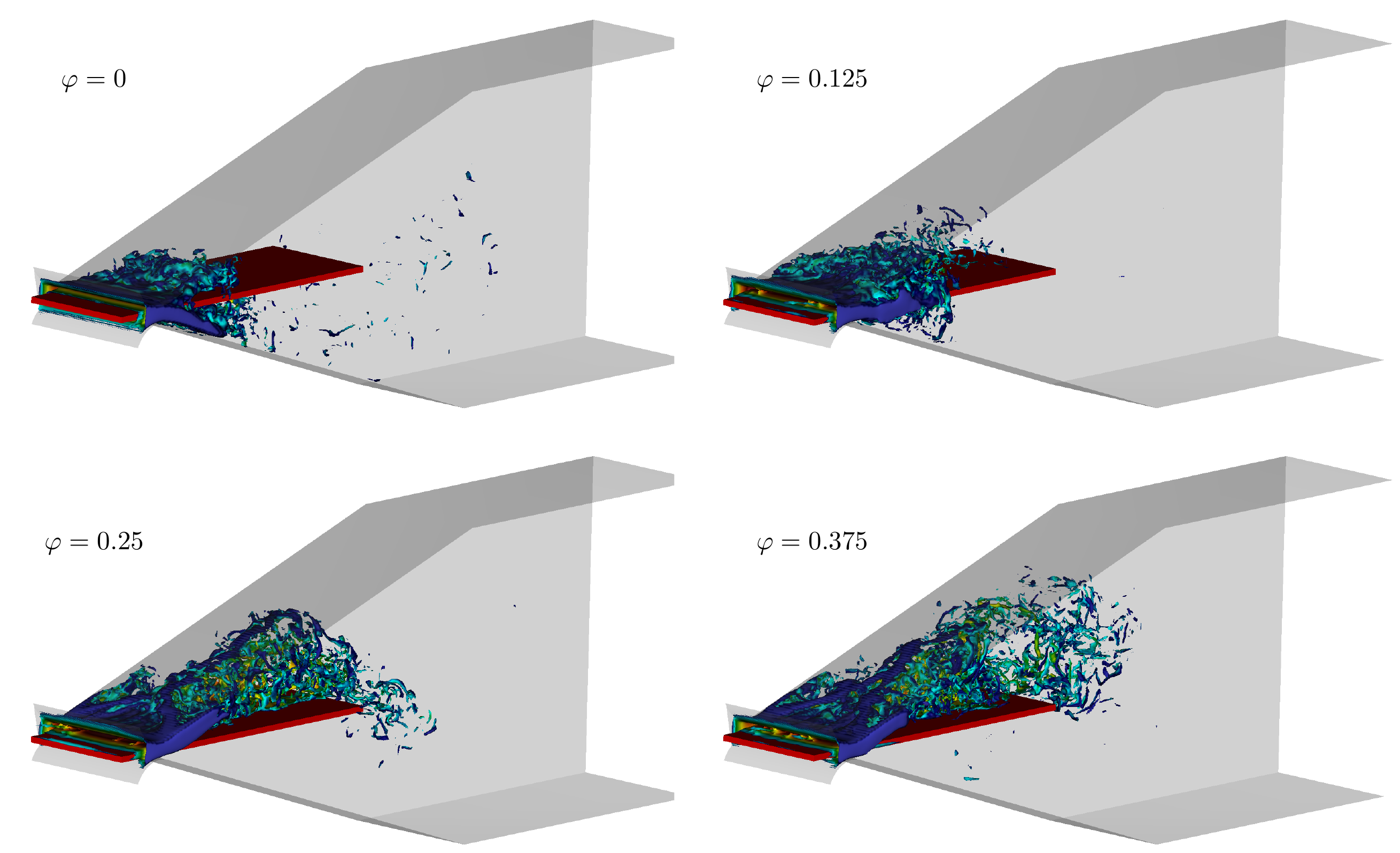}
    \caption{Flow evolution for half a period of the energy harvester, showing
    isosurfaces of vorticity colored by velocity magnitude and zoomed into the
    diffuser region.}
	\label{fig:feh_evolution_all_vorticity}
\end{figure}
The simulation of the full flextensional energy harvester is a
challenging task due to the complex interaction of various physical mechanisms.
We keep the geometry of the fluid channel path identical to the
experimental setup apart from the diffuser exit, which is a sharp edge in the simulation
but smoothed in the experiment. In figure \ref{fig:lbm_setupSnapshot}, the numerical setup is shown. 
As noted in section \ref{sec:flextureMeasurements}, the flexure itself is modeled by a harmonic oscillator to
which a rigid cantilever beam is attached. In the simulations, this is realized by
elastically translating boundary conditions of the beam.  The mass, stiffness
and damping ratio of the harmonic oscillator are prescribed in the simulations
according to the experimental measurements of flex. setting 1 in table \ref{tab:FlexureBoundaryProperties}.

Regarding the rigid-body approximation, as discussed in sections \ref{sec:FEHDesign} and \ref{sec:experiments}, we refrain from modeling the beam bending since 
the most energetic observed mode is primarily a rigid body motion of the entire flexure, and where the damping of the structure as a whole is well approximated by a second-order damped harmonic oscillator. 
We preformed precursor simulations that included elasticity of the beam but neglecting internal damping, and these confirmed the model predictions discussed in section \ref{sec:Model}, namely that higher-order oscillatory beam modes do become unstable in the absence of internal damping.  Based on the experiments, these results are known to be unphysical and we therefore focus our attention to predicting critical properties of the first, primarily rigid, flextensional based mode.   
As discussed in appendix~\ref{sec:CompExp}, we do not account for compressibility and
treat the fluid as incompressible fluid.  
Our simulations are carried out on a uniform Cartesian mesh (with $\Delta=1$ and $\Delta t=1$ in lattice units), 
where we resolve the beam with roughly $250$ lattice points. 
All other dimensions follow from the experimental setup and
a snapshot of the
computational domain is shown in figure \ref{fig:lbm_setupSnapshot}.
Further, the inflow velocity is conservatively set to $u=0.0075$ (in lattice units) to avoid any compressibility effects.
The Reynolds number is set to $Re_h = u_t \bar{h}/\nu \approx 5200$, which
is chosen such
that it is high enough to account for viscous effects but low enough to provide sufficient resolution for all pertinent flow scales.
To that end, convergence of the critical flow rates was verified with coarser meshes.
In addition, the agreement with experiments gives us confidence that all pertinent mechanisms are captured by our simulations.

Figure \ref{fig:feh_evolution_all} shows the evolution of the velocity
magnitude in the mid-plane of the domain for one representative cycle.  In the
beginning of each cycle for a phase angle $\varphi=0$ the beam displacement is
zero and two symmetric jets on the top and bottom of the beam are forming. Note
also that residual turbulence from the previous cycle is  visible in the
bottom half of the diffuser. Subsequently for $\varphi=0.125$, the beam moves
downward, leading to an increase of mass flow through the upper diffuser
channel until the mass flow through the bottom channel almost ceases at
$\varphi=0.25$.  Consequently, the upper jet amplifies and penetrates deeper
into the diffuser until it eventually breaks up into turbulence
beyond the beam. The maximum penetration of the jet into the diffuser is
reached at $\varphi=0.25$. Notably, the jet does not penetrate much beyond the
length of the beam, where it is then expanding into the bottom half of the
domain and rapidly broken up into finer-scale turbulence.
During its upward motion beyond $\varphi=0.25$, the upper jet weakens whereas the 
mass flow rate through the bottom half of the domain gradually increases. 
Finally at $\varphi=0.5$, the process repeats in a symmetric fashion for the 
bottom half of the channel.
In figure \ref{fig:feh_evolution_all_vorticity}, vorticity isosurfaces colored
by velocity magnitude are shown for the first half of the oscillation period.
The behavior is analogous to what was observed for the velocity magnitude.
However, we can additionally observe the effect of spanwise confinement.
Starting from a phase angle of $\varphi=0.25$, one can observe vortical
structures attaching to the side and upper walls of the diffuser geometry.
Downstream of the throat, a large lambda-type vortex structure is formed on the
upper diffuser wall due to vortex rollup from both sides of the beam.
Consequently,  most vorticity is confined in the center region of the beam,
whereas only negligible vorticity is found in regions close to the diffuser
side walls and downstream of the throat.

\begin{figure}[!t]
	\centering					
	\includegraphics[width=0.5\textwidth]{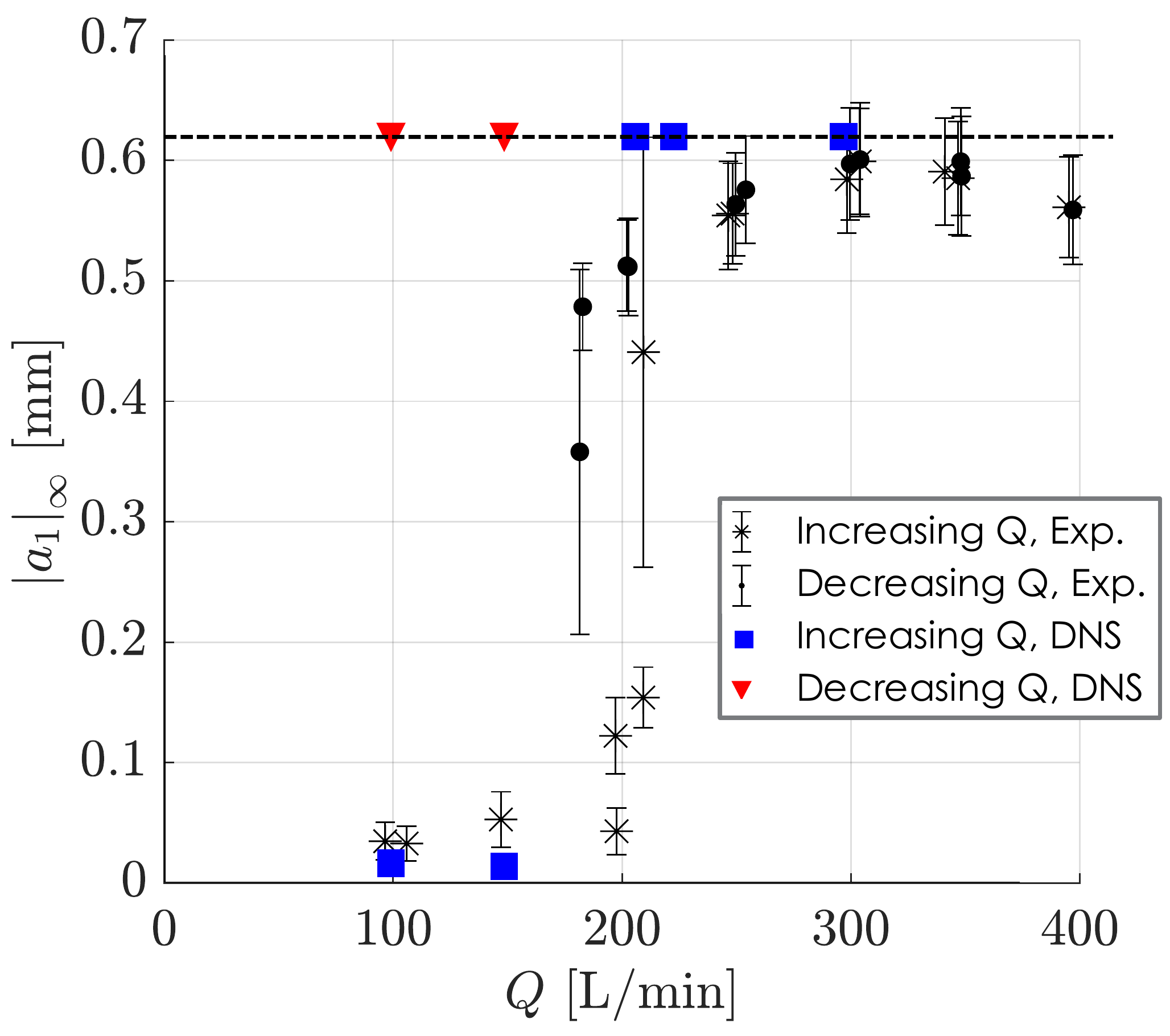}
    \caption{Beam oscillation amplitude as a function of the flow rate for the
    experiments as well as the numerical simulations.}
	\label{fig:lbm_flow_rates}
\end{figure}

To assess the predictive capabilities and validity of our computational model,
we run a series of simulations for flow rates in the range of $Q=100$ to 300 L/min and record the time evolution of the beam displacement.  This allows
us to obtain an estimate of the critical flow rate at which the beam starts to
exhibit self-sustained oscillations.  As shown in figure
\ref{fig:lbm_flow_rates}, the critical flow rates as computed by our simulations
agree well with the experiments.

Note that in the simulation it is not possible to fully resolve the thin fluid
layer between beam and the throat for the experimental geometry.
This is due to the fact that in our experiments, and as indicated in figure \ref{fig:QvsDisp}, the beam oscillation amplitude becomes
large and sometimes collides with the wall (see error bars). 
Such collisions are not explicitly modeled in our numerical model. However, since 
we are only interested in the on-set of self-sustained oscillation we stop the simulation once the beam displacement
reaches the height of the throat.
In addition to recording the oscillation
amplitude for various flow rates, we also probe the hysteresis behavior of the
system to access the bifurcation type.  To that end, the simulation of $Q=208$
L/min is restarted and the flow rate reduced.  As shown in figure
\ref{fig:lbm_flow_rates}, we observe a pronounced hysteresis behavior, again
indicative of a subcritical Hopf bifurcation, and is in agreement with the
experimental findings. The hysteresis is, however, more pronounced in the simulations.  
One potential explanation comes from the perturbation and noise inherent in our experiments (i.e. collision of beam with channel wall), which would tend to push the beam states from the stable limit-cycle basin of attraction to that of the stable equilibrium \emph{earlier} (i.e. at a higher flow rate than the fold point), resulting in a smaller experimental hysteresis loop.  
In figure \ref{fig:displament_evolution}, the evolution of the beam displacement as well as the
power spectral density is shown for the critical flow rate of $Q=208$ L/min.
As expected, the beam displacement undergoes exponential growth and, from 
the power spectral density, oscillates near the natural frequency of the flexure. 
Once again, this is consistent with what is observed in the experiment. 

\begin{figure}[!t]
	\centering					
	\includegraphics[width=0.9\textwidth]{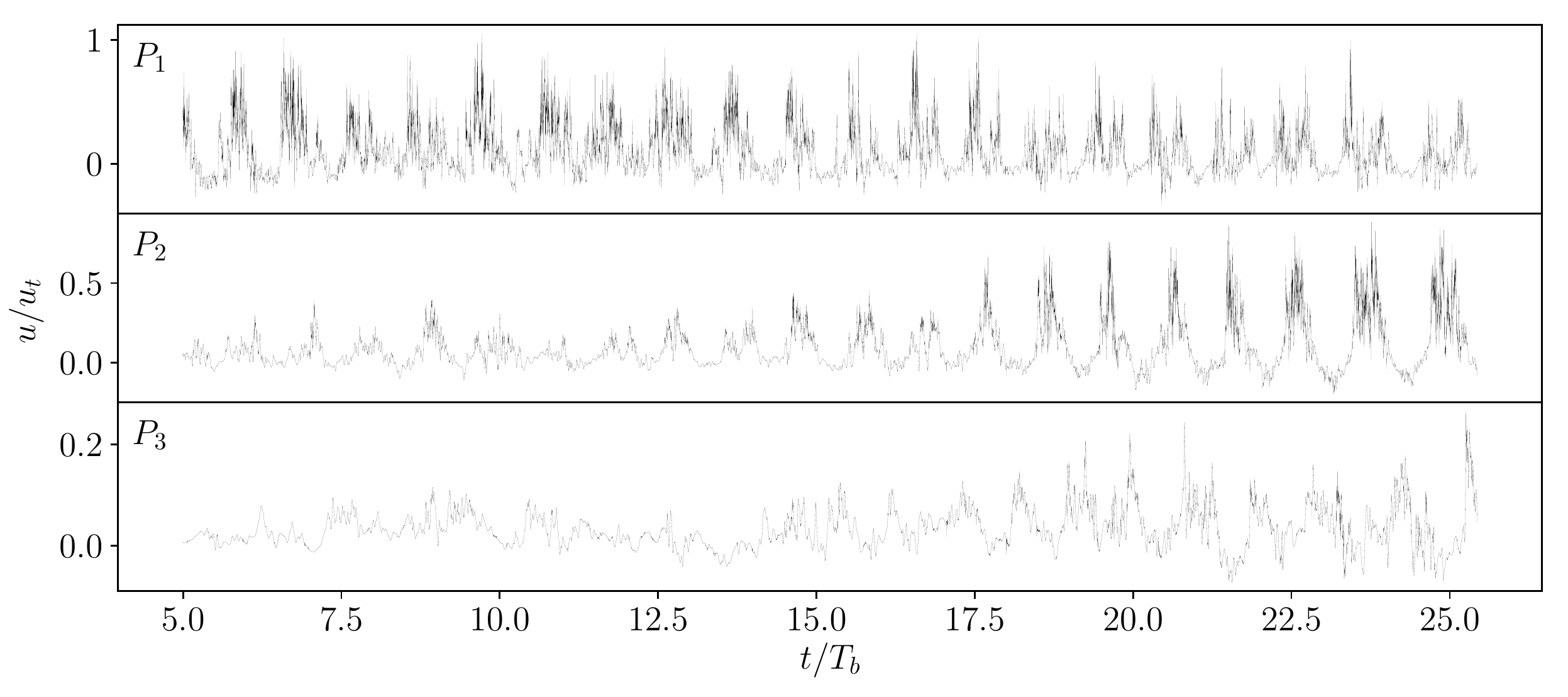}
    \caption{Evolution of the streamwise velocity profile for three probes. }
	\label{fig:obs_evolution}
\end{figure}

In addition to the temporal evolution of the beam displacement, three observer
probes were placed within the domain. In particular, the probes 
were placed in vicinity of the throat, near the trailing edge of the beam
as well as in the far field of the diffuser (see figure \ref{fig:lbm_setupSnapshot}). 
The evolution of the streamwise velocity for all three probes
is depicted in figure \ref{fig:obs_evolution}. 
Probe 1, located near the throat, shows periodic behavior with a largely
constant amplitude and only a slight decrease in amplitude as the oscillation
amplitude of the beam increases due to an increase of the throat gap.  A
different picture is drawn for the two probes downstream. It is apparent that
the amplitude in the initial phase ($t/T_b< 17$) remains relatively low and
increases noticeably afterwards. This is due to the increasing penetration
depth of the jet, which eventually reaches the probe location.  In addition,
the magnitude of the streamwise velocity rapidly decreases as it is diffused
further downstream and diminishes to roughly $20\%$ for probe $2$.
It is further is instructive
to look at the power spectral density plots of the observer probes. It is
noticeable that the most dominant frequency for probe 1 is the beam frequency,
whereas further downstream its first harmonic becomes increasingly larger and
eventually dominates. This can be explained by the fact that further downstream
there is the coupling between jets in the lower as well as in the upper half
of the domain, i.e., the probes feel the influence of both jets and thus
doubling the dominant frequency. 

\begin{figure}[!t]
	\centering
    \begin{subfigure}[b]{0.49\linewidth}
        \centering
        \includegraphics[width=\textwidth]{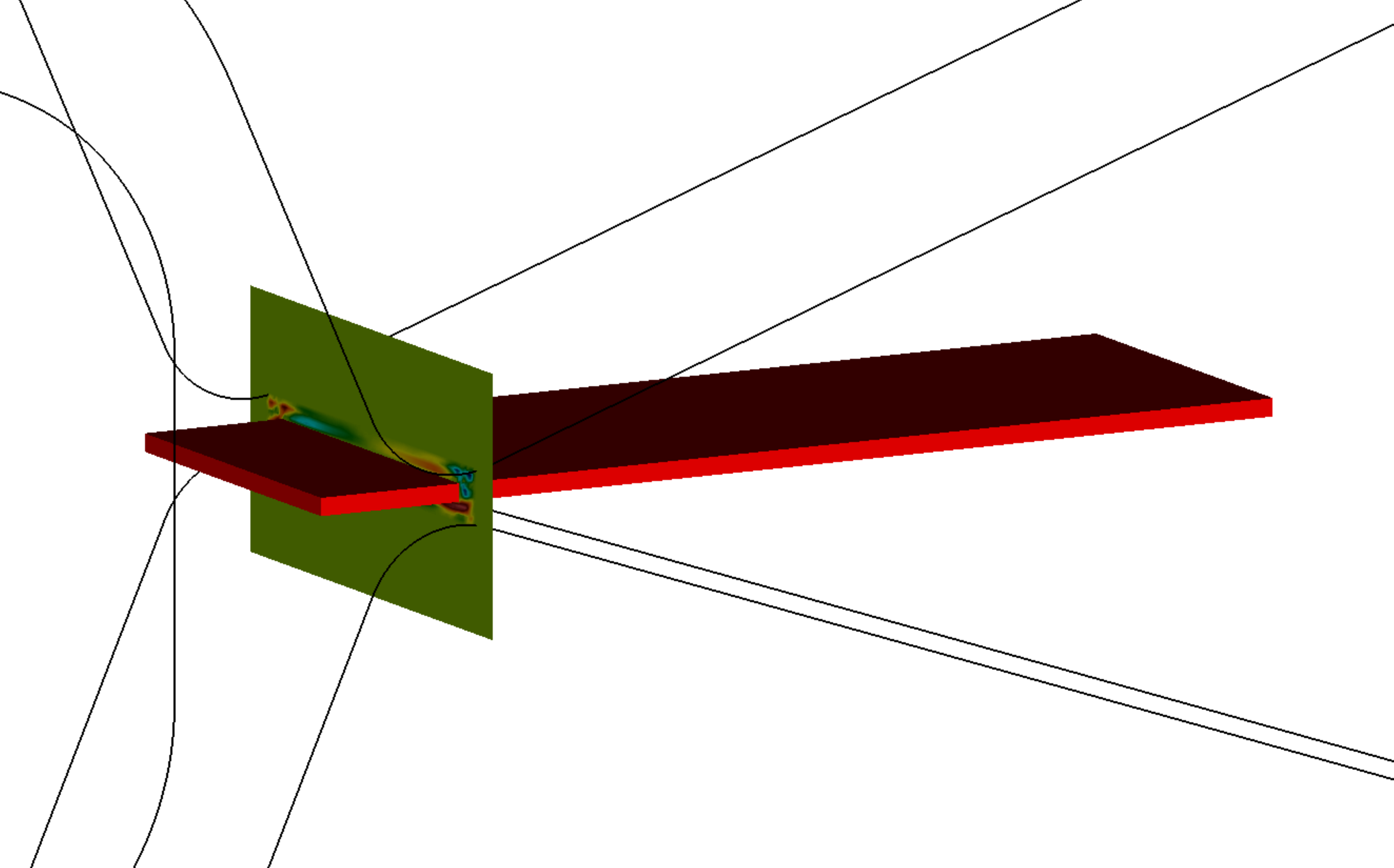}
        \caption{Location of the phase-averaging plane.}
        \label{fig:spanwise_schematic}
    \end{subfigure}
    \begin{subfigure}[b]{0.49\linewidth}
        \centering
        \includegraphics[width=\textwidth]{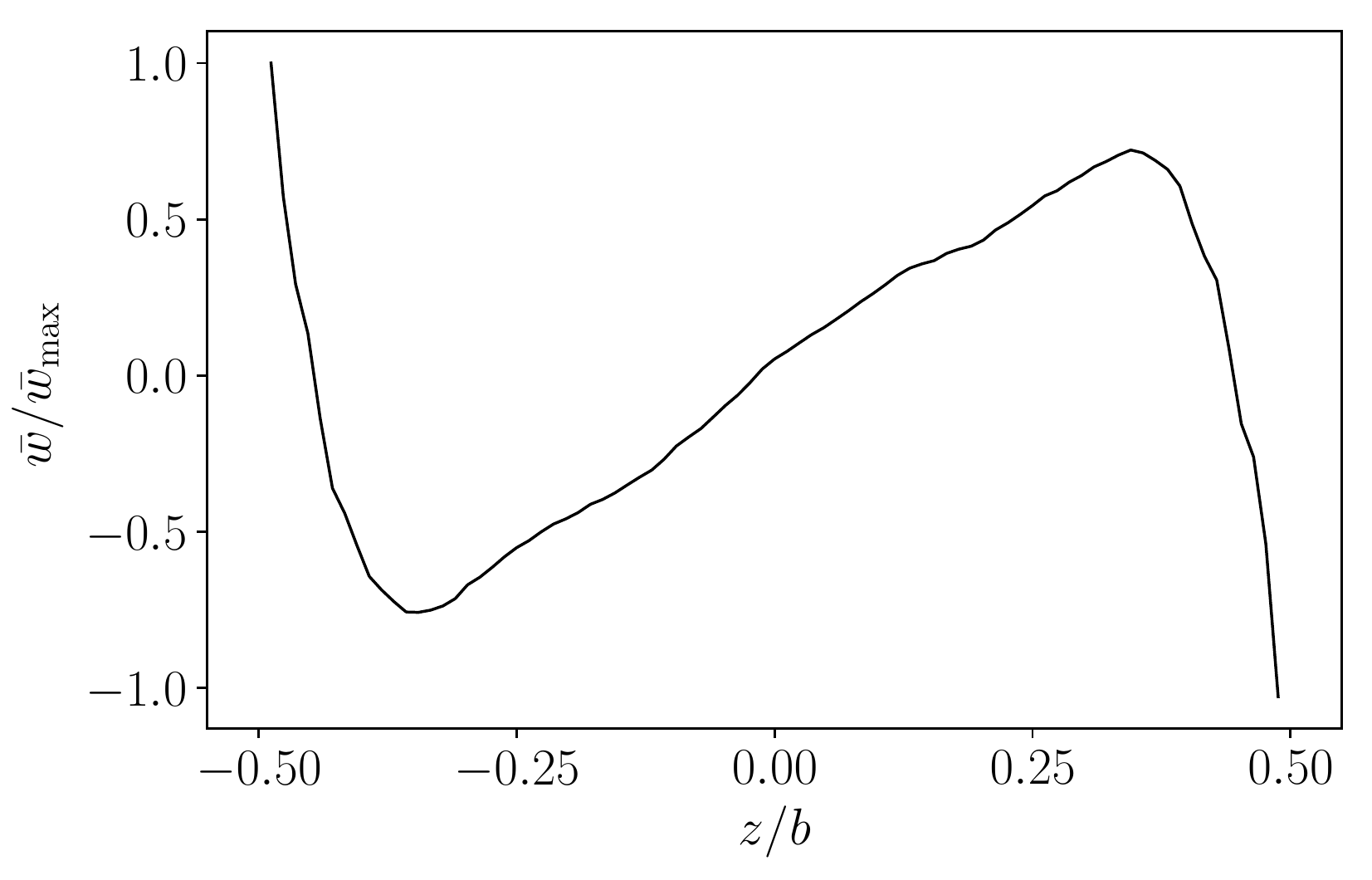}
        \caption{Phase-averaged spanwise velocity profile.}
        \label{fig:spanwiseProfile}
    \end{subfigure}
    \caption{Spanwise velocity profile.}
	\label{fig:spanwise}
\end{figure}

Moreover, we measure the phase-averaged profile at $\varphi=0$ of the spanwise velocity in 
a cross section near the throat as shown in figure \ref{fig:spanwise}. 
The profile is symmetric and linear to a good approximation for most of the span. 
One can also see the effect of spanwise vortices, which are forming due to spanwise confinement at the edge of the beam.
This will be used later as an input and validation of the model assumptions in section \ref{sec:Model}.

Finally, having an indication of how the flow evolves within the internal flow
energy harvester helps support our conjecture that the main driving factor of the
instabilities arising in the flow originates from its modulation due to the
confinement in the channel throat. This is evidenced by the flow field
in figure \ref{fig:feh_evolution_all} and figure
\ref{fig:feh_evolution_all_vorticity} where there are no significant flow
structures that appear able to drive the instabilities in the wake of the beam. 
Further validation is presented next, where we devise a 
reduced-order model that accounts for this modulation phenomenon as the only source of the instability.

\begin{figure}[!t]
	\centering
    \begin{subfigure}[b]{0.49\linewidth}
        \centering
        \includegraphics[width=\textwidth]{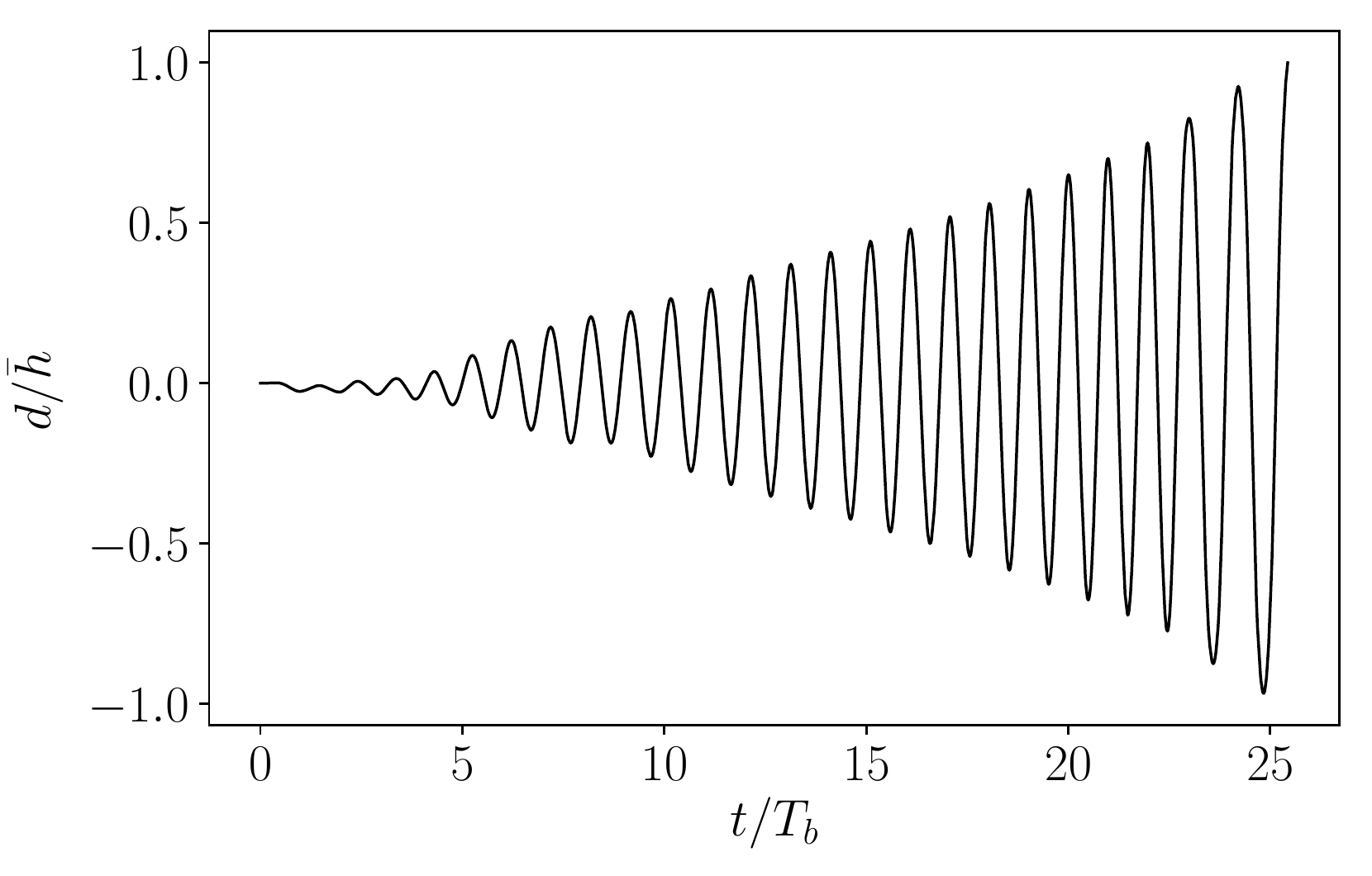}
        \caption{}
        \label{fig:displament_evolution}
    \end{subfigure}
    \begin{subfigure}[b]{0.49\linewidth}
        \centering
        \includegraphics[width=\textwidth]{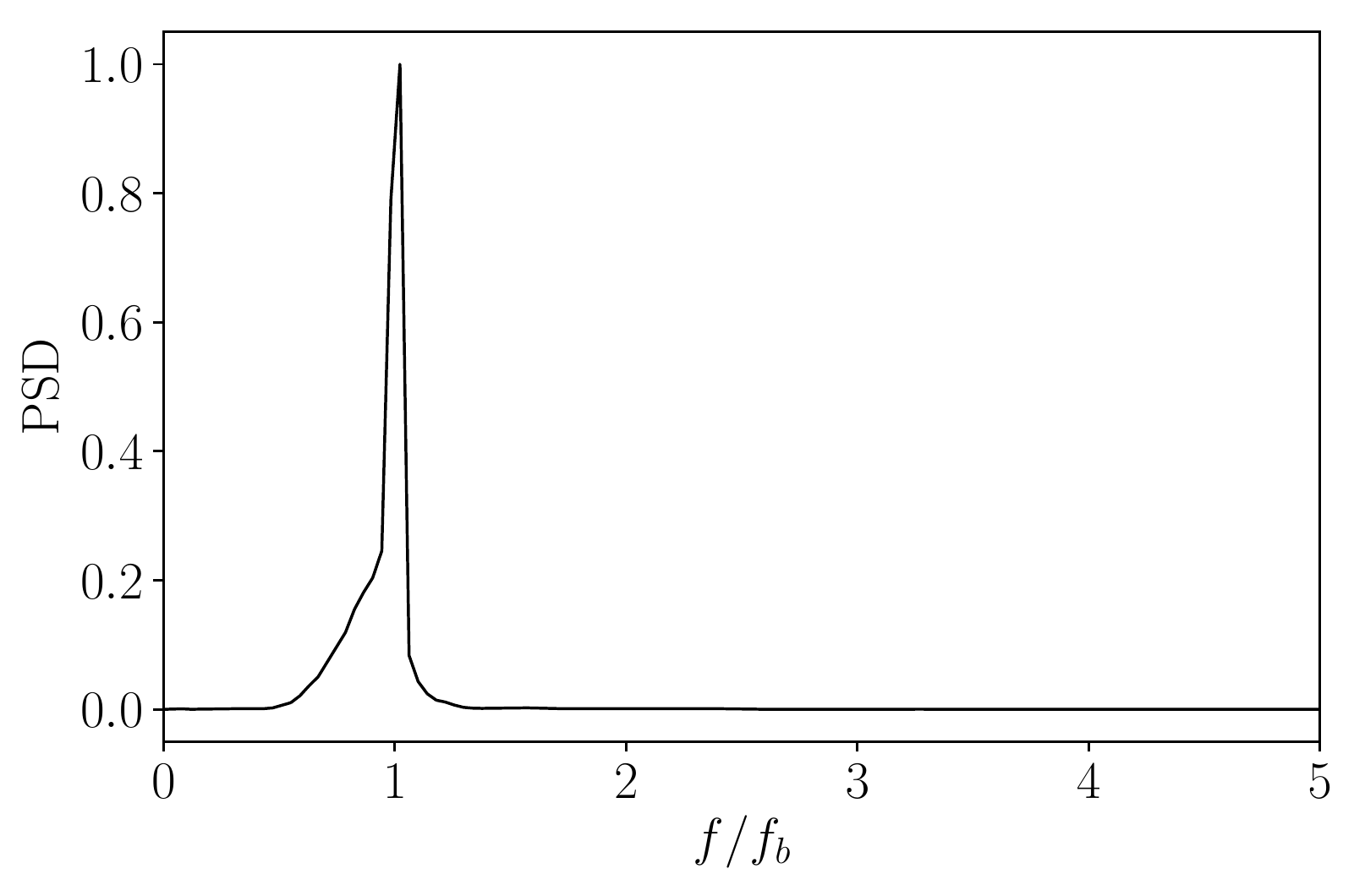}
        \caption{}
        \label{fig:displament_psd}
    \end{subfigure}
    \caption{Evolution of the beam displacement and its power spectral density.}
	\label{fig:displament_evolution}
\end{figure}
\begin{figure}[!t]
	\centering					
	\includegraphics[width=0.5\textwidth]{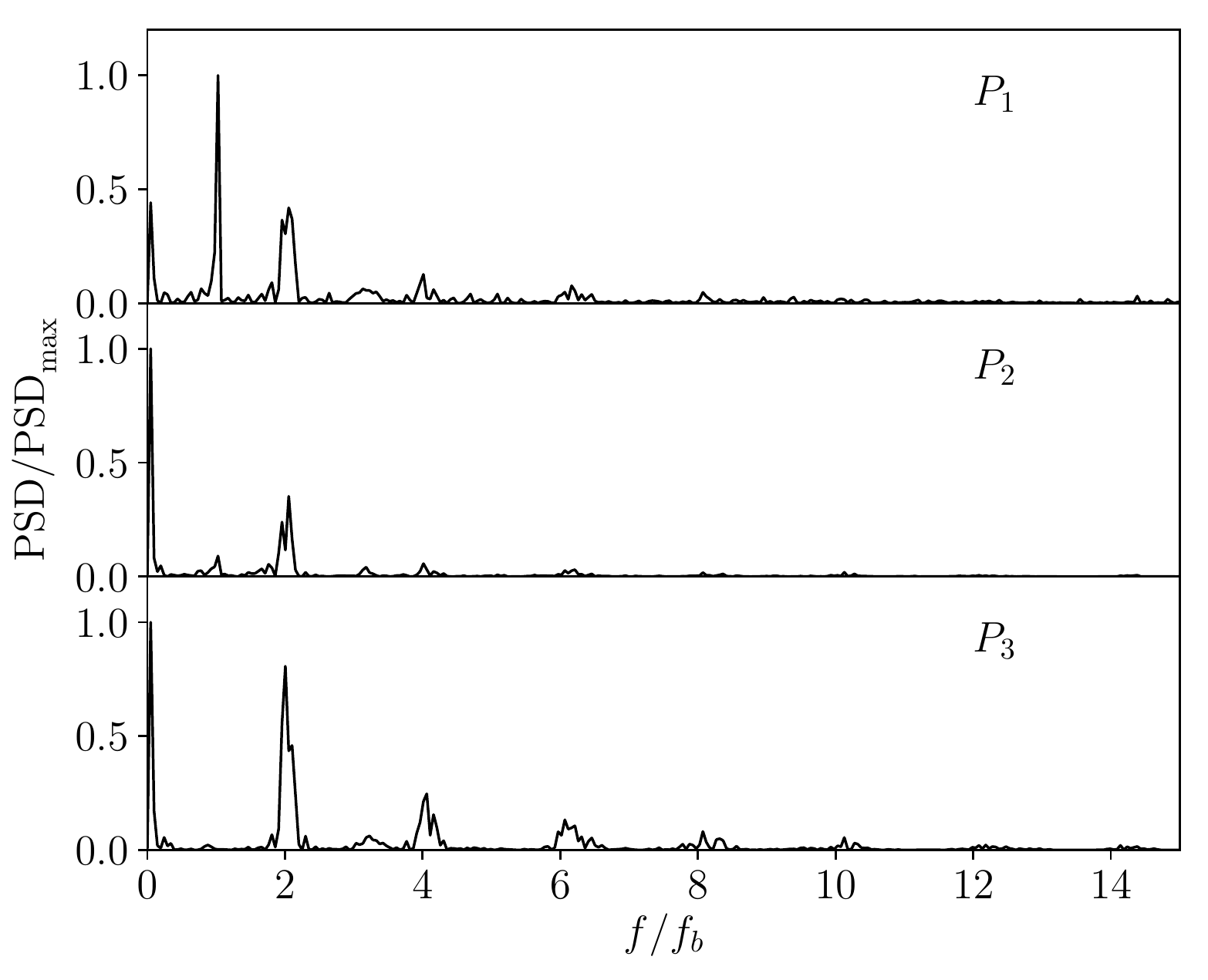}
    \caption{Spectral density of pressure for the observer points.}
	\label{fig:obs_psd}
\end{figure}

\section{Model} \label{sec:Model}
Armed with insights from our numerical simulations, we reduce the fluid-structure equations of motion to terms that are relevant when considering the flow modulation at the channel throat.  In this section, we develop an incompressible fluid-structure model that captures the dynamics of the confined flow as the beam oscillates.  In particular, we formulate the forces on the beam as a function of the channel area as modulated through the beam motion.
The coupled equations of motion, once derived and linearized, represent the modulation of the flow rates through beam motion and confinement.  
The model presented is an extension of a viscous, quasi-one-dimensional model \citep{Tosi2018,Inada1988,Inada1990,Nagakura1991,Fujita1999,Fujita2001,Fujita2007}, where here we include the effect of flow in the spanwise direction as an additional state solved through a spanwise momentum equation.  
The structure model is extended to include a moving beam boundary conditions to account for the motion of the flextensional transducer.  
The solid equations are defined in section \ref{sec:flextureMeasurements} with the appropriate coupling to the fluid pressure.  
We begin by defining the pressure in terms of flow properties and channel geometry.    

\subsection{Fluid equations of motion} \label{sec:SpanwiseLeakage}
\begin{figure}[h]
 \centering
  \begin{subfigure}{.6\textwidth}     %
    \centering
    \includegraphics[width=\textwidth]{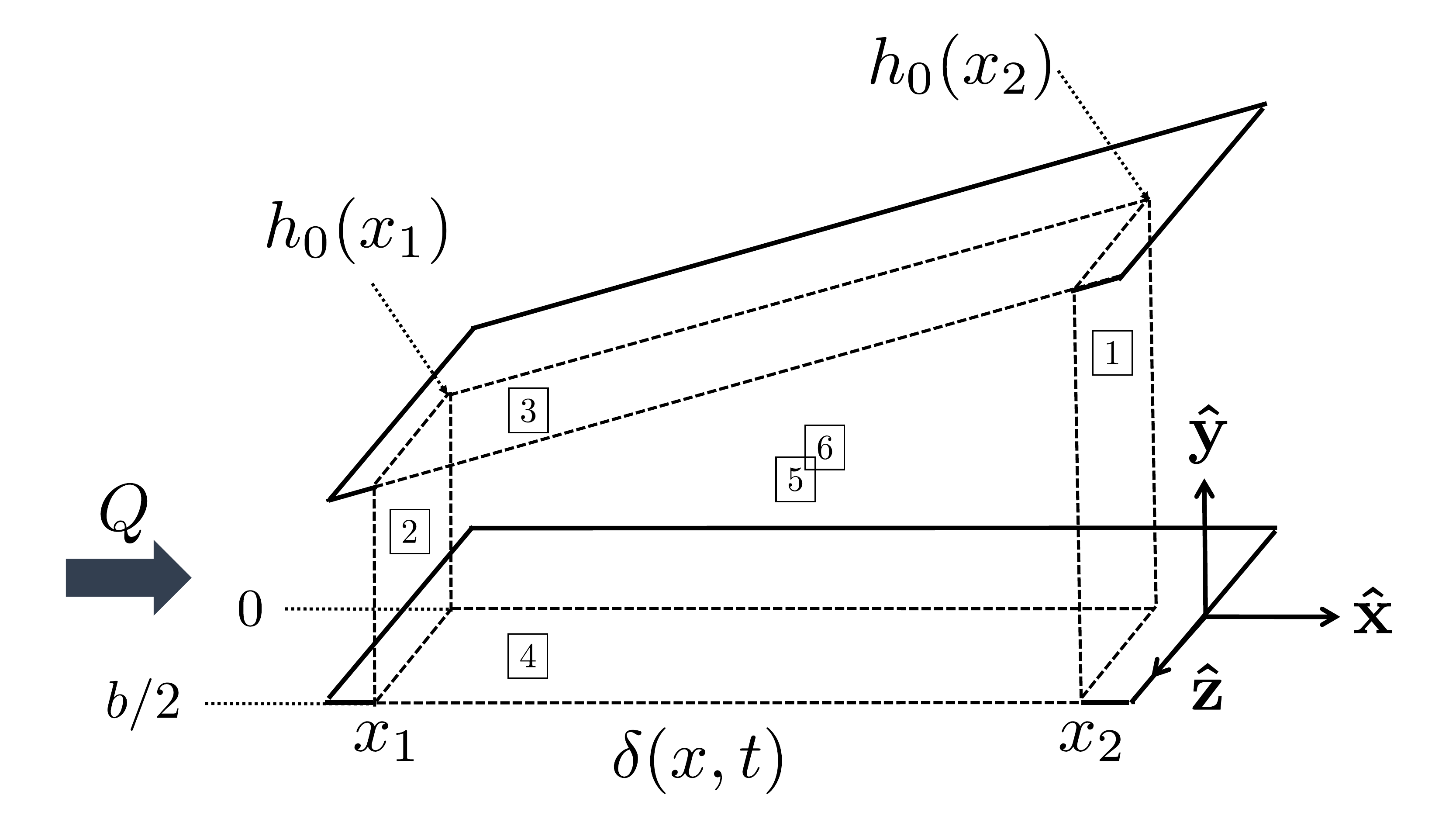}%
    \caption{Projection view of channel control volume.}
  \end{subfigure}%
~
   \begin{subfigure}{.35\textwidth}     %
    \centering
    \includegraphics[width=\textwidth]{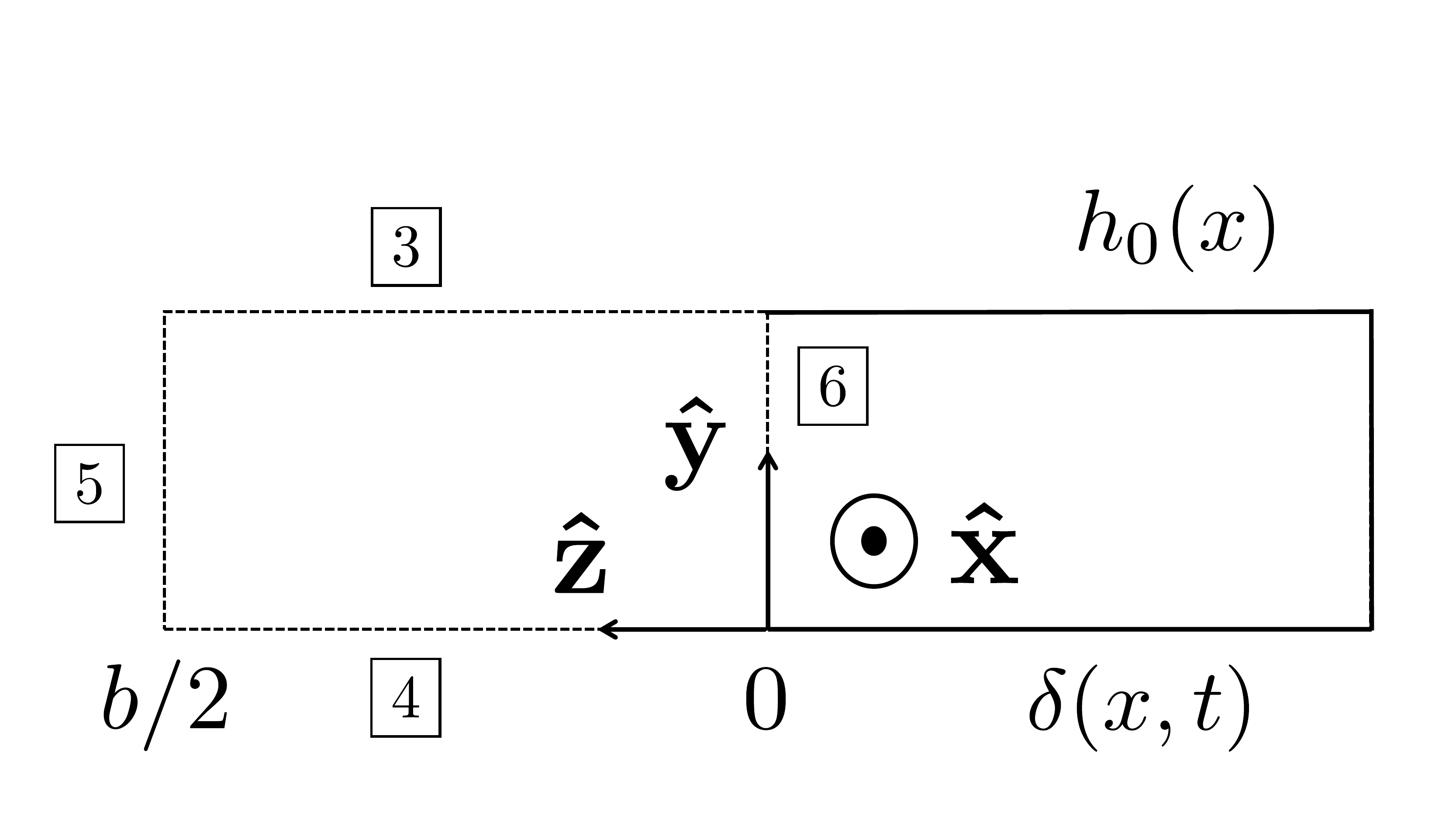}%
    \caption{Front view of channel control volume.}
  \end{subfigure}%
  \caption{Three-dimensional control volume illustration for the spanwise quasi-1D leakage flow model.  }
  \label{fig:3DControlVol}
\end{figure}

We consider a three-dimensional control volume analysis of the half-span section of the channel in order to obtain an expression that contains the $\mathbf{\hat{z}}$ momentum terms\footnote{If the total span is considered, the spanwise flow rates are canceled in momentum conservation because of the symmetry of the flow in the problem.}.  
We would like to obtain an expression of the local pressure to quantify the fluid force onto the flextensional structure.  
Figure \ref{fig:3DControlVol} illustrates the control volume boundaries as a section of the diagram in figure \ref{fig:NozzleDiffuserGeometry}, with only half of the channel demarcating the control surfaces in $\mathbf{\hat{z}}$.  The surface normal vectors are 
\begin{equation} \label{eq:CVsurfaces3D}
\left[ \mathbf{n}_1 \cdots \mathbf{n}_6  \right] = \begin{bmatrix}
    1 & -1 & -\frac{d h_0}{dx} & \frac{d \delta}{dx} & 0 & 0 \\
    0 & 0 & 1 & -1 & 0 & 0 \\
    0 & 0 & 0 & 0 & 1 & -1 \\
\end{bmatrix}.
\end{equation}
We assume the beam is rigid in $z$, such that  $\delta	= \delta(x,t)$.  Solid walls are in $\mathbf{n}_3$ and $\mathbf{n}_4$, with $\mathbf{n}_1$, $\mathbf{n}_2$, $\mathbf{n}_5$, and $\mathbf{n}_6$ representing free surfaces.

We apply mass and momentum conservation to this control volume under the simplifying assumptions of constant fluid density and a gradually-varying channel in the streamwise direction, $h_0'^2 \ll 1$ and $\delta'^2 \ll 1$, such that $\sqrt{1 + h_0'^2} \approx 1$ and  $\sqrt{1 + \delta'^2} \approx 1$ for $x \in [0,L]$. Starting with mass conservation and the three-dimensional velocity vector ${\bf u} = \left[ u, v, w \right]^T$, we have 
\begin{equation} \label{eq:MassConsInt3D_sub}
\frac{\partial }{\partial t} \left( \int_0^{b/2} \int_{\delta}^{h_0} dy dz  \right) + 
\frac{\partial}{\partial x} \left( \int_0^{b/2} \int_{\delta}^{h_0} u dy dz \right) + 
\left. \int_{\delta}^{h_0} w dy \right|_{z=b/2} = 0.
\end{equation}
Integrating in $z$ leads to 
\begin{equation} \label{eq:MassCons3D_subs1}
\frac{\partial Q_x}{\partial x} + \frac{2}{b} \left. Q_z \right|_{z = b/2} = \frac{\partial \delta}{\partial t},
\end{equation} 
where
\begin{equation} \label{eq:FlowRatedef}
Q_x = \int_{\delta}^{h_0} u dy, \
Q_z = \int_{\delta}^{h_0} w dy,
\end{equation}
are the flow rates per unit length in stream- and spanwise directions, respectively.
In a similar manner, the momentum equations in  $\mathbf{\hat{x}}$ can be obtained as, 
\begin{equation} \label{eq:MomxCons3D_subs1}
\begin{aligned}
&\frac{\partial }{\partial t} \left( \int_0^{b/2} Q_x dz \right) +
\int_0^{b/2} \frac{\partial \mathcal{N}_x}{ \partial x} dz +
\left. \mathcal{N}_{xz} \right|_{z = b/2} = \\
&-\frac{1}{\rho_f} \left\{ \int_0^{b/2} \left[
\frac{\partial}{ \partial x} \left( \int_{\delta}^{h_0} P dy \right) - 
h_0' P |_{y=h_0} + \frac{\partial \delta}{\partial x} P |_{y=\delta} \right] dz - 
F_{\mathrm{visc},x} \right\},
\end{aligned}
\end{equation}
and in $\mathbf{\hat{z}}$,
\begin{equation} \label{eq:MomzCons3D_subs1}
\begin{aligned}
\frac{\partial }{\partial t} \left( \int_0^{b/2} Q_z dz \right) + 
 \frac{\partial }{ \partial x} \left( \int_0^{b/2} \mathcal{N}_{xz} dz \right) +
&\left. \mathcal{N}_{z} \right|_{z = b/2} = \\
&\int_{\delta}^{h_0} \left( P|_{z=b/2} - P|_{z=0} \right) dy - 
F_{\mathrm{visc,z}},
\end{aligned}
\end{equation}
where the advection terms are given by
\begin{equation} \label{eq:NLAdvection}
\mathcal{N}_x = \int_{\delta}^{h_0} u^2 dy, \
\mathcal{N}_{z} = \int_{\delta}^{h_0} w^2 dy, \
\mathcal{N}_{xz} = \int_{\delta}^{h_0} u w dy.
\end{equation}
The goal of this analysis is to obtain an expression for the pressure as a function of $\delta$, $Q_x$, and $Q_z$.  To make further progress, we must find a closure for the advection terms $\mathcal{N}_x$, $\mathcal{N}_z$, and $\mathcal{N}_{xz}$, along with $F_{\mathrm{visc,x}}$ and $F_{\mathrm{visc,z}}$ in terms of those variables.  Similarly, we must also relate the local pressure values in $y$ and $z$ to the integrated pressure over the same dimensions.  

We consider the infinitessimal Navier-Stokes equations (NSE) in three dimensions non-dimensionalized similar to lubrication theory \citep{Kundu2012},
\begin{equation} \label{eq:NDgroupsLubrication}
\begin{gathered}
x^* = \frac{x}{L} , \  
y^* = \frac{y}{\bar{h}} , \  
z^* = \frac{y}{b} , \  
u^* = \frac{u}{U_c} , \  
v^* = \frac{L}{\bar{h} U_c} v , \ 
w^* = \frac{w}{U_c} \varepsilon_b , \
t^* = \frac{U_c}{L} t, \\
P^* = \frac{P}{P_{\mathrm{in}}}, \
\hat{h} = \frac{\bar{h}}{L}, \ \varepsilon_b = \frac{b}{L}, \ Re_L = \frac{\rho_f U_c L}{\mu_f}, \ \Lambda = \frac{\mu_f L U_c}{ P_{\mathrm{in}} \bar{h}^2}.
\end{gathered}
\end{equation}
$P_{\mathrm{in}}$ is a constant reference pressure upstream of the channel as defined in equation \ref{eq:PressureInOutBC}. 
For $\hat{h} \rightarrow 0$, the NSE are well approximated by
\begin{equation} \label{eq:MomConsInfin3D_ND_smallchannel}
\begin{aligned}
0 &= -\frac{1}{\Lambda} \frac{\partial P^*}{\partial x^*} + \frac{\partial^2 u^*}{\partial y^{*2}} \\
0 &= -\frac{1}{\Lambda} \frac{\partial P^*}{\partial y^*} \\
0 &= -\frac{1}{\Lambda} \frac{\partial P^*}{\partial z^*} + \varepsilon_b^2 \frac{\partial^2 w^*}{\partial y^{*2}}.
\end{aligned}
\end{equation}

To obtain a simple expression for the flow velocities, we first assume that the flow remains largely one-dimensional in $x$ for $u$ and $v$, and recover the quasi-1 dimensional parabolic profile of $u^* = u^*\left(y^*\left(x^*\right),t^*\right)$, with $v \approx 0$. 
Next we assume \textit{that any $\frac{\partial P^*}{\partial z^*}$ is due to the motion of the channel walls, and that no time-averaged net pressure gradient exists in $z$}.  It follows that, due to the symmetry of the geometry in figure \ref{fig:3DControlVol}, $\left. w^* \right|_{z = 0} = 0$ and $w^*$ odd in $z = \left[-\frac{b}{2},\frac{b}{2} \right]$ with $P^*$ symmetric in the same $z^*$ interval.  
If $\frac{\partial P^*}{\partial z^* } \neq 0$ and $P^* = P^*(x,z,t)$,  the spanwise component in equation \ref{eq:MomConsInfin3D_ND_smallchannel} can be integrated twice (with the no-slip conditions) to also recover a parabolic profile of $w^*$ in $y^*$.  
Combined with a linear function in $z^*$, the simplest that satisfies the specified symmetries, we assume a functional form for the spanwise velocity profile as 
\begin{equation} \label{eq:zvellube}
w^*(x^*,y^*,z^*,t) \sim z^* \left(\frac{\delta(x^*,t^*) }{\bar{h}}- y^* \right) \left(\frac{h_0(x^*) }{\bar{h}} - y^*\right).
\end{equation}
As a first check to the conjecture that $w^* \sim z^*$, we consult the numerical simulation results and compute the phase-averaged $w$ at $\varphi=0$ near the throat as indicated in figure \ref{fig:spanwise_schematic}. 
Figure \ref{fig:spanwiseProfile} shows that $w$ is linear to good approximation for most of the beam span. The deviation from the linear profile can be attributed
to the channel spanwise confinement and the associated vortices forming at the corners of the beam edge. Even considering those effects, the linear approximation appears to be a reasonable trade-off between accuracy and simplicity. 
With an expression for $w^*$ as a function of $y^*$, $\mathcal{N}_z$ and $F_{\mathrm{visc,z}}$ can be defined in terms of $Q_z$, 
\begin{equation}
\mathcal{N}_z  = \xi_z \frac{Q_z^2}{h_0 - \delta}, \ %
F_{\mathrm{visc,z}} =
-12 \mu_f \frac{ Q_z }{\left( h_0 - \delta \right)^2},
\end{equation}
with the latter taking the form for a Newtonian fluid. Here $\zeta_z = 6/5$.

The remaining advection terms in equation \ref{eq:NLAdvection} can be defined in terms of $Q_x$ and $Q_z$, 
\begin{equation} \label{eq:NLz_shapefactor}
\mathcal{N}_{x} = \xi_x \frac{Q_x^2}{h_0 - \delta}, \
\mathcal{N}_{xz} = \xi_{xz} \frac{Q_x Q_z}{h_0 - \delta},
\end{equation}
where $\xi_{x}$ and $\xi_{xz}$ are constant profile ``shape factor'' for axial and axial-spanwise cross-coupling velocities.
$F_{\mathrm{visc,x}}$ takes the form \citep{Tosi2018},
\begin{equation} \label{eq:FrictionTermGen}
F_{\mathrm{visc,x}} =
- \frac{f(Q_x)}{4} \frac{Q_x^2}{\left(h_0 - \delta \right)^2},
\end{equation}
where the Fanning \emph{friction factor}, $f$,
\begin{equation} \label{eq:FrictionTerm_Shimoyama}
f =
\begin{cases}
48 Re_h^{-1} & \text{if } Re_h < 1000\\
0.26 Re_h^{-0.24}& \text{if } Re_h \geq 1000
\end{cases},
\end{equation}  
with $Re_h = \hat{h} Re_L $. We model the profile shape factor as
\begin{equation} \label{eq:NonlinearTerm_general}
\mathcal{\xi}_x = \mathcal{\xi}_{xz} =
\begin{cases}
6/5 & \text{if } Re_h < 1000\\
1 & \text{if } Re_h \geq 1000
\end{cases},
\end{equation}
where the laminar value ($Re_h < 1000$) coincides with the lubrication theory frictional result \citep{Kundu2012, Tosi2018}, and the turbulent case follows from the blunted mean velocity profile in the outer region and neglects the thin inner region.

Next, we define the relation between evaluated and integrated $P$ and $Q_z$ in $y$ and $z$.  
Substituting the form in equation \ref{eq:zvellube} into the spanwise component of \ref{eq:MomConsInfin3D_ND_smallchannel}, we ascertain that $P^* \propto z^{*2}$.
We keep integrated $z$ quantities as model variables, normalizing them such that they represent the spatial average of $P$ and $Q_z$ over $z = \left[0, \frac{b}{2} \right]$, 
\begin{equation} \label{eq:Pavez}
\bar{P} = \frac{2}{b} \int_0^{b/2} P dz, \
\bar{Q}_z = \frac{2}{b} \int_0^{b/2} Q_z dz.
\end{equation}
With the definition of $w$ from $w^*$ in equations \ref{eq:NDgroupsLubrication} and \ref{eq:zvellube}, along with the definitions immediately above, we have, 
\begin{equation} \label{eq:Qeval}
\left. P \right|_{z = 0} = \frac{3}{2} \bar{P}, \
\left. Q_z \right|_{z = b/2} = 2 \bar{Q}_z.
\end{equation}
Mass conservation in equation \ref{eq:MassCons3D_subs1}, and axial and spanwise momentum, equations \ref{eq:MomxCons3D_subs1} and \ref{eq:MomzCons3D_subs1}, can now be simplified to, 
\begin{equation} \label{eq:MassCons3D_final}
\frac{\partial Q_x}{\partial x} + \frac{4}{b} \bar{Q}_z= \frac{\partial \delta}{\partial t}.
\end{equation} 
\begin{equation} \label{eq:MomxCons3D_final}
\frac{\partial Q_x  }{\partial t} +
\frac{\partial }{ \partial x} \left( \xi_x \frac{Q_x^2}{h_0 - \delta} \right) +
4 \xi_{xz} \frac{Q_x \bar{Q}_z}{b \left(h_0 - \delta \right)} = 
-\frac{1}{\rho_f}  
\frac{\partial \bar{P} }{ \partial x} \left( h_0 - \delta \right)  + F_{\mathrm{visc,x}}.
\end{equation}
\begin{equation} \label{eq:MomzCons3D_final}
\begin{aligned}
\frac{\partial \bar{Q}_z  }{\partial t}  +  \frac{\partial }{\partial x} \left(  \xi_{xz} \frac{ Q_x \bar{Q}_z  }{h_0 - \delta } \right) + &
8 \xi_z \frac{ \bar{Q}_z^2 }{b \left( h_0 - \delta \right)} = \\ &
-\frac{2 \left( h_0 - \delta \right)}{ b \rho_f}\left( P|_{z = b/2} - \frac{3}{2} \bar{P} \right) - 
\frac{12 \mu_f}{\rho_f} \frac{ \bar{Q}_z }{\left( h_0 - \delta \right)^2} 
\end{aligned}
\end{equation}

Equations \ref{eq:MassCons3D_final}, \ref{eq:MomxCons3D_final}, and \ref{eq:MomzCons3D_final} comprise the fluid equations of motion that describe the averaged spanwise local pressure in $x$ as a function of the passage shape and dynamics.   
Pressure boundary conditions are required to solve them uniquely. Based on leakage-flow instability work \citep{Nagakura1991,Inada1988,Inada1990,Tosi2018},  
\begin{equation} \label{eq:PressureInOutBC}
\begin{aligned}
\bar{P}(t)|_{x=0}  &= P_{\mathrm{in}} - 
\frac{\zeta_{\mathrm{in}}}{2} \rho_f \left[ \left( \frac{Q_x}{h_0 - \delta} \right)^2 \right]_{x=0}, \\
\bar{P}(t)|_{x=L}  &= P_{\mathrm{out}} + 
\frac{\zeta_{\mathrm{out}}}{2} \rho_f \left[ \left( \frac{Q_x}{h_0 - \delta} \right)^2 \right]_{x=L}.
\end{aligned}
\end{equation}
where $\zeta_{\mathrm{in}} \geq 1 $ and $\zeta_{\mathrm{out}} \geq 0$ are loss coefficients, and the departure from equality represents non-isentropic processes. $P_{\mathrm{in}}$ and $P_{\mathrm{out}}$ are constants.
The boundary value for $P|_{z=b/2}$ appears explicitly in equation \ref{eq:MomzCons3D_final}, and is an additional boundary condition needed for the control volume in figure \ref{fig:3DControlVol}.   
We maintain the same form to define the pressure at the edge surface $z = b/2$,
\begin{equation} \label{eq:PressureOutBCz}
P(x,t)|_{z=b/2}  = p_0(x) + 
\frac{\zeta_{\mathrm{out,z}}}{2} \rho_f \left( \frac{2 \bar{Q}_z(x,t)}{h_0 - \delta} \right)^2 .
\end{equation}
Equation \ref{eq:PressureOutBCz} states that when $\bar{Q}_z = 0$, the pressure at the boundary is the steady pressure of the two-dimensional channel $p_0$.  This is consistent with assumption that no time-averaged net pressure gradient exists in $z$, used to obtain $w^*$.  The pressure loss coefficient is $\zeta_{\mathrm{out,z}} \geq 0$, and it can be used to account for any pressure losses in the movement of the flow between top and bottom channels via surface 5 in figure \ref{fig:3DControlVol}. 

\subsection{Linearized model} \label{sec:LinearizationPressureZ}

The goal of this model is to predict the linear stability (i.e. flutter boundary) of an equilibrium beam shape $\delta_0(x)$, as a function of parameters on table \ref{tab:dimensionpars}.  We begin this process by expanding the dependent variables about their respective equilibrium values in a small parameter, $\varepsilon$, representing the amplitude of the beam displacement.  That is, we take
\begin{align*} 
\delta(x,t) & = \delta_0(x) + \varepsilon \delta_1(x,t) + \ldots \\
\bar{P}(x,t) & = p_0(x) + \varepsilon p_1(x,t)  + \ldots \\
Q_x(x,t) & = q_{x0}(x) + \varepsilon q_{x1}(x,t)  + \ldots \\
\bar{Q}_z(x,t) & = q_{z0}(x) + \varepsilon q_{z1}(x,t) + \ldots,
\end{align*}
as well as the linearized friction factor 
\begin{align*}
 \label{eq:FrictionTermLinearExp}
f(Q_x) & \approx f(q_{x0}) + (Q_x - q_{x0})  \left[ \frac{\mathrm{d} f}{\mathrm{d} Q_x} \right]_{Q_x = q_{x0}} + \ldots \\
& \approx f_0 + \varepsilon \eta  q_{x1}(x,t) + \ldots,
\end{align*}
determined from laminar and turbulent relations in equations \ref{eq:FrictionTermGen} and \ref{eq:FrictionTerm_Shimoyama}.
At zeroth order of $\varepsilon$, we obtain a differential equation describing the equilibrium beam shape
\begin{equation} \label{eq:Eps0BeamShape} 
    \frac{EI}{b} \frac{\mathrm{d}^4}{\mathrm{d} x^4} \delta_0(x) = 
p_0^{\mathrm{bot}} - p_0^{\mathrm{top}}, 
\end{equation}
with homogeneous and elastic boundary condition
\begin{equation} \label{eq:Eps0Struct_BC}
\frac{k_0}{b} \delta_0(0) = \int_0^L \left( p_0^{\mathrm{bot}} - p_0^{\mathrm{top}} \right) dx.
\end{equation}
Once again, the superscripts top and bot refer to parameters associated with $h_0^{\mathrm{top}}$ and $h_0^{\mathrm{bot}}$ as the channel shapes above and below the beam, respectively.
Substituting the expansions into equations \ref{eq:MassCons3D_final}, \ref{eq:MomxCons3D_final} and \ref{eq:MomzCons3D_final}, and applying $q_{z0}(x) = 0$, we recover the same steady pressure and flow rate equations as those in \cite{Tosi2018},
\begin{equation} \label{eq:Eps0Pressure}
p_0(x) = {P_{\mathrm{in}}} - 
{\rho}_{f}\, q_{x0}^2\, \left( \frac{ {f}_{0}\, }{4} \int_{0}^{ {x}} \frac{d  {x_2}}{{h_e\!\left( {x_2}\right)}^3} \, - 
{\xi_x}\, \int_{h_e(0)}^{ h_e(x)} \frac{d h_e}{h_e^3}  + 
\frac{ {\zeta_{ {in}}}}{2\, {h_e\!\left(0\right)}^2} \right),  
\end{equation}
\begin{equation}  \label{eq:Eps0FlowRatex}
q_{x0} =  \left(
\frac{ {P_{\mathrm{in}}} -  {P_{\mathrm{out}}}\,}{   
{\rho}_{f}\, \left[
\frac{ \zeta_{\mathrm{out}}} {2 h_e\!\left(L\right)^2} + 
\frac{ \zeta_{\mathrm{in}}} {2 h_e\!\left(0\right)^2}  -
{\xi_x}\, \left(\int_{h_e(0)}^{h_e(L)} \frac{d h_e}{h_e^3} \right) +
\frac{f_{0}}{4}\, \left(\int_{0}^{L} \frac{d  {x_2}}{{h_e\!\left( {x_2}\right)}^3} \,\right) \right]
} \right)^\frac{1}{2}, 
\end{equation}
where $h_e(x) = h_0(x) - \delta_0 (x)$ is the equilibrium channel height.
The linear order terms are
\begin{equation} \label{eq:Eps1Mass3D}
q_{x1} = \int_0^x \dot{\delta_1} dx_1 - \frac{4}{b} \int_0^x q_{z1} dx_1 + q_{x1}(0,t),
\end{equation}
\begin{equation} \label{eq:Esp1Momx3D}
\begin{aligned}
\dot{q}_{x1} + 2\xi_x q_{x0} \frac{\partial}{\partial x}  \left( \frac{q_{x1}}{h_0} \right) + 
&\frac{q_{x0}}{2 h_0^2}\left( \lambda_0 + \frac{\eta}{2}q_{x0} \right) q_{x1} =  \\
& \xi_x \frac{q_{x0}^2}{h_0^2} \frac{\partial \delta_1}{\partial x}  - \frac{3 }{\rho_f} \frac{\partial p_0 }{\partial x} \delta_1 - 
4\xi_{xz}\frac{q_{x0}}{b h_0} q_{z1} - \frac{h_0}{\rho_f} \frac{\partial p_1}{\partial x} 
\end{aligned}
\end{equation}
\begin{equation} \label{eq:Esp1Momz3D}
\dot{q}_{z1} + \xi_{xz}  q_{x0} \frac{\partial}{\partial x} \left( \frac{q_{z1}}{h_0} \right) + \frac{12 \mu }{\rho_f h_0^2} q_{z1} = \frac{h_0}{3 \rho_f} p_1.
\end{equation}
Manipulation is required to obtain an expression for $p_1$ as a function of $\delta_1$, $q_{z1}$, and their derivatives.  Though it is not useful to show the full form of such an expression because of its length and complexity, the following are the steps carried out in the MATLAB symbolic engine to obtain it: first we differentiate in $x$ equation \ref{eq:Esp1Momx3D}, then substitute equation \ref{eq:Eps1Mass3D} into that result.  Next, we solve equation \ref{eq:Esp1Momz3D} for $\dot{q}_{z1}$ and substitute the resulting expression into the previous result for the combined set of equations.  We can then separate the pressure dependent terms as,
\begin{equation} \label{eq:Eps1LinearPressureODE3D}
\frac{\partial^2 p_1}{\partial x^2} + \left( \frac{h_0'}{h_0} \right) \frac{\partial p_1}{\partial x} - \frac{12}{b^2} p_1 = r(x,t),
\end{equation}
where we have an inhomogeneous differential equation for $p_1$ with the right-hand-side $r(x,t)$ as a forcing term containing $\delta_1$ and its derivatives, along with $q_{z1}$ and its derivatives.  
Equation \ref{eq:Eps1LinearPressureODE3D} cannot be solved analytically for arbitrary forms of $h_0$.  Two solvable forms of $h_0$ are for constant and linear channels.  
For each of those cases, equation \ref{eq:Eps1LinearPressureODE3D} can be solved with variation of parameters. 
The fundamental solutions are found by solving the homogeneous problem ($r(x,t) = 0$), then convolved in the variation of parameters integral to obtain the particular solution.  
Respective coefficients are found by equating the linearly superimposed homogenous and particular pressure solutions to the linearized pressure boundary conditions at $x = 0$ and $x = L$,
\begin{equation} \label{eq:Eps1PressureInBC}
p_1(0,t) =  
\frac{2  \left( {P_{\mathrm{in}}}\, - p_{0}\!\left(0\right)\, \right) }{h_e\!\left(0\right)} \delta_{1}\!\left(0,t\right) -
{\zeta_{\mathrm{in}}}\, \frac{{\rho}_{f}\, q_{x0}\, }{{h_e\!\left(0\right)}^2} q_{x1}\!\left(0,t\right)
\end{equation}
\begin{equation} \label{eq:Eps1PressureOutBC}
p_1(L,t) =
\frac{2\,  \left( {P_{\mathrm{out}}}\, -  p_{0}\!\left(L\right)\, \right)}{{h_e\!\left(L\right)}} \delta_{1}\!\left(L,t\right) +  
{\zeta_{\mathrm{out}}}\, \frac{ {\rho}_{f}\, q_{x0}\, }{{h_e\!\left(L\right)}^2} q_{x1}\!\left(L,t\right).
\end{equation}
Fundamental solutions for a constant channel are two real exponential functions, while those of a linear channel are a set of modified Bessel functions.

Once $p_1$ is defined, two other relations are needed to complete the fluid system of equations.  First, the time evolution of the boundary forcing flow rate $q_{x1}(0,t)$ in equation \ref{eq:Eps1Mass3D} must be defined.  
This is done by substituting $p_1$ into equation \ref{eq:Eps1LinearPressureODE3D} evaluated at $x=0$, and solving for $\dot{q}_{x1}(0,t)$ in terms of $\delta_1$, $q_{z1}$, and their derivatives.  
Lastly, the time evolution of $q_{z1}$ is obtained by substituting $p_1$ into equation \ref{eq:Esp1Momz3D} and also solving for $\dot{q}_{z1}$ in terms of $\delta_1$, $q_{z1}$, and their derivatives. 	

Next we collect and equate coefficients to linear order in $\varepsilon$ for the beam, 
\begin{equation} \label{eq:Eps1BeamShape}
\rho_s h_b\frac{\partial^2 \delta_1}{\partial t^2}  + 
\frac{EI}{b}\frac{\partial^4 \delta_1}{\partial x^4}   =
p_1^{\mathrm{bot}} - p_1^{\mathrm{top}},
\end{equation}
together with homogeneous and elastic boundary condition, 
\begin{equation} \label{eq:Eps1MovingBC}
\frac{m_0}{b} \ddot{\delta}_1(0,t) + \frac{c_0}{b} \dot{\delta}_1(0,t) + \frac{k_0}{b}  \delta_1(0,t) = 
\int_0^L \left( p_1^{\mathrm{bot}} - p_1^{\mathrm{top}} \right) dx.
\end{equation}

To numerically solve the linear system of PDEs given by equations \ref{eq:Eps1Mass3D} to \ref{eq:Eps1MovingBC}, we expand the first-order beam displacement in a series of basis functions
\begin{equation} \label{eq:delta_exp}
    \delta_1(x,t) = \sum_{i=0}^{n} a_i(t) g_i(x) 
\end{equation}
where 
\begin{equation} \label{eq:ClampedFreeBC_BasisExpansion}
g_i(x) =
\begin{cases}
1 & \text{for } i = 0 \\
\phi_i(x) & \text{for } i = [1, n]
\end{cases}, 
\end{equation}
and $\phi_i(x)$, defined in equation \ref{eq:ClampedFreeEigFun}, are solutions of the homogeneous (unforced) beam equation in the domain $x \in [0,L]$. The constant $g_0=1$ base accounts for the elastic boundary condition via equation \ref{eq:Eps1MovingBC}. 
Because $\phi_i$ does not enforce the boundary values for $q_{z1}$ at $x = 0$ and $x = L$, we seek another basis expansion that does.  Specifically, $q_{z1}|_{x = 0,L}$ are determined by equation \ref{eq:Esp1Momz3D} when evaluated at $x = 0$ and $x = L$, with pressure boundary condition at $x=0$ and $x=L$ in equations \ref{eq:Eps1PressureInBC} and \ref{eq:Eps1PressureOutBC} applied,
\begin{equation} \label{eq:Esp1Momz3D_x0}
\left. \dot{q}_{z1} \right|_{x=0,L} = 
\left. \frac{h_0}{3 \rho_f} p_1 \right|_{x=0,L} -
\xi_{xz}  \left. q_{x0} \frac{\partial}{\partial x} \left( \frac{q_{z1}}{h_0} \right) \right|_{x = 0,L} - 
\left. \frac{12 \mu }{\rho_f h_0^2} q_{z1}\right|_{x=0,L}.
\end{equation}
We use the linear superposition of solutions that satisfy the inhomogeneous boundary conditions, but homogenous equation, and those that satisfy the homogeneous boundary condition, but inhomogeneous problem to solve the full inhomogeneous boundary value problem.
A sine series expansion, truncated at $m$ terms, is chosen for the latter since homogeneous Dirichlet boundaries are present. Hence, for the expansion,
\begin{equation} \label{eq:qz1Exp}
q_{z1}(x,t) = \sum_{i=0}^{m} \tilde{q}_{i}(t) \psi_i(x) 
\end{equation}
we have,
\begin{equation} \label{eq:QzSeries}
\psi_i(x) =
\begin{cases}
\left( 1 - \frac{x}{L} \right) & \text{for } i = 0 \\
\tilde{\psi}_i(x) & \text{for } i = [1, m-1] \\
\left( \frac{x}{L} \right) & \text{for } i = m \\
\end{cases}, 
\end{equation}
where   
\begin{equation} \label{eq:PsiBasis}
\tilde{\psi}_i(x) = \sin\left( \frac{i \pi x }{L}  \right),
\end{equation} 
for $i \in \mathbb{Z} : [0,m]$.

\subsection{Fluid-structure equations for symmetric channels} \label{sec:FSIOperator3D}
The model developed includes the analytical formulation of distinct constant or linear top and bottom channel geometries.
Here we write the coupled equations for a symmetric channel relevant to the flextensional geometry in figure \ref{fig:Flowpath}.  
We would like to understand the dynamics around the equilibrium $\delta_0 = 0$, which is a solution to equations \ref{eq:Eps0BeamShape} and \ref{eq:Eps0Struct_BC} when $p_0^{\mathrm{top}} = p_0^{\mathrm{bot}}$. 
Two formulations of the structure are considered.  
For the Euler-Bernoulli (EB) beam formulation, we apply the expansion of $\delta_1$ in $g_i(x)$ and $q_{z1}$ in $\psi_i(x)$ via steps in section \ref{sec:LinearizationPressureZ} to obtain the fluid-structure coupled equations, 
\begin{equation} \label{eq:Eps1FluidStructureOpsZ}
\begin{aligned}
\sum_{i = 0}^{n} & \Big( M_{\mathrm{s} i}(x) \ddot{a}_i(t) + 
C_{\mathrm{s} i }(x) \dot{a}_i(t) +
K_{\mathrm{s} i }(x) a_i(t) \Big) = 
- 2T_{\mathrm{f}}(x) q_{x1}(0,t) \  - \\
2\sum_{i = 0}^{n} & \Big(  
M_{\mathrm{f} i}(x)  \ddot{a}_i(t) + 
C_{\mathrm{f} i}(x) \dot{a}_i(t) +
K_{\mathrm{f} i}(x)  a_i(t) \Big) - 
2\sum_{i = 0}^{m} H_{\mathrm{f} i}(x) \tilde{q}_{i}(t),
\end{aligned}
\end{equation}
with $M_{\mathrm{s} i }$, $C_{\mathrm{s} i }$, $K_{\mathrm{s} i }$ defined in the appendix by equations  \ref{eq:Eps1MassCoeff_MovingBC}, \ref{eq:Eps1DampCoeff_MovingBC}, and \ref{eq:Eps1StiffCoeff_MovingBC}, respectively. 
Coefficients $T_{\mathrm{f}}$, $M_{\mathrm{f} i}$, $C_{\mathrm{f} i}$, $K_{\mathrm{f} i }$, and $H_{\mathrm{f} i }$ are obtained through equation \ref{eq:Eps1BeamShape} for $i = [1,n]$, and equation \ref{eq:Eps1MovingBC} for $i=0$ (i.e. boundary term), both via procedures in section \ref{sec:LinearizationPressureZ} to solve for $p_1$.  
In the rigid-body (RB) beam formulation, $\delta = \delta(t)$, and only equation \ref{eq:Eps1MovingBC} for $i=0$ boundary term is considered in equation \ref{eq:Eps1FluidStructureOpsZ}.  

The dynamics of the axial boundary flow rate %
are given by
\begin{equation} \label{eq:Eps1FlowRateForcingZ}
\dot{q}_{x1}(0,t) = G_{\mathrm{q}} q_{x1}(0,t) + 
\sum_{i = 0}^{n} \Big( B_{\mathrm{q} i} \ddot{a}_i(t) +  D_{\mathrm{q} i} \dot{a}_i(t) + E_{\mathrm{q} i} a_i(t) \Big) + 
\sum_{i = 0}^{m} H_{\mathrm{q} i} \tilde{q}_{i}(t),
\end{equation}
and the spanwise boundary flow rate dynamics as
\begin{equation} \label{eq:Eps1FlowRateOpsZ}
\begin{aligned}
\sum_{i = 0}^{m}\dot{\tilde{q}}_{i}(t) \psi_i(x) = \tilde{G}_{\mathrm{q}}(x) q_{x1}(0,t)  + & 
\sum_{i = 0}^{n} \Big( \tilde{B}_{\mathrm{q} i}(x) \ddot{a}_i(t) +  \tilde{D}_{\mathrm{q} i}(x) \dot{a}_i(t) + \tilde{E}_{\mathrm{q} i}(x) a_i(t) \Big) + \\ & 
\sum_{i = 0}^{m} \tilde{H}_{\mathrm{q} i}(x) \tilde{q}_{i}(t). 
\end{aligned}
\end{equation}
Coefficients for $a_i$, $\dot{a}_i$, $q_{x1}(0,t)$, and $\tilde{q}_{i}$ in equations \ref{eq:Eps1FlowRateForcingZ} are determined following steps in section \ref{sec:LinearizationPressureZ} for equation \ref{eq:Eps1LinearPressureODE3D}. Coefficients in equation \ref{eq:Eps1FlowRateOpsZ} are produced from equation  \ref{eq:Esp1Momz3D_x0} evaluated at $x = 0$ and $x = L$ for the boundary terms at $i = 0$ and $i = m$, respectively, and from equation \ref{eq:Esp1Momz3D} for $i = [1,m-1]$.   
We obtain the semi-continuous system in time via projection of equations \ref{eq:Eps1FluidStructureOpsZ} and \ref{eq:Eps1FlowRateOpsZ} onto test functions. For the EB formulation, we write the solution vector as,
\begin{equation} \label{eq:statevector}
{\bf x} = \left[
a_0 \ \ a_1 \ \ \ldots \ \ a_n \ \ 
{\dot a}_0 \ \ {\dot a}_1 \ \ \ldots \ \ {\dot a}_n \ \ 
q_{x1}(0,t) \ \ \tilde{q}_{0} \ \ \tilde{q}_{1} \ \ \ldots \ \ \tilde{q}_{m}
\right]^T,
\end{equation}
and the resulting ODE system is,
\begin{equation} \label{eq:Eps1LinearEq}
{\dot {\bf x}} = {\bf A} {\bf x}.
\end{equation}
The entries of ${\bf A}$ and the projection test functions are given in the appendix.  The eigenvalues and eigenvectors of ${\bf A}$ are computed to determine the flutter boundary for the coupled FSI system. For the RB formulation, components of $\mathbf{x}$, $a_i$ and  $\dot{a}_i$ for $i = [1,n]$, are removed since the beam motion is driven by the boundary equation only.

\subsection{Modeling flow separation}

\begin{figure}[H]
    \centering
   \includegraphics[width=.65\textwidth]{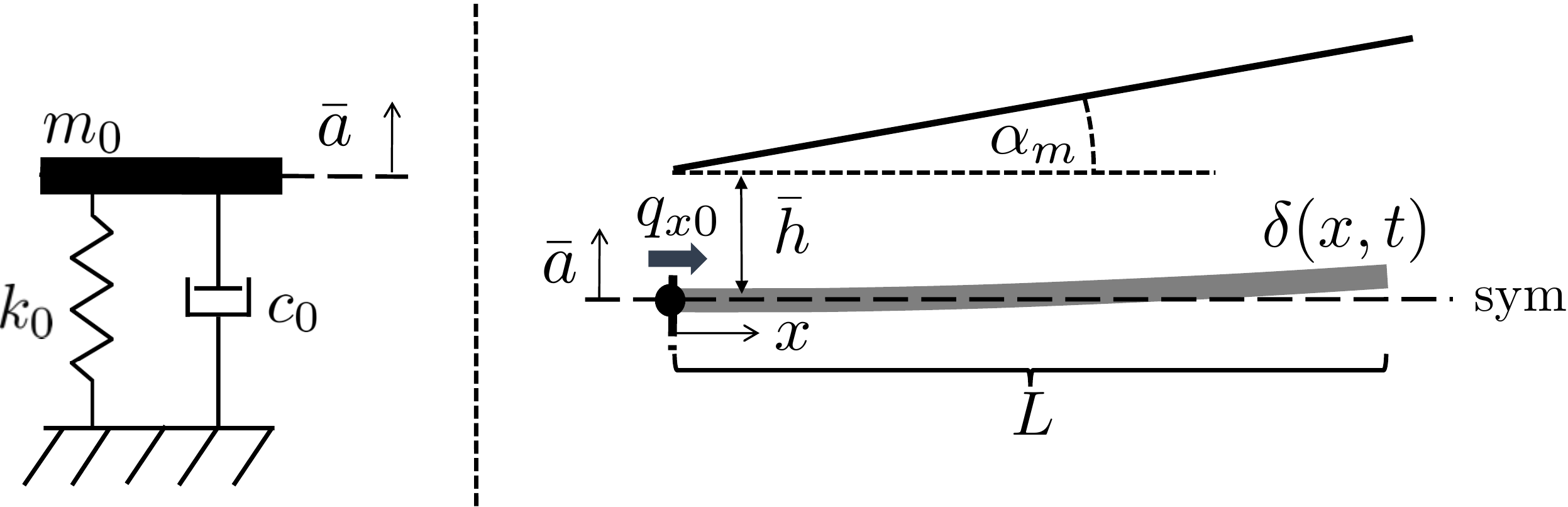}
        \caption{Illustration of spanwise quasi-1D geometry for comparison to experimental results.} 
     \label{fig:LeakageQzGeometry}
\end{figure}

Results from numerical simulations in section \ref{sec:3DDNSFEH} show the flow is separated from the top wall as it enters the diffusing part of the channel.  
In order to account for flow separation within the model framework, we conjecture that the pressure distribution over the beam surface behaves approximately as that of attached flow within plane-asymmetric diffuser of angle $\alpha_{\mathrm{m}}$.  
High Reynolds number numerical and experimental studies of plane-asymmetric diffusers suggest that flow separation from the diffusing wall happens for $\alpha_{\mathrm{m}} \gtrsim 7^{\circ}$, and is independent of Reynolds number for turbulent flows \citep{Kaltenbach1999,Lan2009, Tornblom2009, Chandavari2014}.
Hence, we solve equation \ref{eq:Eps1LinearEq} for the simplified geometry in figure \ref{fig:LeakageQzGeometry}.
To summarize, the separation bubble over the diffusing channel walls effectively serve as a secondary diffuser boundary at an effective expansion angle of $\alpha_m < \theta$. The pressure distribution on the beam surface behaves as if the flow had been attached and expanding at $\alpha_m$. At the end of the effective diffuser expansion ($x=L$), we assume that the outlet boundary pressure variation behaves as an abrupt expansion at the outlet, where $\zeta_{\mathrm{out}} = 1$.

\subsection{Model-experiment comparison} \label{sec:ModelResults}
We now use the model to assess critical parameters at the on-set of flutter.  
Critical flow rates and frequencies are calculated over integer values of $\alpha_m = [1 - 8]^{\circ} $ for the flexible Euler-Bernoulli (EB) and a rigid-body (RB) beam formulations.
As in pre-cursor numerical simulations, the EB modeled cantilever modes are quickly (or immediately) unstable once flowing in the absence of internal material damping terms.  This unphysical behavior precludes their direct comparison to experimental results (i.e. mode 2 in figure \ref{fig:SPODProcessing_Flex1}).   
However, the EB model still allows us to extract the flexure/flextensional mode where damping has been accounted and experimentally measured, and compare its critical flow rate, frequency, and corresponding shape to those observed experimentally (i.e. mode 1 in figure \ref{fig:SPODProcessing_Flex1}). Hence, only the primary flexure mode is considered for comparison in the EB formulation.  

\begin{figure}[H]
    \centering
    \begin{subfigure}[b]{0.31\textwidth}
        \centering
        \captionsetup{width=.8\linewidth}
   	    \includegraphics[width=1\textwidth]{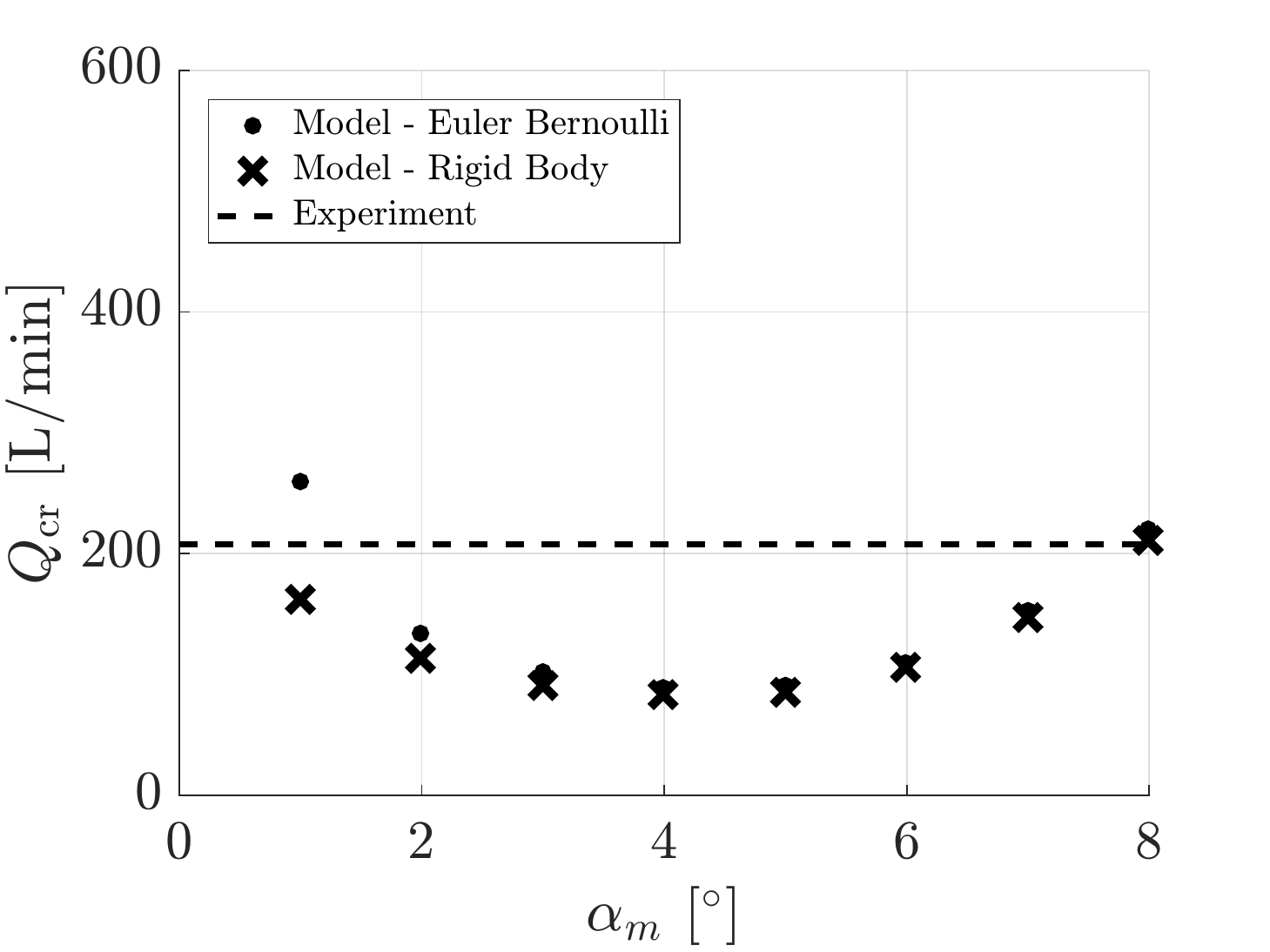}
        \caption{Flex. 1.}  
    \end{subfigure}%
    ~ 
    \begin{subfigure}[b]{0.31\textwidth}
        \centering
        \captionsetup{width=.8\linewidth}
   	    \includegraphics[width=1\textwidth]{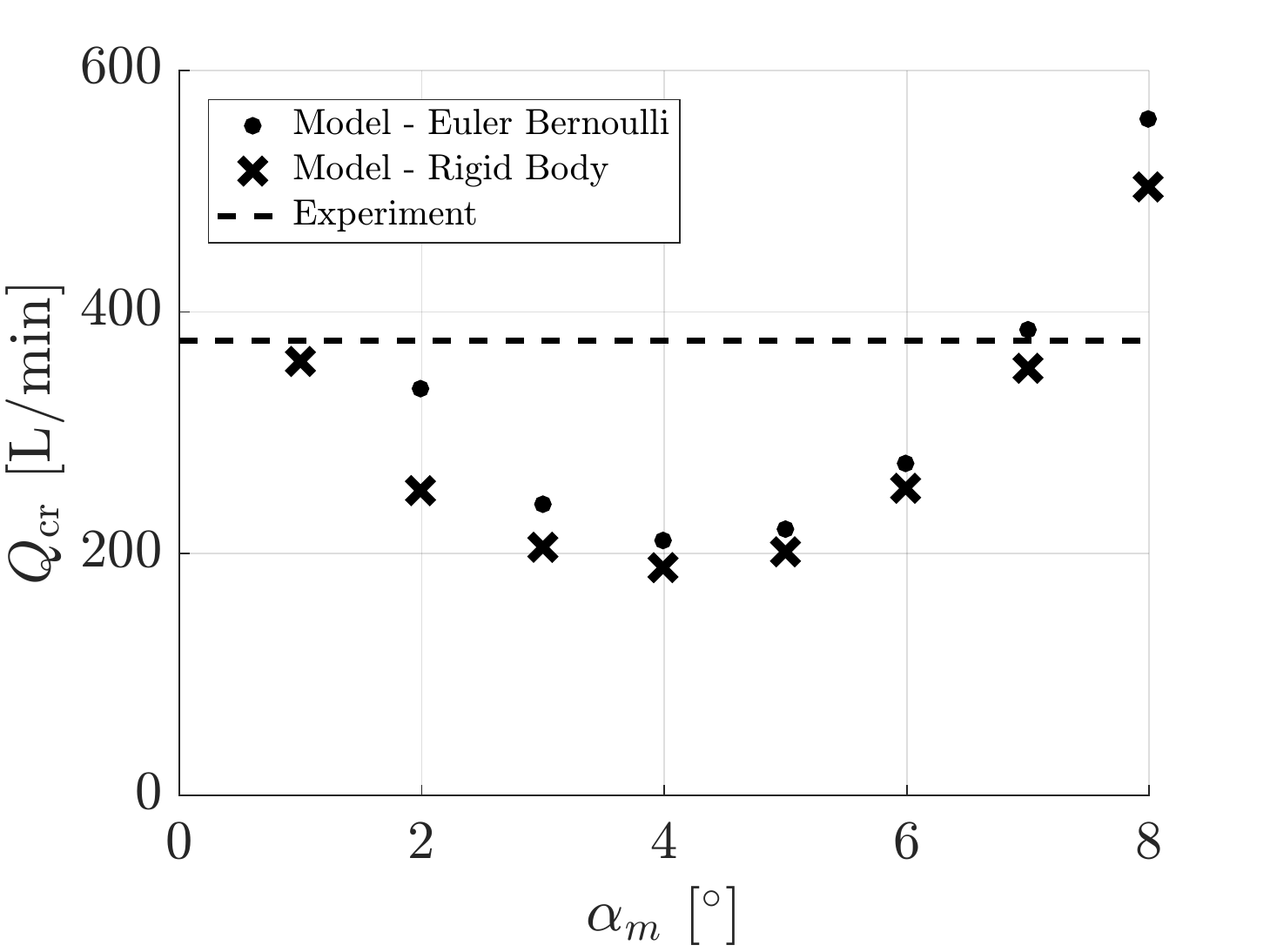} 
        \caption{Flex. 2.}  
    \end{subfigure}
     ~ 
    \begin{subfigure}[b]{0.31\textwidth}
        \centering
        \captionsetup{width=.8\linewidth}
   	    \includegraphics[width=1\textwidth]{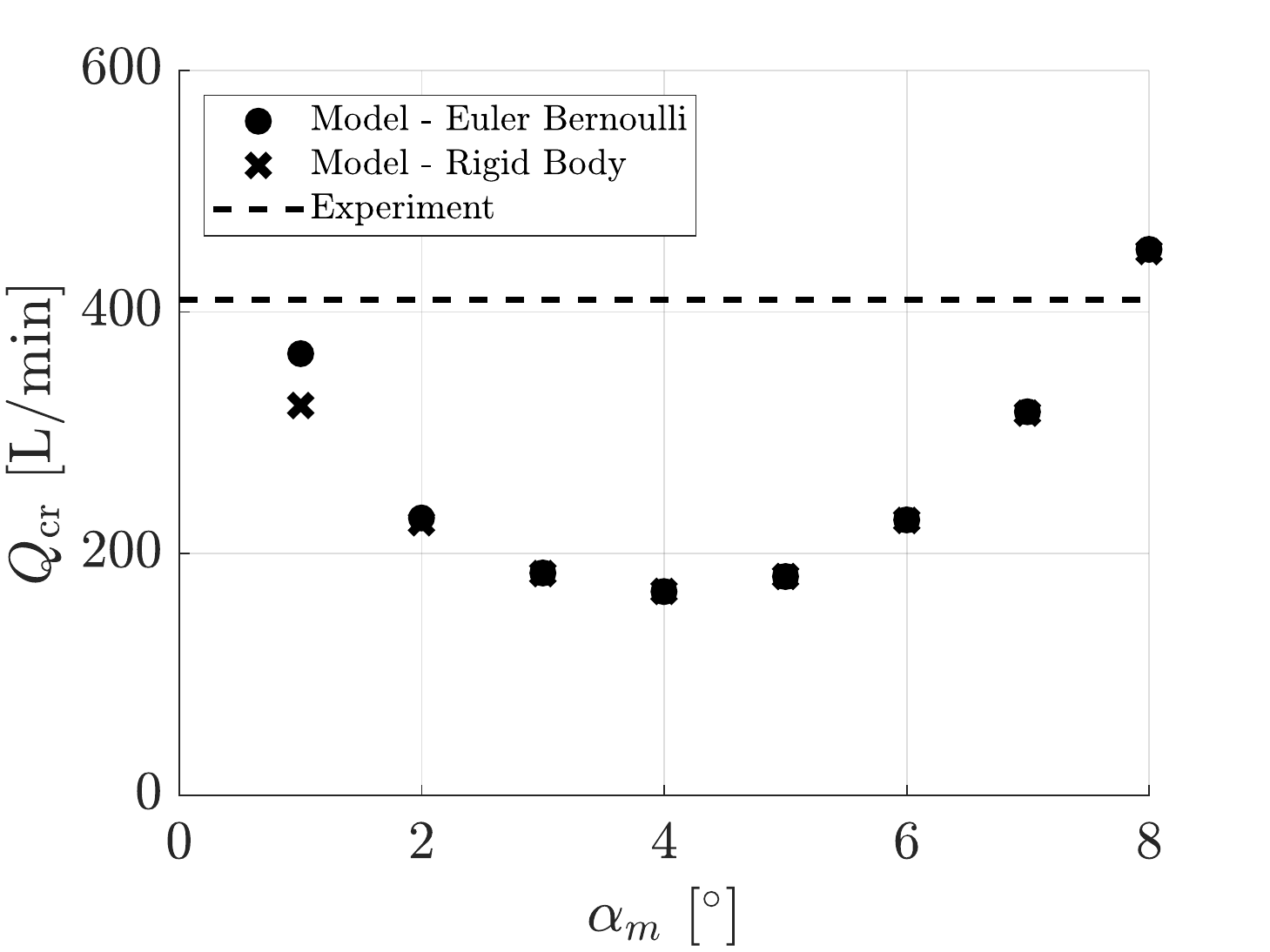} 
        \caption{Flex. 3.}  
    \end{subfigure}
    
    \caption{Flutter boundary defined by critical flow rate vs. diffuser expansion angle for all three flextensional settings.}  \label{fig:ExpModelAlpmvsQ}
\end{figure}

\begin{figure}[H]
    \centering
    \begin{subfigure}[b]{0.31\textwidth}
        \centering
        \captionsetup{width=.8\linewidth}
   	    \includegraphics[width=1\textwidth]{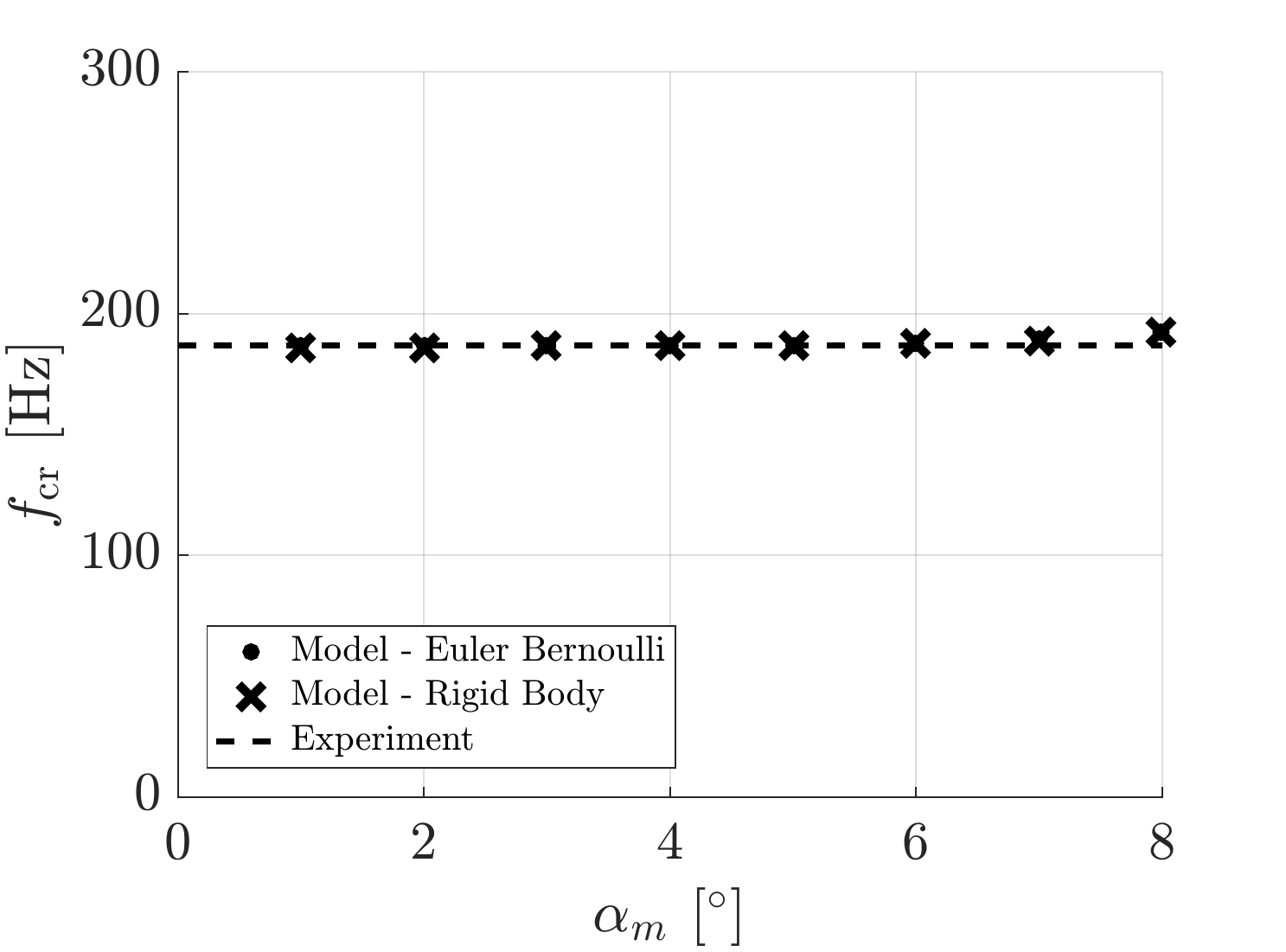}
        \caption{Flex. 1.}  
    \end{subfigure}%
    ~ 
    \begin{subfigure}[b]{0.31\textwidth}
        \centering
        \captionsetup{width=.8\linewidth}
   	    \includegraphics[width=1\textwidth]{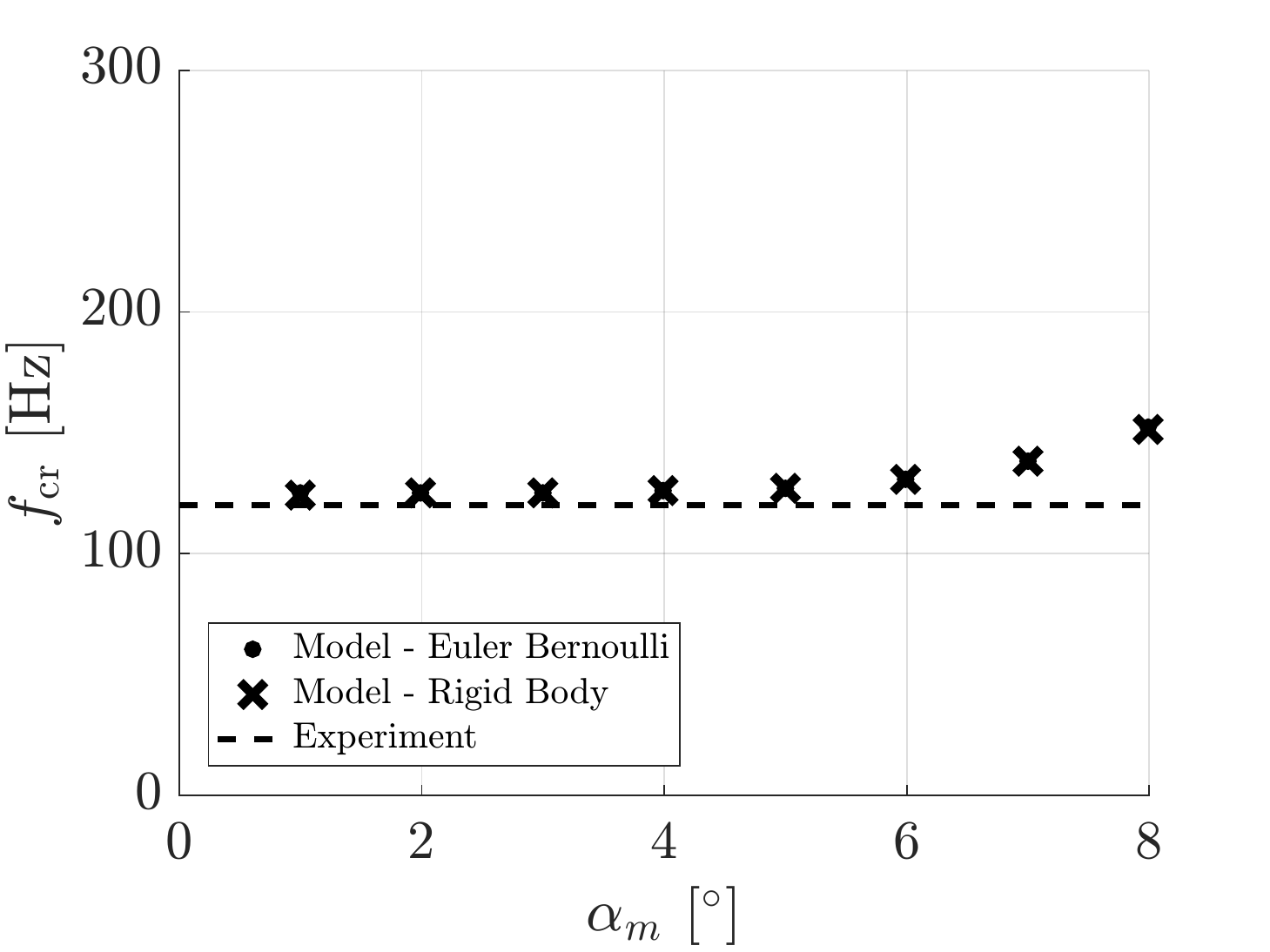}
        \caption{Flex. 2.}  
    \end{subfigure}
     ~ 
    \begin{subfigure}[b]{0.31\textwidth}
        \centering
        \captionsetup{width=.8\linewidth}
   	    \includegraphics[width=1\textwidth]{Flex3_freqvsalp}
        \caption{Flex. 3.}  
    \end{subfigure}
    
    \caption{Frequencies at critical flow rates vs. diffuser expansion angle for all three flextensional settings. }  \label{fig:ExpModelAlpmvsfreq}
\end{figure}

Beginning with critical flow rates, figure \ref{fig:ExpModelAlpmvsQ} shows their calculated values for the EB and RB model formulations 
compared with experimental values for the three flextensional settings.  
The corresponding critical frequencies are shown in figure \ref{fig:ExpModelAlpmvsfreq}. 
Both model critical flow rate trends are convex, with $\alpha_{\mathrm{m}} \approx 4^{\circ}$ representing the least stable configuration over the diffuser angles tested. 
$\alpha_{\mathrm{m}} = 1^{\circ}$ critical values are not shown for  flex. 2 as the mode was stable for tested flow rates ([0-500] L/min).
Though EB model critical flow rate values tend to be higher than those predicted by the RB model, they are close to one another at $\alpha_{\mathrm{m}} > 3^{\circ}$.
Both EB and RB model predictions match experimental critical flow rates near $\alpha_m \approx 7^{\circ}$ for all three flextensional settings.  This suggests that the critical diffuser angle for plane-asymmetric diffusers may be dictating the flow expansion and pressure distribution over the flow energy harvester channel.  
Critical frequency trends in figure \ref{fig:ExpModelAlpmvsfreq} are largely constant, with a slight increase as $\alpha_{\mathrm{m}}$ increases in both EB and RB models.  Predicted frequency values are also close between both model formulations and to those observed experimentally.  

\begin{figure}[H]
    \centering
    \begin{subfigure}[b]{0.31\textwidth}
        \centering
        \captionsetup{width=.8\linewidth}
   	    \includegraphics[width=1\textwidth]{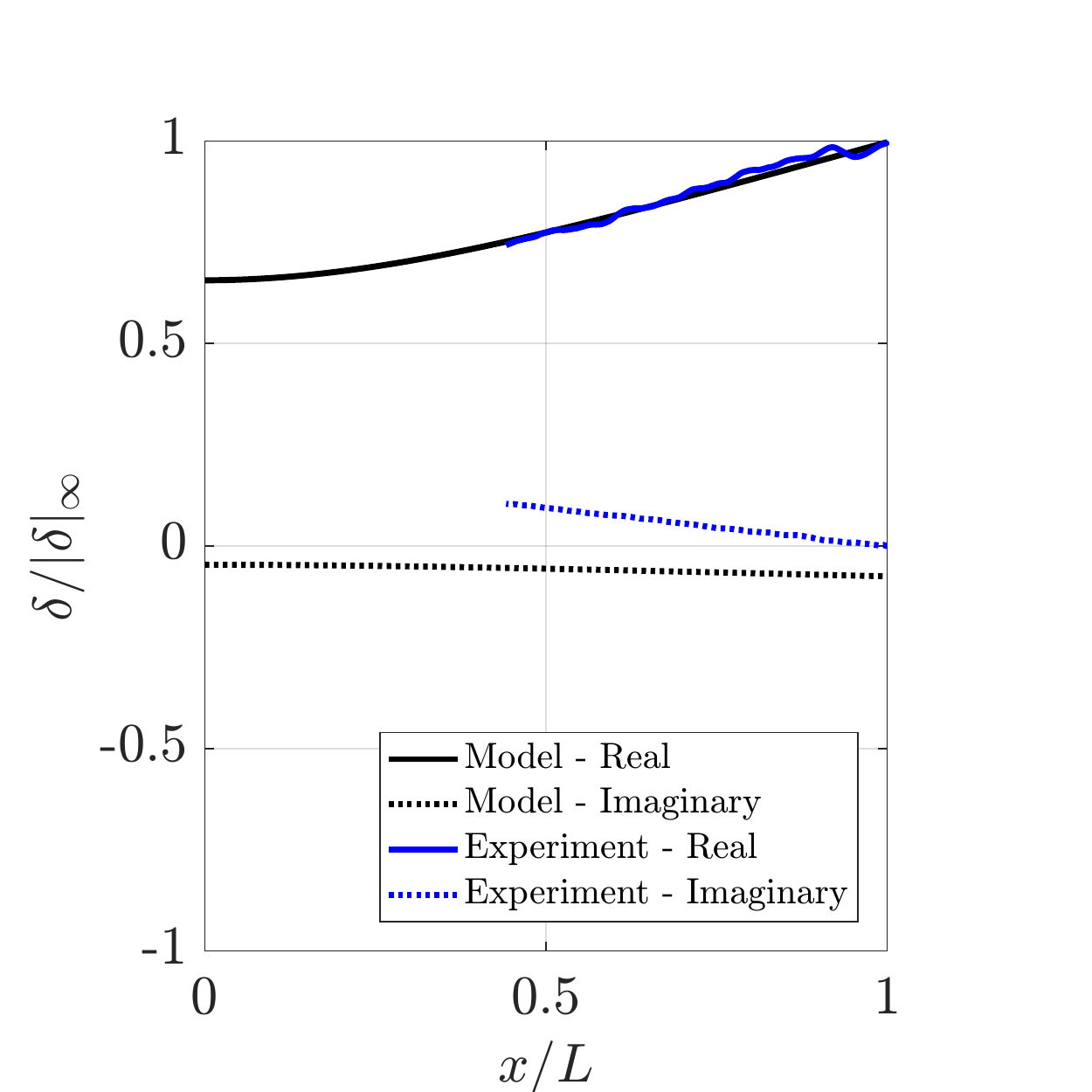}
        \caption{Flex. 1.}  
    \end{subfigure}%
    ~ 
    \begin{subfigure}[b]{0.31\textwidth}
        \centering
        \captionsetup{width=.8\linewidth}
   	    \includegraphics[width=1\textwidth]{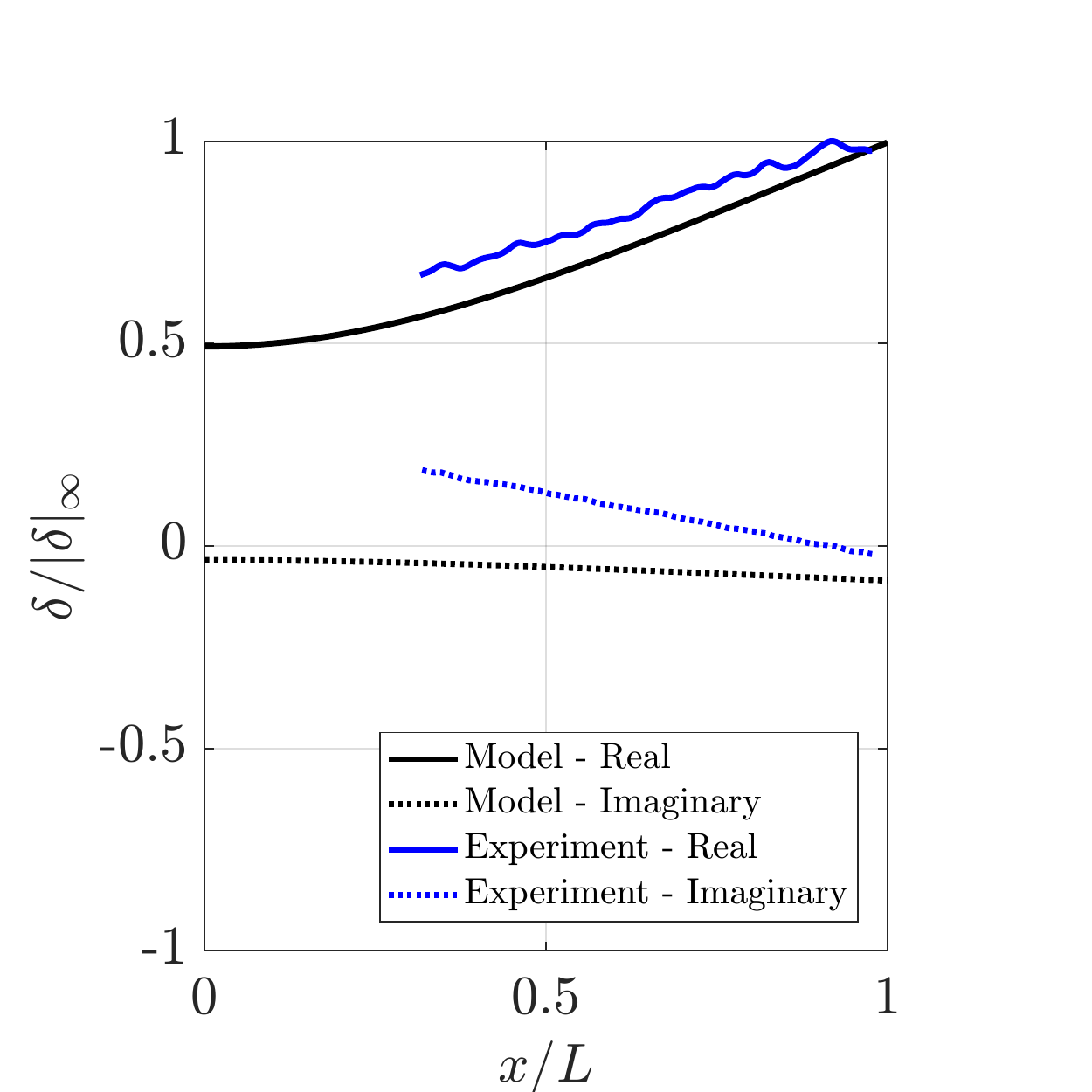}
        \caption{Flex. 2.}  
    \end{subfigure}
     ~ 
    \begin{subfigure}[b]{0.31\textwidth}
        \centering
        \captionsetup{width=.8\linewidth}
   	    \includegraphics[width=1\textwidth]{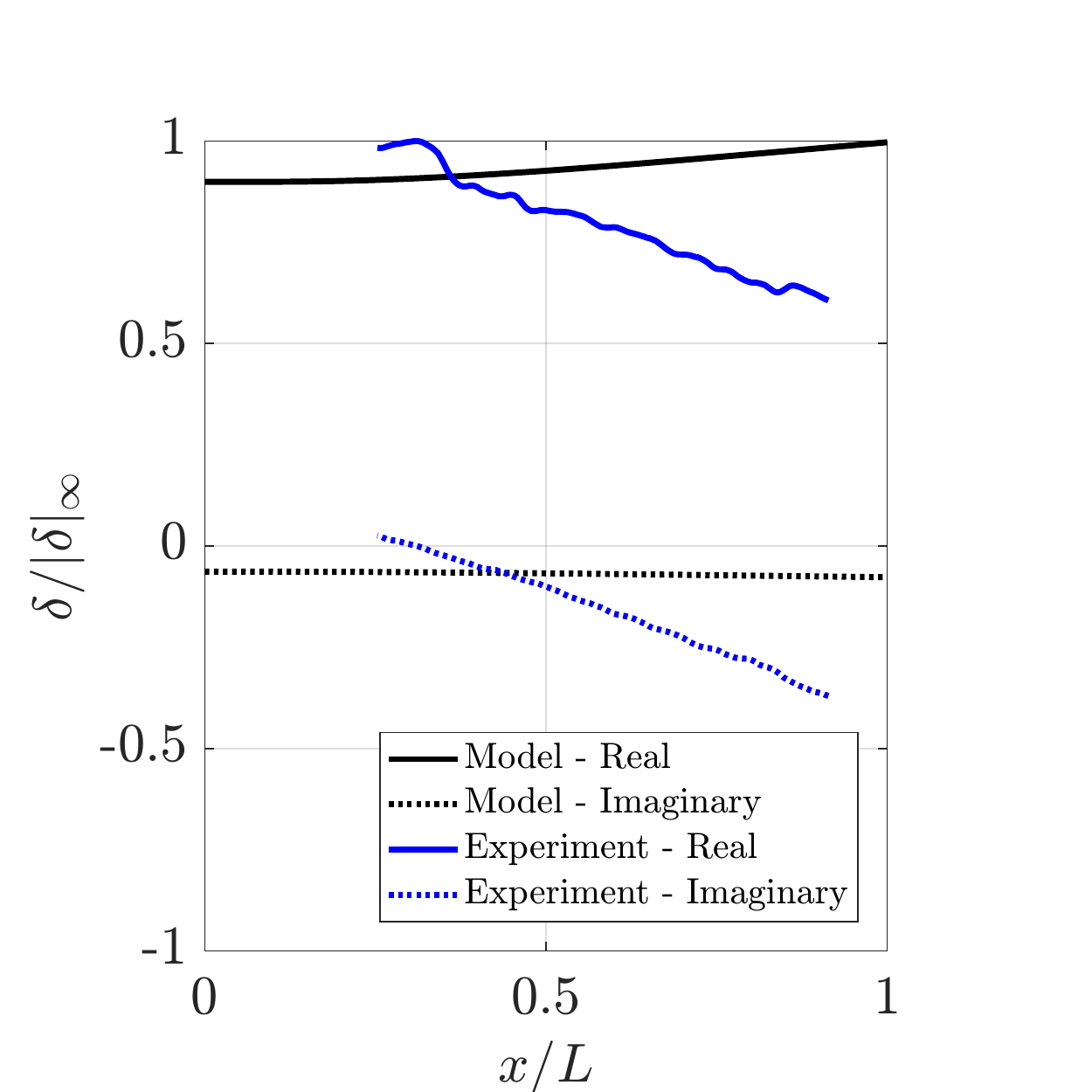}
        \caption{Flex. 3.}  
    \end{subfigure}
    
    \caption{Mode shape comparison between Euler-Bernoulli model and experimental SPOD results.}  \label{fig:FlexModesComp}
\end{figure}

Figure \ref{fig:FlexModesComp} shows the unstable EB flexure eigenvectors and experimental SPOD modes closest to the flutter bifurcation point for each of the three flextensional settings.  Predicted EB model mode shapes are similar to SPOD modes for flextensional settings 1 and 2, but captures only the rigid body motion, missing the beam shape for flex. 3.  
The primary motion seen in the modes shown is associated with the translation of the flextensional boundary condition, and are largely captured by the RB formulation.       

\section{Conclusions}
This paper explored the fluid-structure instability that drives the dynamics of a flextensional based flow-energy harvester.  In particular, we sought to elucidate the mechanisms that drive the system into flutter, which represents the transition between low and high power extraction regimes for the device.   

First, we experimentally assessed the dynamics of the flextensional based flow-energy harvester in air flow.  Experiments characterized the device's mechanical properties, then appraised the system dynamics in flowing conditions.  
Critical flow rates and frequencies were measured for three different flextensional settings, with self-sustaining oscillations reached in the flextensional mode (translational motion at the base of the beam) in all three cases tested. 
Hysteresis was observed as the flow rate direction is reversed, indicating a bi-stable region and a subcritical Hopf bifurcation at the critical point, also in all three settings.  

Numerical simulations were then carried out in three-dimensions to characterize the flow field and edify the experimentally observed flutter.  
Structure equations for a rigid beam were coupled with a lattice-Boltzmann flow solver to characterize the motion of the flextensional base, and any ensuing coherent structures within the velocity field.  
The incompressible formulation of flow equations and rigid body structure were able to replicate the critical point and the bi-stable region of the subcritical Hopf bifurcation. Results showed that the flow rate modulation due to confinement at the channel throat  largely drives the velocity fluctuations observed at downstream stations.  They pointed at a confinement based instability where flow compressibility does not play a significant role.   

Finally, an incompressible quasi-one dimensional fluid-structure model based on flow rate modulation due to confinement in the axial and spanwise directions was developed.  Flow equations were derived for small throat-to-beam length ratios and defined the pressure on the structure surface as a function of beam displacement and velocity.  Results showed that the flutter on-set is captured for a linear diffuser channel, matching experimental values near the typical separation angle for plane-asymmetric diffusers of $7^{\circ}$.  The resulting model mode shapes agreed well with experimental SPOD modes for two of the three flextensinal settings tested, with all three capturing the primary base translation motion.    

We believe that this work provides tantalizing evidence that the positive feedback between beam displacement (and velocity) and the flow modulation due to confinement is likely the dominant mechanism that drives the flutter instability within the flextensional flow energy harvester system.  
Flow compressibility and beam flexibility do not appear to significantly impact the fluid-structure dynamics on the current design.  
Agreement between model-predicted critical properties and experimental results suggest that the framework developed can be used to assess not only future flow energy harvester designs, but fluid-structure systems where small throat-to-beam length dominate the dynamics.     

\section*{Acknowledgements}
The authors would like to acknowledge the help, insight, and facilities support of Stewart Sherrit, Hyeong Jae Lee, and Yosi Bar-Cohen at the Jet Propulsion Laboratory.  %
We would also like to acknowledge Bosch Energy Research Network (BERN) grant 13.01.CC17, the Stanback Space Innovation Program, and the NASA Jet Propulsion Laboratory for their support of this research. B.D. also gratefully acknowledges the support of SNF under the Grant No. P2EZP2\_178436.

\section*{Declaration of interests}
The authors report no conflict of interest.

\appendix
\clearpage
\section{Dimension, Material, and Measurement Tables} \label{sec:MatTabs}

Table \ref{tab:FlowPathGeom} shows dimensions associated with the flextensional design highlighted in figure \ref{fig:Flowpath}.
\begin{table}[hbt!] 
\centering
\caption{Table of flow path parameter dimensions illustrated in figure \ref{fig:Flowpath}.} \label{tab:FlowPathGeom}
\begin{tabular}{ c c c  }  \hline 
\textbf{Variable}	&	\textbf{Value}	&	\textbf{Units}	\\ \hline
$	L_1	$ & 	70	&	mm	\\
$	L_2	$ & 	5	&	mm	\\
$	L_3	$ & 	19.4	&	mm	\\
$	L	$ & 	40.7	&	mm	\\
$	h_1	$ & 	9.8	&	mm	\\
$	h_2	$ & 	2.3	&	mm	\\
$	h_3	$ & 	9	&	mm	\\
$\bar{h}$ & 	0.62	&	mm	\\
$	b	$ & 	14	&	mm	\\
$	b_1	$ & 	16.5	&	mm	\\
$	h_b	$ & 	0.7	&	mm	\\
$	\theta	$ & 	19	&	deg	\\ \hline
\end{tabular} 
\end{table}

Material and electrical properties for flexure and PZT stacks are shown in tables \ref{tab:MaterialProps} and \ref{tab:PZTproperties}.
\begin{table}[H] 
\centering
\caption{Table of structural material properties \citep{MatWebAlum,PZTProp}.} \label{tab:MaterialProps}
\begin{tabular}{ c c c c }  \hline 
\textbf{Variable}	&	Aluminum	&	Steel & PZT 	\\ \hline
Density [kg/m$^3$] & 2700	&	8000  & 7500 \\
Young's modulus [GPa] & 68.9	&	193	 & 64.5 \\
Poisson's ratio [ND] & 0.33 & 0.29 & 0.31 	\\ \hline
\end{tabular} 
\end{table}

\begin{table}[H] 
\centering
\caption{Table of relevant piezoelectric stack properties.} \label{tab:PZTproperties}
\begin{tabular}{ c c c c}  \hline 
\textbf{Variable}	&	\textbf{Value}	&	\textbf{Units} & \textbf{Description}	\\ \hline
$	C_p^*	$ & 	3.6	&	$\mu$F & stack capacitance	\\
$	K_s	$ & 	2.8E7	&	N/m	 & stack stiffness in 33 mode \\
$	L_s\times W_s \times H_s	$ & 	5x5x36	&	mm & dimensions 	\\
$	R_e	$ & 	68	&	k$\Omega$ &  circuit resistor 	\\
$	\tau	$ & 	0.245	&	s	& RC time constant \\ 
$	D_{\mathrm{m}}	$ & 	40	&	$\mu$m	& maximum displacement \\ \hline
\end{tabular} 
\end{table}

Results from the static displacement-force and dynamic tests for three flextensional configurations are shown in tables \ref{tab:StiffnessFlexture} \ref{tab:FlexureRawDampValues}.  
\begin{table}[H] 
\centering
\caption{Table of mean force [N] at different displacement values for three flexure settings, with a linear best fit slope describing the stiffness $k_0$ [N/m].} \label{tab:StiffnessFlexture}
\begin{tabular}{c c c c }  \hline 
$\bar{a}$ [mm]	&	\textbf{Flex. 1} $F_a$ [N]	&	\textbf{Flex. 2} $F_a$ [N]	&	\textbf{Flex 3} $F_a$ [N] \\ \hline
0	&	0	&	0	&	0		\\
0.127	&	4.86	&	5.40	&	2.80 	\\
0.254	&	9.12	&	10.04	&	5.44 	\\
0.381	&	14.44	&	16.00	&	8.24 	\\ \hline \hline
$k_0$ [N/m]	&	3.73E4	&	4.12E4	&	2.16E4 	\\ \hline
\end{tabular}
\end{table} 
\begin{table}[H] 
\centering
\caption{Table of experimental flexure values for flexure dynamic test.} \label{tab:FlexureRawDampValues}
\begin{tabular}{ c c c c }  \hline 
\textbf{Variable}	&	\textbf{Flex. 1}	&	\textbf{Flex. 2}	&	\textbf{Flex. 3}	\\ \hline
$\omega$ - mean [rad/s]	&	1167.6	&	1423.1	&	767.7	\\
$\omega$ - STD [rad/s]	&	0.348	&	4.055	&	18.821	\\
$\zeta$ - mean [1/s]	&	-2.471	&	-7.718	&	-3.834	\\
$\zeta$ - STD [1/s]	    &	0.071	&	0.076	&	0.271	\\ \hline
\end{tabular}
\end{table} 

\section{Compressibility at Critical Flow Rates} \label{sec:CompExp}
Given $Q_{\mathrm{cr}}$ results in table \ref{tab:FlexureCriticalProperties}, it is plausible that throat velocities may reach a considerable fraction of the sound speed.  
To assess whether the critical flow rates may present compressible effects, we estimate the Mach number at the channel throat $\mathcal{M}_{\mathrm{t}}$.
Assuming the flow accelerates isentropically over the converging section of the flow path in figure \ref{fig:Flowpath} ($x \le L_2$), we take isentropic relations for stagnation (subscript $\mathrm{o}$) and throat (subscript $\mathrm{t}$) quantities,
\begin{equation} \label{eq:StagTempRatio}
\frac{T_o}{T_{t}} = 1 + \frac{\gamma_g - 1 }{2} \mathcal{M}_{\mathrm{t}}^2,
\end{equation}
\begin{equation} \label{eq:StagDensRatio}
\frac{\rho_o}{\rho_{t}} = \left( 1 + \frac{\gamma_g - 1 }{2} \mathcal{M}_{\mathrm{t}}^2 \right)^{\frac{1}{\gamma_g - 1}},
\end{equation}
\begin{equation} \label{eq:MachNum}
\mathcal{M}_{\mathrm{t}} = \frac{U_{\mathrm{cr}}} {\sqrt{\gamma_{\mathrm{g}} R_{\mathrm{g}} T_{\mathrm{t}}}},
\end{equation}
where $R_{\mathrm{g}}$ is the specific gas constant, $\gamma_{\mathrm{g}}$ is the ratio of specific heats, $T$ is the temperature, and $\rho$ is the density. With the definition of the critical mass flow rates as 
\begin{equation} \label{eq:massflowratedef}
\dot{m}_{\mathrm{cr}} = \rho_{\mathrm{STP}} Q_{\mathrm{cr}} = \rho_t A_{\mathrm{t}} U_{\mathrm{cr}},
\end{equation}
we can combine equations \ref{eq:StagTempRatio}, \ref{eq:StagDensRatio}, and \ref{eq:MachNum} to represent an implicit relation between the fluid flow properties and $\mathcal{M}_{\mathrm{t}}$ valid for $\mathcal{M}_{\mathrm{t}} \le 1$,
\begin{equation} \label{eq:MassFlowRateComp} 
\frac{\dot{m}_{\mathrm{cr}}}{\sqrt{\frac{\gamma_g R_g T_o}{1 + \frac{\gamma_g - 1 }{2} \mathcal{M}_{\mathrm{t}}^2}} \mathcal{M}_{\mathrm{t}} A_{\mathrm{t}}  } = \frac{\rho_o}{\left( 1 + \frac{\gamma_g - 1 }{2} \mathcal{M}_{\mathrm{t}}^2 \right)^{\frac{1}{\gamma_g - 1}}}.
\end{equation}
Values for $T_o = T_{\mathrm{STP}} = 295$ [K],  $\rho_o = \rho_{\mathrm{STP}} = 1.20$ [kg/m$^3$], $R_{\mathrm{g}} = 287.0$ [kg/J/K], $\gamma_{\mathrm{g}} = 1.40$ (per \citep{Moran2010}) and
\begin{equation}
    A_{t} = b_1 (2 \bar{h} + h_b) - b h_b
\end{equation}
as the throat flow area . 
Table \ref{tab:FlexureCriticalProperties} lists results of $\mathcal{M}_{\mathrm{t}}$ in the three flextensional settings.      

\begin{table}[H] 
\centering
\caption{Table of critical values for flexure settings listed.} \label{tab:FlexureCriticalProperties}
\begin{tabular}{c c c c c}  \hline 
	\textbf{Critical Properties }	&	\textbf{Flex. 1	}&	\textbf{Flex. 2}	&	\textbf{Flex. 3}	&	\textbf{Description}	\\ \hline
$	Q_{\mathrm{cr}}$ [L/min]	&	208	&	376	&	410	&	critical flow rate	\\
$	\mathcal{M}_{\mathrm{t}}$	&	0.53	&	1	&	1	&	throat Mach number est. \\
 \hline
\end{tabular}
\end{table}

The chocked flow rate is $Q_{\mathrm{ch}} \approx 267$ [L/min] for flow path with dimensions in table \ref{tab:FlowPathGeom}.  $\mathcal{M}_{\mathrm{t}}$ values suggest that flextensional settings 2 and 3 are chocked, while flextensinal 1 is not.  The possibility of having $Q_{\mathrm{cr}} > Q_{\mathrm{ch}}$ for the former two settings is due to the increase in stagnation pressure downstream of the needle valve:   
the flow meter measurements represent a mass flow rate rather than a purely volumetric one. Since the flow control (needle) valve is upstream of the flow meter and the test section, by further opening the valve,  the upstream flowing and stagnation pressures are increased, which in turn increase the density at throat and allows for the higher mass flow rate through the system. This happens despite the \textit{volumetric} flow rate remaining constant in the choked condition.

\section{Spectral Proper Orthogonal Decomposition}  \label{sec:SPOD}

To define the spectral proper orthogonal decomposition, we choose the transverse displacement $\delta$ as the primary quantity to characterize the fluid-structure system dynamics.  The inertial coordinate $x$ spans the length of the beam, with $y$ displacement at discrete $x_i$ ($i \in \mathbb{Z}: [1,p]$) and time $t_j$  ($j \in \mathbb{Z}: [1,n]$ ) as $\delta(x_i,t_j) = \delta^{(j)}_i$.
We define the data matrix $\mathbf{X}$, 
\begin{equation} \label{eq:DataMatrix}
\mathbf{X} = \begin{bmatrix}
 	\delta_1^{(1)} & \delta_1^{(j)} & \dots  & \delta_1^{(n)} \\
    \delta_i^{(1)} & \delta_i^{(j)} & \dots  & \delta_i^{(n)} \\
    \vdots & \vdots & \ddots & \vdots \\
    \delta_p^{(1)} & \delta_p^{(j)} & \dots  & \delta_p^{(n)}
\end{bmatrix} \in \mathbb{R}^{p\times n}.
\end{equation}  
The rows of $\mathbf{X}$ are measurements of points along the beam, and the columns are the time series for each point with size $\Delta t$. 

Assuming that the system is stationary and consistent with the procedure in \citep{Towne2018, Schmidt2017},  the DFT of each row of our $\mathbf{X}$ is carried out using Welch's method \citep{Welch1967}.  In the procedure, each discrete time series is segmented into 50\% overlapping blocks of size $n_f \le n$, Fourier transformed, and assembled into a Fourier domain data matrix $\tilde{\mathbf{X}}_{f_l}$ at each discrete frequency $f_l$, 
\begin{equation} \label{eq:SpecDataMatrix}
\tilde{\mathbf{X}}_{f_l} = \begin{bmatrix}
 	\tilde{\delta}_1^{(1)} & \tilde{\delta}_1^{(k)} & \dots  & \tilde{\delta}_1^{(N)} \\
    \tilde{\delta}_i^{(1)}  & \tilde{\delta}_i^{(k)} & \dots  & \tilde{\delta}_i^{(N)} \\
    \vdots  & \vdots & \ddots & \vdots \\
    \tilde{\delta}_p^{(1)} & \tilde{\delta}_p^{(k)} & \dots  & \tilde{\delta}_p^{(N)}   
\end{bmatrix}_{f_l} \in \mathbb{C}^{p \times N} ,
\end{equation}
where $l \in \mathbb{Z}: [1,n_f]$, $N \ge 1 \in \mathbb{Z}$ is the total number of blocks in Welch's method, $k \in \mathbb{Z}: [1,N]$ is a Fourier realization of the data and block number index.  Elements in $\tilde{\mathbf{X}}_{f_l}$ are 
\begin{equation} \label{eq:DFTDefinition}
\tilde{\delta}_{i}^{(k)} = 
\frac{1}{\sqrt{n_f}} \sum_{j = \frac{n_f}{2}\left( k - 1 \right) + 1}^{\frac{n_f}{2}\left( k + 1 \right) } 
\delta_{i}^{(j)} \text{e} ^{-2 \pi \sqrt{-1}  \left( l - 1 \right) \frac{j - 1}{ n_f } },
\end{equation}
for a rectangular windowing function, and discrete frequencies 
\begin{equation} \label{eq:DFTFrequenciesDefinition}
f_l = 
\begin{cases}
\frac{l-1}{n_f \Delta t} & \text{for } l \le n_f/2 \\
\frac{l-1-n_f}{n_f \Delta t} & \text{for } l > n_f/2
\end{cases}.
\end{equation}
We build the cross-spectral density matrix at each $f_l$, 
\begin{equation} \label{eq:CrossSpectralDensityMatrix}
\tilde{\mathbf{S}}_{f_l} = \frac{\Delta t}{n_f N} \tilde{\mathbf{X}}_{f_l} \tilde{\mathbf{X}}_{f_l}^* \in \mathbb{C}^{p \times p},
\end{equation}
where $\tilde{\mathbf{X}}_{f_l}^*$ is the conjugate transpose of $\tilde{\mathbf{X}}_{f_l}$ and $\Delta t$ is the time increment for the series.  $\tilde{\mathbf{S}}_{f_l}$ is Hermitian and represents the cross-correlation of measurement $i$ Fourier coefficients with all other measurements, averaged over all realizations. We can eigendecompose $\tilde{\mathbf{S}}_l$,  
\begin{equation} \label{eq:PODDecompositionS}
\tilde{\mathbf{S}}_l = \hat{\mathbf{U}}_l \bm{ \Sigma}_l \hat{\mathbf{U}}_l^*
\end{equation}
where $\hat{\mathbf{U}}_l$ is unitary (along with its conjugate transpose $\hat{\mathbf{U}}_l^*$) and its columns $(\hat{\mathbf{u}}_i)_l$ are orthonormal eigenvectors of $\tilde{\mathbf{S}}_l$.  $\bm{\Sigma}_l \in \mathbb{R}^{p \times  p}$ is a diagonal matrix with its entries as the eigenvalues $(\sigma_i)_l$ in descending order. $(\sigma_i)_l$ can be interpreted as the amount of energy its pair $(\hat{\mathbf{u}}_i)_l$ contains at $f_l$.  The cross-spectral density at each $f_l$ is tensor invariant tr$( \hat{\mathbf{S}}_l) = \text{tr} \left( \bm{\Sigma}_l \right)$, and represents the total energy at each frequency. The fraction of energy each mode contains is 
\begin{equation} \label{eq:SPODModeEnergy}
\left(\hat{\sigma}_i\right)_l = \frac{ \left( \sigma_i \right)_l }{\text{tr}( \hat{\bm{\Sigma}}_l)}.  
\end{equation}

The system may be reduced further if a single $\left(\hat{\sigma}_i\right)_l$, $(\hat{\mathbf{u}}_i)_l$ pair contains most of the energy at these peak frequencies. In systems where both holds true, it is often useful to understand the dynamics of these predominant modes. Frequencies where $\text{tr}(\hat{\bm{\Sigma}}_l)$  peaks indicate periodic behavior, but do not discern between periodic oscillations characteristic of a limit-cycle, or intermittent periodic behavior associated with a stochastically forced under-damped system. However, the SPOD modes provide a means to filter the original time domain data and discern those states exactly. In \citep{Schmidt2017b} first explored this by projecting time domain pressure data onto the leading SPOD modes to find intermittent behavior of noise in a turbulent jet.  Here, we would like to do the same by projecting the time domain beam displacement data onto the leading SPOD beam shapes.  

Suppose the system has $m < n_f$ peak frequencies in $\text{tr}(\hat{\bm{\Sigma}}_l)$. To explore the time behavior of the most energetic modes at each peak frequency, we build a basis, 
\begin{equation} \label{eq:ProjEqBasis}
\hat{\bm{\Phi}} = \big[ (\hat{\mathbf{u}}_1)_{1}, \cdots, (\hat{\mathbf{u}}_1)_{m} \big] \in \mathbb{C}^{ p \times m},
\end{equation}
where subscript 1 in $\hat{\mathbf{u}}_1$ indicates the leading mode.  We would like to approximate the time domain data $\mathbf{X}$ as 
\begin{equation} \label{eq:SPODDataApprox}
\mathbf{X} \approx \hat{\bm{\Phi}}\mathbf{A}
\end{equation}
where $\mathbf{A}$ is the matrix with coefficients of each basis (rows) in $\hat{\bm{\Phi}}$ over time (columns).  To solve for $\mathbf{A}$,   
\begin{equation} \label{eq:ProjEq}
\mathbf{A} =  \left( \hat{\bm{\Phi}}^* \hat{\bm{\Phi}} \right)^{-1} \hat{\bm{\Phi}}^* \mathbf{X}.
\end{equation} 
where $\hat{\bm{\Phi}}^*$ is the conjugate transpose of $\hat{\bm{\Phi}}$.  The columns of  $\hat{\bm{\Phi}}$ are not orthogonal, and $\left( \hat{\bm{\Phi}}^* \hat{\bm{\Phi}} \right)^{-1}$ accounts for the cross-coupling between the modes.  By construction, modes are orthonormal within a single frequency, but not across frequencies when only considering the spatial norm\footnote{Modes across frequencies are orthogonal in the temporal sense.  However, if the spatial modes are considered in the projection framework here, they are not orthogonal in that the norm $(\mathbf{\hat{u}}_1)_i^* (\mathbf{\hat{u}}_1)_j \neq 0$ for $i \neq j$. }.  

The map between $\mathbf{A}$ and $\mathbf{X}$ is, in essence, a spatial filter that when applied to the time-domain data elucidates how each shape $\hat{\mathbf{u}}_1$ behaves in time. With $\mathbf{X}$ built as transverse displacement $\delta_p^{(j)}$, each basis in $\hat{\bm{\Phi}}$ represents a beam mode shape and the columns of $\mathbf{A}$ their amplitudes at a particular instance in time. 

Since $\mathbf{A}$ represents beam displacement over time, the velocity of each shape can be defined as $\frac{\mathrm{d} \mathbf{A}}{\mathrm{d} t}$ and estimated through a discrete time derivative for the data set.  We can access a two-dimensional phase-portrait of each mode, and discern their individual dynamics: periodic orbits will be closed orbits (donut shape), while amplifier states as points clumped around the origin, as the mode is perturbed stochastically, but decays back to its equilibrium.  

\section{Euler-Bernoulli Beam Fundamental Frequency} \label{sec:EBbeamFreq}

From classical Euler-Bernoulli beam theory, we can calculate the theoretical clamped-free beam frequencies as 
\begin{equation}
  f_i = \frac{\left( \beta_i L \right)^2}{2 \pi L^2} \sqrt{\frac{E I}{\rho_s b h_b}}.
\end{equation}
$I$ is the square cross-section moment of inertia for the beam in three dimensions, 
\begin{equation} \label{eq:MomofIntertiaI}
I = \frac{h_b^3 b}{12}.
\end{equation}
The eigenfunctions $\phi_k$, $k \in \mathbb{Z}:[1,\infty]$, when subject to the clamped-free boundary conditions, are
\begin{equation} \label{eq:ClampedFreeEigFun}
\phi_k(x)=\cosh\!\left(\beta_k\, x\right) - \cos\!\left(\beta_k\, x\right) + \left[ 
\frac{ \cos\!\left(\beta_k\,  {L}\right) + \cosh\!\left(\beta_k\,  {L}\right)}{\sin\!\left(\beta_k\,  {L}\right) + \sinh\!\left(\beta_k\,  {L}\right)} \right] \Big( \sin\!\left(\beta_k\, x\right)\, - \sinh\!\left(\beta_k\, x \right)\, \Big),
\end{equation}
with characteristic equation
\begin{equation} \label{eq:ClampedFreeCharEq}
\cosh \left(\beta_k L \right) \cos \left(\beta_k L \right) + 1 = 0.
\end{equation}
The first six corresponding eigenvalues are listed in table \ref{tab:BLvecvalues}

\begin{table}[h] 
\centering
\caption{Table of solutions to the characteristic equation for clamped-free Euler-Bernoulli Beam.} \label{tab:BLvecvalues}
\begin{tabular}{c  c  c  c  c c}  \hline 
$\beta_1 L$ &$\beta_2 L$ & $\beta_3 L$ & $\beta_4 L$ & $\beta_5 L$ & $\beta_6 L$  \\ 
\hline
 1.8751 &    4.6941 &    7.8548 &   10.9955&   14.1372   & 17.2788 
  \\ \hline
\end{tabular} 
\end{table}

\section{Fluid-structure coefficients} \label{sec:FSIAppendix}

\begin{equation} \label{eq:Eps1FSISystem}
\mathbf{A}
= \mathbf{
\begin{bmatrix}
    0 & \mathbf{1} & 0 & 0 \\
    M^{-1} K & M^{-1} C & M^{-1} T & M^{-1} H  
    \\
    E_{\mathrm{q}} + B_{\mathrm{q}} (M^{-1} K) & D_{\mathrm{q}} + B_{\mathrm{q}} (M^{-1} C) & G_{\mathrm{q}} + B_{\mathrm{q}} (M^{-1} T ) & H_{\mathrm{q}} + B_{\mathrm{q}} (M^{-1} H )  
    \\
    \tilde{E}_{\mathrm{q}} + \tilde{B}_{\mathrm{q}} (M^{-1} K) & \tilde{D}_{\mathrm{q}} + \tilde{B}_{\mathrm{q}} (M^{-1} C) & \tilde{G}_{\mathrm{q}} + \tilde{B}_{\mathrm{q}} (M^{-1} T ) & \tilde{H}_{\mathrm{q}} + \tilde{B}_{\mathrm{q}} (M^{-1} H )
\end{bmatrix} }
\end{equation}
where, 
\begin{equation} \label{eq:Eps1MassCoeff_MovingBC}
{M}_{\mathrm{s} i}  =
\begin{cases}
\frac{m_0}{b} & \text{for } i = 0 \\
\rho_s h_b g_i(x) & \text{for } i = [1, n]
\end{cases},
\end{equation}

\begin{equation} \label{eq:Eps1DampCoeff_MovingBC}
{C}_{\mathrm{s} i} =
\begin{cases}
\frac{c_0}{b} & \text{for } i = 0 \\
0 & \text{for } i = [1, n]
\end{cases},
\end{equation}

\begin{equation} \label{eq:Eps1StiffCoeff_MovingBC}
{K}_{\mathrm{s} i} =
\begin{cases}
\frac{k_0}{b} & \text{for } i = 0 \\
\frac{E I}{b} \frac{\mathrm{d}^4}{\mathrm{d} x^4} g_i(x) & \text{for } i = [1, n]
\end{cases}.
\end{equation}
and 
\begin{equation} \label{eq:Eps1MCKCoeffs}
\begin{aligned}
M_{ji} &=  \int_0^L\left(  M_{\mathrm{s} i}(x) + 2M_{\mathrm{f} i}(x) \right) g'_j(x) dx, \ 
C_{ji} =  -\int_0^L\left(  C_{\mathrm{s} i}(x) + 2C_{\mathrm{f} i}(x) \right) g'_j(x) dx, \\ 
K_{ji} &= -\int_0^L\left(  K_{\mathrm{s} i}(x) + 2K_{\mathrm{f} i}(x) \right) g'_j(x) dx,  \
\end{aligned}
\end{equation}
exist in $\mathbb{R}^{n+1 \ \times \ n+1}$,  
\begin{equation} \label{eq:Eps1HCoeffs}
T_j = -2\int_0^L T_{\mathrm{f}}(x)  g'_j(x) dx, \  
\end{equation}
exists in $\mathbb{R}^{n+1 \ \times \ 1}$, and
\begin{equation} \label{eq:Eps1TCoeffs}
H_{ji} = -2\int_0^L H_{\mathrm{f} i}(x)  g'_j(x) dx, \  
\end{equation}
exists in $\mathbb{R}^{n+1 \ \times \ m+1}$.  The test functions are 
\begin{equation} \label{eq:testfunctionsBeam}
g'_i(x) =
\begin{cases}
\delta(x) & \text{for } i = 0 \\
\phi_i(x) & \text{for } i = [1, n]
\end{cases}, 
\end{equation}
where $\delta$ is the Dirac delta function and $\phi_i$ is defined in \ref{eq:ClampedFreeEigFun}.    
Coefficients for the spanwise terms
\begin{equation} \label{eq:Eps1MCKCoeffs}
\begin{aligned}
\tilde{B}_{\mathrm{q} ji} &= N_{ji} \int_0^L  B_{\mathrm{q} i}(x)  \psi'_j(x) dx, \ 
\tilde{D}_{\mathrm{q} ji} = N_{ji} \int_0^L D_{\mathrm{q} i}(x) \psi'_j(x) dx, \\ 
\tilde{E}_{\mathrm{q} ji} &= N_{ji} \int_0^L  E_{\mathrm{q} i}(x)  \psi'_j(x) dx, \ 
\tilde{G}_{\mathrm{q} ji} = N_{ji} \int_0^L G_{\mathrm{q} i}(x) \psi'_j(x) dx, \\ 
\tilde{H}_{\mathrm{q} ji} &= N_{ji} \int_0^L  H_{\mathrm{q} i}(x)  \psi'_j(x) dx, \ 
N_{ji} = \left(\int_0^L \psi'_i(x) \psi'_j(x) dx \right)^{-1}. 
\end{aligned}
\end{equation}
 The test functions are 
\begin{equation} \label{eq:testfunctionsBeam}
\psi'_i(x) =
\begin{cases}
\delta(x) & \text{for } i = 0 \\
\tilde{\psi}_i(x) & \text{for } i = [1, m-1] \\
\delta(x-L) & \text{for } i = m \\
\end{cases}, 
\end{equation}
where $\delta$ is the Dirac delta function and $\tilde{\psi}$ is defined in equation \ref{eq:PsiBasis}.

\bibliographystyle{jfm}
\bibliography{bibliography.bib}

\end{document}